\documentclass[numberedappendix]{emulateapj}
\usepackage{amssymb,amsmath}
\def\bs{\boldsymbol}

\def\gsim{\;\rlap{\lower 2.5pt
\hbox{$\sim$}}\raise 1.5pt\hbox{$>$}\;}
\def\lsim{\;\rlap{\lower 2.5pt
\hbox{$\sim$}}\raise 1.5pt\hbox{$<$}\;}
\makeatletter
\newcommand{\vast}{\bBigg@{3}}
\newcommand{\Vast}{\bBigg@{5}}
\makeatother

\newcommand{\Rmnum}[1]{\expandafter\@slowromancap\romannumeral #1@}

\begin{document}

\title{Turbulence-Induced Relative Velocity of Dust Particles III: The Probability Distribution}

\author{Liubin Pan}
\affil{Harvard-Smithsonian Center for Astrophysics,
60 Garden St., Cambridge, MA 02138; lpan@cfa.harvard.edu}
\author{Paolo Padoan}
\affil{ICREA \& ICC, University of Barcelona, Marti i Franqu\`{e}s 1, E-08028 Barcelona, Spain; ppadoan@icc.ub.edu}
\and
\author{John Scalo}
\affil{Department of Astronomy, University of Texas, Austin, TX 78712; parrot@astro.as.utexas.edu}

\begin{abstract}
Motivated by its important role in the collisional growth of dust particles in protoplanetary disks, we investigate the probability distribution function (PDF) 
of the relative velocity of inertial particles suspended in turbulent flows. Using the simulation from our previous work, we compute the relative velocity PDF as a function of the friction timescales, $\tau_{\rm p1}$ and $\tau_{\rm p2}$, of two particles of arbitrary sizes. The friction time of particles included in the simulation ranges from  $0.1\tau_\eta$ to $54T_{\rm L}$, with $\tau_\eta$ and $T_{\rm L}$ the Kolmogorov time and the Lagrangian correlation time of the flow, respectively.  The relative velocity PDF is generically non-Gaussian, exhibiting fat tails. For a fixed value of $\tau_{\rm p1}$, the PDF is the fattest for equal-size particles ($\tau_{\rm p2} = \tau_{\rm p1}$), and becomes thinner at both $\tau_{\rm p2}<\tau_{\rm p1}$ and $\tau_{\rm p2}>\tau_{\rm p1}$. Defining $f$ as the friction time ratio of the smaller particle to the larger one, we find that, at a given $f$ in 
$ \frac{1}{2} \lsim f \lsim 1$, the PDF fatness first increases with the friction time, $\tau_{\rm p, h}$, of the larger particle, peaks at $\tau_{\rm p, h}\simeq\tau_\eta$, and then decreases as $\tau_{\rm p,h}$ increases further.  For $0\le f\lsim\frac{1}{4}$, the PDF shape becomes continuously thinner with increasing $\tau_{\rm p,h}$. 
%Interestingly, if $\tau_{\rm p, h} \simeq T_{\rm L}$, the PDF tails can be well fit by a $4/3$ stretched exponential function for any value of $f$. 
The PDF is nearly Gaussian only if $\tau_{\rm p, h}$ is sufficiently large ($\gg T_{\rm L}$). These features are successfully explained by the Pan \& Padoan model. 
Using our simulation data and some simplifying assumptions, we estimated the fractions of collisions resulting in sticking, bouncing, and fragmentation as a function of the dust size in protoplanerary disks, and argued that accounting for non-Gaussianity of the collision velocity may help further alleviate the bouncing barrier problem.

\end{abstract}

\section{Introduction}

This is the third paper of a series on turbulence-induced relative velocity of dust particles. 
The study is mainly motivated by the problem of dust particle growth and planetestimal formation 
in protoplanetary disks (e.g., Dullemond and Dominik 2005; Zsom et al.\ 2010, 2011; Birnstiel et al.\  2011; 
Windmark et al.\ 2012a, 2012b; Garaud et al.\ 2013, Testi et al.\ 2014). In the first two papers of the series 
(Pan \& Padoan 2013, Pan, Padoan \& Scalo 2014; hereafter Paper I and Paper II, respectively), 
we conducted an extensive investigation of the root-mean-square or variance of the  relative velocity of inertial particles in turbulent flows using both analytical 
and numerical approaches. In particular, we showed 
that the prediction of the Pan \& Padoan (2010) model for the rms relative velocity is in satisfactory 
agreement with the simulation data, confirming the validity of its physical picture. 
In Paper I, we also explored the collision kernel for the case of equal-size particles, 
known as the monodisperse case, and analyzed the probability distribution function (PDF) 
of the turbulence-induced relative velocity as a function of the particle inertia.  
%and eThe case of equal-size particles is usually referred to the monodisperse case.  
The main goal of the current paper is to study the relative velocity PDF in the bidisperse case for 
particles of arbitrarily different sizes.

The collision velocity of dust particles in protoplanetary disks plays an important 
role in coagulations models of particle growth, which is crucial for 
understanding the challenging problem of planetesimal formation. 
The collision velocity determines not only the collision rate, but also the collision 
outcome.    
%As described in Paper I, the probability distribution function (PDF) of the collision velocity 
%of dust particles is crucial for the modeling of the particle growth in protoplanetary disks. 
As the size grows, the particles become less sticky, and, depending 
on the particle properties and the collision energy, the collisions may lead to 
bouncing or fragmentation (Blum \& Wurm 2008, G\"uttler et al.\ 2010). Due to the stochastic 
nature of turbulence-induced relative velocity,  the collision outcomes for dust particles with 
exactly the same sizes and properties may be completely different. It is thus not sufficient 
to use an average or mean collision velocity to predict the size evolution of dust particles. 
Instead, an accurate prediction would require the PDF of the collision velocity to 
evaluate the fractions of collisions resulting in sticking, bouncing or fragmentation. 

The effect of the collision velocity PDF on particle coagulation has been considered by 
several recent studies (Windmark et al.\ 2012b, Garaud et al.\ 2013). These studies assumed a 
Maxwellian distribution for the 3D amplitude of the  collision velocity, or equivalently a Gaussian distribution for each component\footnote{For simplicity, 
we will not distinguish ``Maxwellian"  for the 3D amplitude and ``Gaussian" for one component, and refer to both as Gaussian.}.
Implementing the distribution into coagulation models, they found important differences in the prediction of the dust size evolution.
In particular, they showed that accounting for the distribution of the collision velocity 
can soften the fragmentation barrier and overcome the bouncing barrier (Windmark et al.\ 2012b). 
The maximum particle size that can be reached by 
collisional growth appears to be significantly larger than the case using only a mean collision velocity. This alleviates the so-called meter-size 
barrier for planetesimal formation via  collisional growth of dust particles. %(see, e.g., the review by Chang \& Youdin 2010).  

The assumption of a Gaussian PDF made in the above-mentioned papers is not 
justified for the collision velocity induced by turbulent motions.  For equal-size particles, 
turbulence-induced relative velocity has been found to be highly non-Gaussian, exhibiting extremely fat tails 
(Sundaram \& Collins 1997, Wang et al.\ 2000, Cate et al.\ 2004, Gustavsson et al.\ 2008, Bec et al.\ 2010, 2011, 
de Jong 2010, Gustavsson et al.\ 2012, Gustavsson \& Mehlig 2011, 2014,  Lanotte et al.\ 2011, 
Gualtieri et al.\ 2012, Hubbard 2012,  Salazar \& Collins 2012).  The earlier works showed that the PDF for 
equal-size particles may be fit by exponential or stretched exponential distributions (e.g., Sundaram \& Collins 1997, 
Wang et al.\ 2000, Cate et al.\ 2004).  A stretched exponent PDF with an index of -4/3 was 
predicted theoretically for inertial-range particles under the assumption of exactly Gaussian flow 
velocity and Kolmogorov scaling (Gustavsson et al.\ 2008, Paper I). %Paper I The applicability of theprediction  
Lanotte et al.\  (2011) found that the relative velocity for small particles of equal size shows a 
power-law PDF in the limit of small particle distance. A power-law PDF was predicted by recent theoretical 
models based on a Gaussian smooth velocity field with rapid temporal decorrelation (e.g., Gustavsson \& Mehlig 2011, 2014).

In Paper I, we conducted a systematic study for the PDF of equal-size particles as a function 
of the particle inertia, and showed that the behavior of the PDF shape was successfully 
explained by the model of Pan \& Padoan (2010, hereafter PP10).
%for the generalized acceleration and shear effects. 
In this work, we further analyze the PDF for particles of different sizes, and show that non-Gaussianity 
is a generic feature of turbulence-induced relative velocity, which should be incorporated into 
coagulation models for an accurate prediction of the dust size evolution.  

Despite extensive studies on the variance or rms of the relative velocity of different-size particles (e.g., V\"olk et al.\ 1980, 
Markiewicz et al.\ 1991, Zhou et al.\ 2001, Cuzzi \& Hogan 2003, Ormel \& Cuzzi 2007, Zaichik et al.\ 2006, 2008, Zaichik \& Alipchenkov 2009),  
the PDF in the general bidisperse case has received little investigation (see, however, Johansen et al.\ 2007). 
A recent work by Hubbard (2013) explored the PDF for particles of different sizes using a synthetic or ``model"  flow 
velocity filed. However, the results obtained from such 
an approach are clearly not accurate because the artificial velocity field does not correctly account for the 
non-Gaussianity or intermittency of turbulent flows, which does leave an imprint on the relative 
velocity of inertial particles (Paper I).  The commonly-adopted models  for the rms relative 
velocity (V\"olk et al.\ 1980, Ormel \& Cuzzi 2007) in the astronomy community are based on 
the responses of particles to turbulent eddies in Fourier space, and do not provide adequate insight for 
understanding or predicting  the distribution of the relative velocity. A physical weakness of these models was 
pointed out and discussed in PP10 and Paper I.  On the other hand, 
the PP10 model is based on the statistics of the flow velocity structures in 
real space, and we have shown in Paper I that the physical picture revealed 
by the model offers a satisfactory explanation for the PDF behavior in the case of 
equal-size particles. In Paper II, we further developed the model, and established physical 
connections between the particle relative velocity 
and the temporal and spatial velocity structures of the carrier flow. In this paper, we will
continue to use the picture of PP10 to interpret the trend of the relative velocity PDF 
shape as a function of the particle friction times in the bisdisperse case. 

%The trend of the PDF fatness with varying Stokes number pairs 
%is systematically explored and successfully explained using our physical picture for the generalized acceleration and 
%shear effects.  
%For particles of different sizes 
%the PDF shape is typically  thinner than the monodisperse case. 
%for the PDF to take a nearly Gaussian shape.
%the larger of the two particles must have a friction time much larger than the Lagrangian correlation 
%time or the large-eddy turnover time of the turbulent flow. 

We use the same simulation as in Papers I and II. It is carried out using the Pencil code 
(Brandenburg \& Dobler 2002) with a periodic $512^3$ box.  The simulated flow is driven and 
maintained by a large-scale force, $f_i$, which is generated in Fourier space using all modes 
with wave length $\ge$ half box size.  Each mode gives an independent 
contribution to the driving force.  At each time step of the simulation, 
the direction of each mode is random,  and the amplitudes of each mode in three spatial 
directions are independently drawn from a Gaussian distribution. 
This conventional method of driving produces a turbulent flow with the broadest inertial range at a given resolution and with 
the maximum degree of statistical isotropy. 
At steady state, the 1D rms  flow velocity is $u'\simeq 0.05$, 
in units of the sound speed, corresponding to a (3D) rms Mach number of 0.085. 
The Taylor Reynolds number of the flow is estimated to be $Re_{\lambda}\simeq 200$, 
and the regular Reynolds number is $Re \simeq1000$.  The integral length, $L$, of the flow is 
1/6 of the box size, and the Komlogorov scale is $\eta\simeq 0.6$ times the
size of a computational cell, so that $L \simeq 140 \eta$. The Kolmogorov scale, $\eta$, is 
computed from its definition, $\eta \equiv (\nu^3/\bar{\epsilon})^{1/4}$,  using the viscosity, $\nu$, and the average 
dissipation rate, $\bar{\epsilon}$, in the simulated flow. Using tracer particles, we estimated the Lagrangian correlation
time of the flow, $T_{\rm L}$, which is relevant  for the particle dynamics  than 
the large-eddy turnover time, $T_{\rm eddy}$($\equiv L/u'$). We found $T_{\rm L} \simeq 14.4 \tau_\eta$, where the 
Kolmogorov timescale, $\tau_\eta$ ($\equiv (\nu/ \bar{\epsilon})^{1/2}$), corresponds to the turnover time of 
the smallest eddies. {The large eddy turnover time, $T_{\rm eddy}$, is estimated to $T_{\rm eddy} = 20 \tau_\eta$.}
%is  $2 \pi$. The simulated flow is fully developed and reaches a steady state at $t_{\rm dev} \simeq 5-10 T_{\rm eddy}$. 
%The Kolmogorov timescale is estimated to be $\tau_\eta \equiv (\nu/ \bar{\epsilon})^{1/2} = 1.04$, 
The Kolmogorov velocity scale, $u_\eta$ ($\equiv (\nu \bar{\epsilon})^{1/4}$), is related to the 1D rms velocity of the flow, $u'$, 
by $u' = (Re_\lambda/\sqrt{15})^{1/2} u_\eta \simeq 7 u_\eta$.  

%Both structure functions are expected to become smooth at $\Delta \tau \ll \tau_\eta$.
%The circles in the figure correspond to the Eulerian time structure function.  
%As discussed in \S 2, the particle-flow relative velocity can be roughly estimated as the 
%flow velocity difference seen by the particle at a temporal separation of $\tau_{\rm p}$. 
%Therefore, an examination of  $\Delta { \bs u}_{\rm L} (\Delta \tau)$ would be helpful to 
%understand the relative velocity between the flow and small particles, assuming that these particles more 
%or less follow Lagrangian trajectories. 

%For a large particle with $\tau_{\rm p} \gsim T_{\rm L}$, the temporal 
%series of the flow velocity along its trajectory may be better described as Eulerian 
%(Paper I). We thus also consider the temporal velocity difference, 
%as the contribution of the acceleration term is related to the 
%temporal flow velocity difference seen by the particles. 

The simulated flow evolved 14 species of inertial particles of different sizes, 
each containing 33.6 million particles. We integrated the trajectory of each particle 
by solving the momentum equation for the particle velocity, 
\begin{equation}
\frac {d {\bs v} } {dt} = \frac {1} {\tau_{\rm p}} [{\bs u} \left( {\bs X} (t), t \right) - {\bs v}]      
\label{particlemomentum}     
\end{equation}
where ${\bs u}({\bs X}(t), t)$ is the flow velocity  at the particle position, ${\bs X}$, 
at time $t$. The friction timescale, $\tau_{\rm p}$, of the smallest particles in the 
simulation is $\simeq 0.1\tau_\eta$, while that of the largest particles is $\simeq 54 T_{\rm L}$ 
%($  \tau_{\rm p}  =  41T_{\rm eddy }$). 
Defining a Stokes number as $St \equiv \tau_{\rm p}/\tau_\eta$,  this range corresponds to $ 0.1\le St \le795$.  %The friction timescales of the 
%14 species are equally spaced, increasing by a factor of two in each successive 
%species. 
Spanning 4 orders of magnitude, this range of $\tau_{\rm p}$ covers 
all the length scales of interest in the simulated turbulent flow.  
%Initially, the particles were randomly distributed in the 
%simulation box, and each velocity component of a particle was independently drawn from a 
%uniform distribution between -0.01 and 0.01 sound speed. 
When integrating the particle trajectories, we interpolated the flow velocity inside computational cells using 
the triangular-shaped-cloud (TSC) method %, already implemented in the Pencil code 
(Johansen and Youdin 2007).  Our simulation run lasted $26 T_{\rm eddy}$ (or $35 T_{\rm L}$). 
%In Paper I, we analyzed the temporal evolution of the 1-particle rms velocity, 
%the number counts of particle pairs at given distances ($r\lsim 1\eta$), and the rms relative velocity of the pairs. 
At the end of the run, all the statistical measures reached a quasi steady-state and 
the dynamics of all particles was relaxed. %Even for the largest particles whose friction 
%time ($\tau_{\rm p}  = 41 T_{\rm eddy}$ or $54 T_{\rm L}$) is larger than the run time of the simulation, we found sufficient evidence 
%for their dynamical relaxation (Paper I). 
%As in Paper I, we use three snapshots at $t= 21.5$, 24, and 26 $T_{\rm eddy}$ 
%in our analysis for the velocity statistics of all particles.  
For our statistical analysis, we use three well-separated snapshots toward the end of the run. 

%We denote the flow velocity field and particle velocity as ${\bs u}({\bs x}, t)$ and $\bs {v}$, respectively. 
Following Paper II, we name two nearby particles under 
consideration as particles (1) and (2), and denote their friction times as 
$\tau_{\rm p1}$ and $\tau_{\rm p2}$. 
The particle position, their velocity and the flow velocity at the particle position 
are denoted by ${\bs X}^{(1, 2)}(t)$, ${\bs v}^{(1, 2)} (t)$,  and ${\bs u}^{(1, 2)} (t) (\equiv {\bs u} ({\bs X}^{(1, 2)}(t), t))$, 
respectively. We set the time at which the relative velocity is measured to be time zero, i.e., $t=0$.  
The particle separation and the relative velocity at $t=0$ are denoted 
by ${\bs r}$($= {\bs X}^{(2)} (0)-  {\bs X}^{(1)}(0)$) and ${\bs w}$($ = {\bs v}^{(2)} (0) - {\bs v}^{(1)} (0)$). 
%We denote the particle separation at time zero as ${\bs r} $.  
%using superscripts (1) and (2) to distinguish the two particles. 
The separation of the two particles as a function of time is  ${\bs d}(t)= {\bs X}^{(2)} (t)-  {\bs X}^{(1)}(t)$, 
and ${\bs d}(0) ={\bs r}$. The spatial flow velocity difference across the particle distance is 
denoted by $\Delta {\bs u} ({\bs d}(t)) = {\bs u}^{(2)}(t) - {\bs u}^{(1)}(t)$. 
We denote as $\Delta_{\rm T}^{(1,2)} {\bs u} (\Delta \tau) ={\bs u}^{(1,2)}(t+\Delta \tau) - {\bs u}^{(1,2)}(t)$ the 
temporal flow velocity difference at a time lag of $\Delta \tau$ along the trajectories of particles (1) and (2), 
respectively.  
%and ${\bs w} = {\bs v}^{(2)} - {\bs v}^{(1)}$. 
We also consider the relative velocity, ${\bs w}_{\rm f} = {\bs v} (0) - {\bs u} ({\bs X}(0), 0)$, between a 
particle and the flow element at the particle location. %will measure the relative velocity of particle pairs from all the species at given separations, ${\bs r}$. 
For convenience, we denote the friction times (Stokes numbers) of the smaller and larger 
particles as $\tau_{\rm p, l}$ and $\tau_{\rm p, h}$ ($St_{\ell}$ and $St_{h}$), where the subscripts ``l" (or ``$\ell$") and  ``h" (or  ``$h$") stand for low and high, respectively. 
We define a friction time or Stokes ratio as $f\equiv \tau_{\rm p, l}/\tau_{\rm p, h} =St_{\ell}/St_{h}$. 
By definition, $0\le f\le 1$. %Throughout the paper, We will examine the PDF of the particle relative velocity as a function of the friction time or Stokes number pairs, 
%($\tau_{\rm p1}$, $\tau_{\rm p2}$) or ($St_1$, $St_2$). 
It is also convenient to define the ratio, $\Omega  \equiv \tau_{\rm p}/T_{\rm L}$, 
of the friction time to the Lagrangian correlation time. 

In \S 2, we review the physical picture of the PP10 model for 
turbulence-induced relative velocity in the bidisperse case. 
In \S 3, we discuss several interesting limits, and, 
in particular, we present our simulation result for the PDF of the 
particle-flow relative velocity, ${\bs w}_{\rm f}$.
%A brief summary of our 
%numerical simulation is given in \S 4. 
 %we show the statistics of the particle-flow relative velocity. 
%In \S 6, we show simulation results for the rms relative velocity, against 
%which we test the prediction of the model by Pan and Padon (2010). 
In \S 4, we compute  the particle relative velocity  
PDF for all friction time (or Stokes number) pairs available in our simulation, 
and interpret the results using the physical picture of PP10.  The implications of our results for dust particle collisions in protoplanetary disks 
are discussed in \S 5.
We summarize the main conclusions of this study in \S 6. 
%For small particles,  we expect our simulations and theoretical model to apply for the 
%turbulence-induced collision speed of small dust particles in protoplanetary disks, because 
%the statistical isotropy is likely to be restored the scale of the particle size, which is below the 
%Kolmogorov scale of the turbulent flow, Finally, the idealized sitation isolates various 
%complexities in a protoplanetary disk, and  is  thus a useful tool to reveal the fundamental 
%physics of turbulence-induced relative speed of inertial particles.   

\section{The Physical Picture of the PP10 Model}

In Papers I and II, we introduced and further developed the formulation of Pan \& Padoan (2010, PP10) 
for the relative velocity of inertial particles induced by turbulent motions. We showed in Paper II that the model 
prediction for the root-mean-square relative velocity 
agrees well with our simulation results for particles of arbitrarily different sizes. 
Here, we briefly review the physical picture revealed by the PP10 model, and refer the reader to Papers I 
and II for further details. The physical picture will be used to {\it qualitatively} interpret 
the trend of the relative velocity PDF measured from our simulation data.

%which will then be used to interpret our simulation data for the probability distribution function of the relative velocity. 
   
In the formulation of PP10, the relative velocity, ${\bs w}$, for particles of different sizes
can be approximately written as the sum of two terms, ${\bs w} = {\bs w}_{\rm a} + {\bs w}_{\rm s}$, 
where ${\bs w}_{\rm a}$ and ${\bs w}_{\rm s}$ are named the generalized acceleration and shear 
contributions, respectively.  The two contributions reduce to the corresponding terms in the 
formula of Saffman \& Turner (1956) in the limit of small particles with $\tau_{\rm p}$ much smaller than 
the Kolomogorov time of the flow, $\tau_\eta$. The generalized acceleration term reflects different 
responses of particles of different sizes to the flow velocity. In Paper II, we showed that ${\bs w}_{\rm a}$ is 
physically associated with the {\it temporal} flow velocity difference, $\Delta_{\rm T} {\bs u}$, on individual particle 
trajectories (see Fig.\ 1 of Paper II for an illustration). Based on a calculation of the contribution 
of the acceleration term  to the rms relative velocity, we established in Paper II an approximate 
expression for ${\bs w}_{\rm a}$, 
\begin{equation}
{\bs w}_{\rm a} \simeq \frac{1-f}{[(1+\Omega_{\rm l} )]^{1/2}} \Delta {\bs u}_{\rm T} (\tau_{\rm p,h}) 
\label{approxaccel}
\end{equation} 
where $f$ is the friction time ratio and $\Omega_{\rm l} = \tau_{\rm p, l}/T_{\rm L}$.  
The factor, $1-f$, indicates that  ${\bs w}_{\rm a}$ depends on the friction time 
difference, and it vanishes for equal-size particles.  
The equation connects ${\bs w}_{\rm a}$ to $\Delta_{\rm T} {\bs u} (\Delta \tau)$ 
at a time lag, $\Delta \tau$, equal to the friction time, $\tau_{\rm p,h}$, of the larger 
particle. The probability distribution of the acceleration contribution is 
thus related to the PDF of $\Delta {\bs u}_{\rm T}$. In eq.\ (\ref{approxaccel}), we ignored the 
possible difference in the statistics of $\Delta {\bs u}_{\rm T}^{(1)}$ and $\Delta {\bs u}_{\rm T}^{(2)}$ 
along the trajectories of two different particles, and $\Delta_{\rm T} {\bs u}$ here should be 
viewed as the average of $\Delta {\bs u}_{\rm T}^{(1)}$ and $\Delta {\bs u}_{\rm T}^{(2)}$ (Paper II).  
In the limit $\tau_{\rm p,h} \to 0$, the particle trajectories are close to the Lagrangian trajectories, and 
we have $\Delta {\bs u}_{\rm T} (\tau_{\rm p,h}) \to \Delta {\bs u}_{\rm L} (\tau_{\rm p,h}) \to {\bs a} \tau_{\rm p,h}$, 
where  $\Delta {\bs u}_{\rm L}$ is the Lagrangian temporal flow velocity difference and ${\bs a}$ is 
the local flow acceleration. Therefore, eq.\ (\ref{approxaccel}) gives ${\bs w}_{\rm a} \simeq {\bs a} (\tau_{\rm p,h} - \tau_{\rm p,l})$ 
for small particles with $\tau_{\rm p,l}, \tau_{\rm p,h} \ll \tau_\eta$, consistent with the  Saffman-Turner 
formulation (see also Weidenschilling 1980).   
 
Physically, the generalized shear term represents the particles' memory of the {\it spatial} 
flow velocity difference, $\Delta {u} (d(t))$, across the distance, $d(t)$, of the two 
particles at given times $t \le 0$ in the past (see Fig.\ 1 of Paper I and Fig.\ 2  of Paper II 
for illustrations for the mono- and bi- disperse cases, respectively).
In the case of equal-size particles, the generalized shear term is the only contribution to 
the relative velocity. As discussed in Paper I,  the relative velocity of 
a particle pair of equal size can be approximated by $w \simeq \Delta {u} (r_{\rm p}) [t_{\rm p}/(t_{\rm p} + \tau_{\rm p,h})]^{1/2}$, 
where $r_{\rm p}$, named the primary distance, is defined as the separation of 
the particle pair at a friction timescale ago, i.e., $r_{\rm p} \equiv d(-\tau_{\rm p})$. 
This distance is of particular importance because it is the largest distance before 
the particle memory cutoff takes effect at $t \lsim -\tau_{\rm p}$. 
The estimate of $r_{\rm p}$ depends on how the particle pair separates backward in time. 
Due to the particle inertia, an initial ballistic separation is expected, and, assuming that the duration of the 
ballistic behavior is $\simeq \tau_{\rm p}$, 
the primary distance is estimated as $r_{\rm p} = (r^2 + w^2 \tau_{\rm p}^2)^{1/2}$, 
where $r$ is the particle distance at the current time. Note that  the primary distance $r_{\rm p}$ of a particle 
pair depends on their relative velocity, $w$, at $t=0$, which has important implications for the relative velocity 
PDF (see Paper I and discussions in \S 3.2). The timescale $t_{\rm p}$ is the correlation time of the flow velocity structures at the scale $r_{\rm p}$, 
which is essentially the turnover time of turbulent eddies of size $r_{\rm p}$. 
%The correlation time of turbulent structures as a function of the length scale can be estimated using eq.\ (23) 
%in Paper I (XX). 
The $[t_{\rm p}/(t_{\rm p} + \tau_{\rm p})]^{1/2}$ factor is due to the cutoff by the flow 
``memory", which in effect causes a reduction in the time range of the particle memory that can 
contribute to the relative velocity when $t_{\rm p} \lsim \tau_{\rm p}$ (see Paper I).
  
Similar to the monodisperse case, Paper II proposed an approximate 
equation for ${\bs w}_{\rm s}$ for particles of different sizes. Based on eq.\ (37) in 
Paper II, we have\footnote{Here, we replaced  $R_{\rm p}$ and $T_{\rm p}$ in eq. (37) of Paper II 
by $r_{\rm p}$ and $t_{\rm p}$, respectively.  In Paper II, $R_{\rm p}$ denotes 
the average primary distance over all particle pairs at a given distance, $r$, and the same 
average is implied in the timescale $T_{\rm p}$.  Therefore, 
eq.\ (37) of Paper II  is an approximation for the overall average shear contribution. Here, in order to understand the PDF of the relative velocity, 
we are interested in individual particle pairs, and thus used the primary distance $r_{\rm p}$ 
and the timescale $t_{\rm p}$ for a single particle pair to replace $R_{\rm p}$ and $T_{\rm p}$. 
With this replacement, ${\bs w}_{\rm s}$ in eq.\ (\ref{approxshear}) corresponds to the generalized
shear contribution for each individual pair.}, 
\begin{equation}
{\bs w}_{\rm s} \simeq \Delta {\bs u} ({\bs r}_{\rm p}) \left(\frac{t_{\rm p}}{t_{\rm p} + \tau_{\rm p,h} }\right)^{1/2}. 
\label{approxshear} 
\end{equation} 
As discussed in Paper II,  the primary distance $r_{\rm p}$ 
for the bidisperse case is mainly determined by the smaller particle, and a rough estimate is $r_{\rm p} = (r^2 + w^2 \tau_{\rm p, l}^2)^{1/2}$, 
where $w = (w_{\rm a}^2 + w_{\rm s}^2)^{1/2}$ is the 3D amplitude of the relative velocity of the particle pair 
including both the acceleration and shear contributions\footnote{Since the rate of the 
ballistic backward separation in the bidisperse case has a contribution from the 
acceleration term, the distribution of ${\bs w}_{\rm s}$ is complicated.  
Rigorously, for particles of different sizes, it depends not only on the sptatial flow velocity structures but also on
the statistics of ${\bs w}_{\rm a}$. On the other hand, the distribution of ${\bs w}_{\rm a}$ is simpler, as it is 
completely controlled by $\Delta_{\rm T} {\bs u}$ (see eq.\ (\ref{approxaccel})).}
% independent of the shear term.}. %A more accurate method to evaluate $r_{\rm p}$ 
%and $t_{\rm p}$ in the bidisperse case can be found in Footnote 4 of Paper II. 
Since $r_{\rm p}$ depends on the amplitude of ${\bs w}_{\rm s}$, eq.\ (\ref{approxshear}) 
is an implicit equation.  As in the monodisperse case, the last factor in eq.\ (\ref{approxshear}) represents 
the effect that the flow memory cutoff may reduce the range of the larger particle's memory  (around the memory time of the smaller particle) 
that can contribute to the relative velocity. The direction of ${\bs r}_{\rm p}$ with respect to ${\bs r}$ is 
stochastic due to the ``turbulent" separation of the particle pair backward in time. 
In the small particle limit $\tau_{\rm p, h}, \tau_{\rm p, l} \to 0$, we have ${\bs r}_{\rm p} \to {\bs r}$, 
and ${\bs w}_{\rm s}$ is  approximately equal to $\Delta {\bs u} ({\bs r})$, meaning 
that the shear term follows the spatial flow velocity difference across the particle 
distance (Saffman \& Turner 1956).  For $r$ below the Kolmorgorov scale, 
the flow velocity is smooth and $\Delta {\bs u} ({\bs r})$ is linear with $r$, 
i.e., $\Delta u_i \simeq \partial_j  u_i r_j$, with  $\partial_j  u_i$ the local 
velocity gradients. The distribution of ${\bs w}_{\rm s}$ in the small particle limit is thus related 
to the PDF of  $\Delta {\bs u} ({\bs r})$ or equivalently the PDF of the velocity gradients.
%\footnote{However, the estimate ${\bs w}_{\rm s} \to \Delta {\bs u} ({\bs r})$ becomes 
%invalid at places with large flow velocity gradients $\partial_j  u_i  \gsim 1/\tau_{\rm p}$. 
%In these locations,  the so-called sling effect occurs.}  
%Clearly, a $r-$dependence exists for ${\bs w}_{\rm s}$ in this limit.  

There are several interesting qualitative differences 
between the generalized acceleration and shear terms. A fundamental difference is that 
the acceleration term is determined by the flow velocity along individual particle trajectories, 
while the shear term depends on the trajectories of the two particles relative to each other, 
as seen from the dependence of ${\bs w}_{\rm s}$  on $r_{\rm p}$. 
The shear contribution is thus more complicated. For example, for small particles, ${\bs w}_{\rm s}$  
has a $r-$dependence, and it also 
depends on whether the particles are approaching or separating from each other. 
In contrast, the generalized acceleration term is independent of the particle distance $r$ or 
the relative motion of the two particles.
%Reconsider \footnote
Also, the direction of $\Delta {\bs u}_{\rm T}$ in eq.\ (\ref{approxaccel}) 
is random with respect to ${\bs r}$, and thus ${\bs w}_{\rm a}$ provides 
equal contributions to the radial ($w_{\rm r}$) and tangential ($w_{\rm t}$) components of the 
relative velocity along and perpendicular to ${\bs r}$, respectively. 
On the other hand,  the shear contributions to the two components may differ for small 
particles because the longitudinal  ($\Delta u_{\rm r}$) and transverse  ($\Delta u_{\rm t}$) 
velocity difference in a turbulent flow across a small distance are  nonequal (Papers I and II).    

\section{Three Limiting Cases}

In this section, we consider three special cases of interest, i.e., 
the particle-flow relative velocity, the relative velocity between equal-size particles, 
and a case where one of the particles is extremely large with $\tau_{\rm p} \to \infty$.  
These limiting cases are helpful for the understanding of the 
general behavior of the relative velocity PDF in the bidisperse case.

\subsection{The Particle-flow Relative Velocity}

The relative velocity, ${\bs w}_{\rm f}$, between an inertial particle and the 
instantaneous local flow velocity is a special bidisperse case with one of the particles, 
say particle (2), being a tracer particle with $\tau_{\rm p2} =0$, and at the same position as particle (1). 
We refer to this limiting case with $St_2 \to 0$ as Limit I. In this limit, only the generalized acceleration term contributes\footnote{
The generalized shear term is determined by the flow velocity differences the two particles saw within their friction times.   
Due to the zero memory time of the tracer particle, the shear term depends on the flow velocity difference 
across the particle distance at time zero.  It thus vanishes if the tracer particle 
is at the same position as the inertial particle at time zero. This can also be also be seen from 
eq.\ (\ref{approxshear}), as the primary distance $r_{\rm p}  \to 0$ in this case.}, 
and it thus gives useful information for the distribution of ${\bs w}_{\rm a}$ for particles of different sizes. 
As discussed in Paper II, for a particle with a friction time $\tau_{\rm p}$,  ${\bs w}_{\rm f}$ 
can be approximately estimated by the temporal flow velocity difference, $\Delta_{\rm T} {\bs u} (\Delta \tau)$, 
on the particle trajectory at $\Delta \tau \simeq \tau_{\rm p}$. This can be seen from 
eq.\ (\ref{approxaccel}) with $f=0$, $\Omega_{\rm l} =0$, and $\tau_{\rm p,h} =\tau_{\rm p}$.

%In this limit, particles (2) are essentially tracers, and their relative velocity 
%with particles (1) reduces to the particle-flow relative velocity, ${\bs w}_{\rm f}$, of particles (1). 
%In this section, we consider Limit I, i.e.,  the particle-flow relative velocity, ${\bs w}_{\rm f}$. 

\begin{figure*}[t]
\includegraphics[height=2.9in]{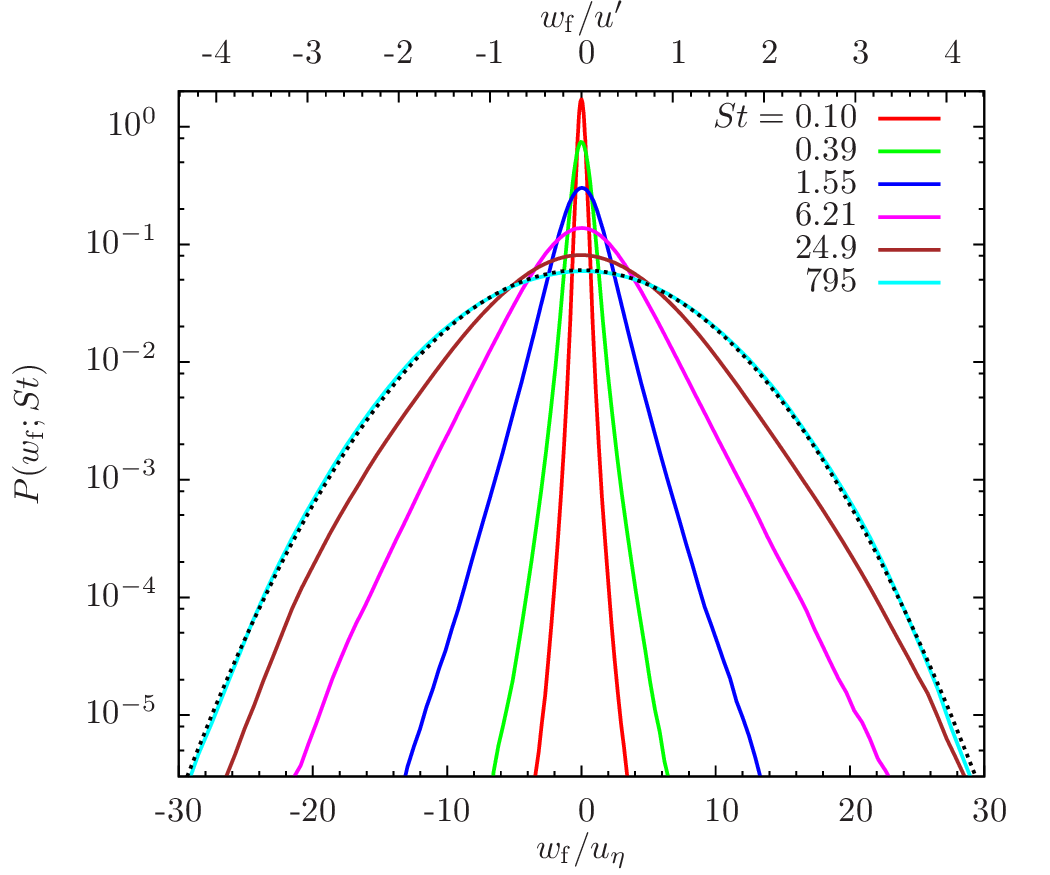}
\includegraphics[height=2.9in]{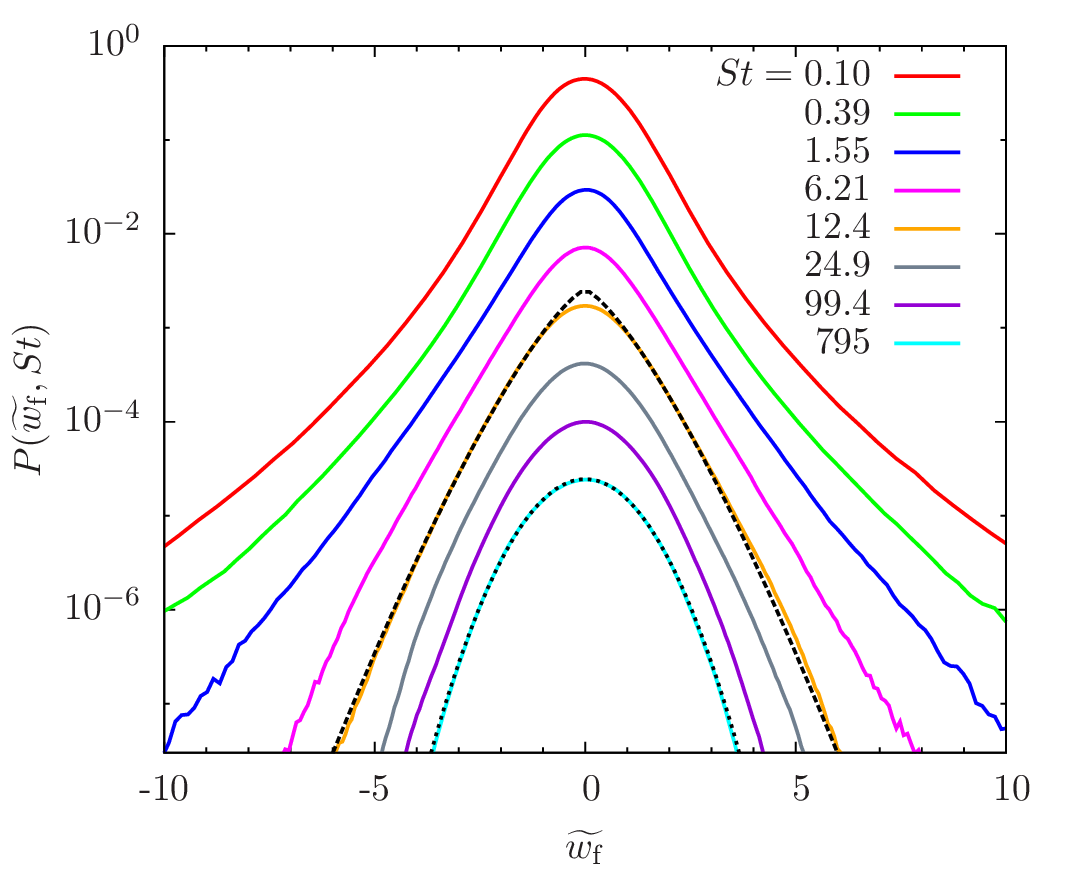}
\caption{The PDF of one component, $w_{\rm f}$, of the particle-flow relative velocity as 
a function of $St$.  In the left panel,  $w_{\rm f}$ is normalized to $u_\eta$ and $u'$ on the bottom 
and top X-axes, respectively. In the right panel,   $w_{\rm f}$ is normalized to its rms value 
(i.e., $\widetilde{w_{\rm f}}\equiv w_{\rm f}/w_{\rm f}^{ \prime}$), so that each PDF has a variance of unity. 
The top curve in this panel plots the actual PDF values for $St =0.1$, and, for clarity, each curve 
below is shifted downward by a factor of 4. The black dashed line is 
the stretched exponential fit with $\alpha = 4/3$ to the PDF tails of $St =12.4$ particles.  
In both panels, the black dotted line is the Gaussian fit for the largest particle ($St =795$) in our simulation.
%With increasing $St$, the PDF shape becomes less fat  and finally approaches Gaussian (dotted black line) at 
%$\tau_{\rm p} \gsim 7 T_{\rm L}$. 
}
\label{gaspartPDF} 
\end{figure*}

Using our simulation data, we computed the PDF of ${\bs w}_{\rm f}$. 
The flow velocity at the position of each particle was obtained by the same 
interpolation method (TSC) used in the simulation. The computed relative 
velocity is at zero particle-flow distance. In the left panel of Fig.\ \ref{gaspartPDF}, 
we plot the PDF, $P(w_{\rm f}, \tau_{\rm p})$, of one component, $w_{\rm f}$, of the particle-flow 
relative velocity for 6 particles with $St = 0.1$, 0.39, 1.55, 6.21, 24.9, and $795$.
Note that here $w_{\rm f}$ is not the 3D amplitude of ${\bs w}_{\rm f}$. 
Each curve shows the PDF averaged over the three components of ${\bs w}_{\rm f}$ 
along the base directions of the simulation grid. The PDF of $w_{\rm f}$ is 
symmetric for all particles, as expected from isotropy for a 1-point statistical 
quantity. The width or rms of the PDF increases with $St$, corresponding to 
the increase of $\Delta {\bs u}_{\rm T}$ with the time lag (see Paper II).  
%However, the shape of the PDF become less fat at larger $St$. 
The dotted black line is the Gaussian fit to the $St =795$ case.  

%like the particle-flow relative velocity at zero distance. 
% and  corresponds to the average PDF of the three components. 
%function of $St$ has been shown in Fig.\ \ref{gaspartrms}.  %Since at $r=0$ the generalized acceleration term 
%is the only contribution to the particle-flow relative velocity, this suggests that the PDF from 
%the acceleration contribution is alway symmetric. 
%The right panel shows the PDF of the radial component of the particle-flow relative velocity at a distance of $1\eta$. 
%We first focus on the PDF at $r=0$ and discuss the radial PDF at $r=1\eta$ at the end of this subsection. 

To see the PDF shape more clearly,  the right panel of Fig.\ \ref{gaspartPDF} plots the normalized 
PDF, $P(\widetilde{w_{\rm f}}, \tau_{\rm p})$, with unit variance, 
where $\widetilde{w_{\rm f}}\equiv w_{\rm f}/w_{\rm f}^{ \prime}$, with $w_{\rm f}^{ \prime}$ the rms of 
$w_{\rm f}$. For clarity, we briefly introduce our terminology for the description of  the PDF shape.  
Following Paper I, we use ``fat" or ``thin" to specifically describe the shape of the PDF, while the extension or width of 
the PDF, corresponding to the rms, will be described as ``broad"  or ``narrow". Conventionally, the fatness of 
a PDF is used as a measure of the departure from a Gaussian distribution, and in particular, it refers to higher probabilities than a Gaussian distribution 
at the \emph{tail} part of the PDF. The fatness of a PDF may be quantified, e.g., by its  kurtosis, and a fatter PDF 
has a larger kurtosis. Typically, a larger kurtosis corresponds to a fatter tail part, and at the same time 
a sharper or thinner shape at the inner part of the PDF. Despite a sharper inner part,  the convention is to describe the PDF with 
a larger kurtosis as fatter based on the tail part.  In the right panel of Fig.\ \ref{gaspartPDF}, we see that, for small particles, the PDF is highly non-Gaussian 
with fat tails.  For example, for  $St =0.1$, the kurtosis of the PDF is $\simeq 5.6$. 
%indicating high non-Gaussianity.  
As $St$ increases,  the PDF shape becomes thinner and thinner. 
The bottom line in the right panel corresponds to the largest particles ($St =795$) in our simulation, and the 
PDF for these particles is Gaussian (the black dotted line). In fact,  $P(w_{\rm f}, \tau_{\rm p})$ becomes 
close to Gaussian at $St \gsim 99.4$ (or $\Omega \gsim 6.9$). 

%Based 
%on the assumption that small particles more or less follow Lagrangian trajectories, 
From the PP10 picture, the PDF of ${\bs w}_{\rm f}$ is controlled by the distribution 
of $\Delta_{\rm T} {\bs u}$, which, however, is unknown. For very small particles with 
$\tau_{\rm p} \ll \tau_\eta$, one may approximate $\Delta_{\rm T} {\bs u} (\Delta \tau)$ 
by the temporal velocity difference, $\Delta_{\rm L} {\bs u} (\Delta \tau) $, along Lagrangian trajectories of tracers. 
On the other hand, due to their slow motions, the flow velocity seen by large particles with $\tau_{\rm p} \gg T_{\rm L}$ 
may be better described as Eulerian, and $\Delta_{\rm T} {\bs u} (\Delta \tau)$ is likely close to the 
Eulerian temporal flow velocity difference $\Delta_{\rm E} {\bs u}(\Delta \tau)$ at fixed points.  
In Appendix A, we compute the PDFs of $\Delta_{\rm L} {u}$ and $\Delta_{\rm E} {u}$ in 
our simulated flow and show that both PDFs are non-Gaussian at small time lags $\Delta \tau$.  
The PDF tails of both $\Delta_{\rm L} {u}$ and $\Delta_{\rm E} {u}$ %The PDF tails of both $P(\Delta_{\rm L} {u}, \Delta \tau)$ and $P(\Delta_{\rm E} {u}, \Delta \tau)$  
have a thinning trend with increasing $\Delta \tau$ and reach Gaussianity at $\Delta \tau \gg T_{\rm L}$ (see, e.g., Mordant et al.\ 2001, Chevillard et al.\ 2005). 
The non-Gaussianity of $\Delta_{\rm L} {u}$ and $\Delta_{\rm E} {u}$ and their trend 
with $\Delta \tau$ is a well-known phenomenon in turbulent flows, usually referred to as intermittency (e.g., Frisch 1995). 
The behavior of the PDFs of $\Delta_{\rm L} {u}$ and $\Delta_{\rm E} {u}$ 
explains the non-Gaussianity of $w_{\rm f}$ and the thinning of  $P(w_{\rm f}, \tau_{\rm p})$ 
with increasing  $St$.  In this sense,  the particle-flow relative velocity 
``inherits" non-Gaussianity from the turbulent flow. 
From eq.\ (\ref{approxaccel}), the PDF trend of $w_{\rm f}$ suggests 
that the distribution of the acceleration contribution ${\bs w}_{\rm a}$ 
becomes thinner and thinner with increasing $\tau_{\rm p, h}$.

A comparison of the normalized PDF of $w_{\rm f}$ with that of $\Delta_{\rm L} {u}$ (Appendix A) 
at a time lag $\Delta \tau =\tau_{\rm p}$ shows that, for $St  \lsim 6.21$, 
the tails of $P(\widetilde{w_{\rm f}}, \tau_{\rm p})$ are thinner than $P(\widetilde{\Delta_{\rm L} {u}}, 
\tau_{\rm p})$. The physical reason is that inertial particles do not exactly follow the fluid elements. 
In particular, inertial particles tend to be pushed out of regions of high vorticity by a centrifugal force arising 
from the rotation (e.g., Pan et al.\ 2011). Therefore, they do not experience the flow velocity in strong 
vortex tubes at small scales, which are intense structures responsible for the high non-Gaussianity (or intermittency) 
of $\Delta_{\rm L} {u}$ in turbulent flows (Bec et al.\ 2006). This effect contributes to make the PDF of $w_{\rm f}$ thinner. 
For particles with $St \gsim  6.21$, the PDF shape for $w_{\rm f}$ and $\Delta_{\rm L} {u} (\Delta \tau =\tau_{\rm p})$ 
are close to each other, and this is because both the dynamics of these larger particles and the temporal flow 
velocity structures at larger timescales are less sensitive to the intense, coherent vortices at small scales. 
%Consequently, the PDF of  $w_{\rm f}$ is less fat than $\Delta_{\rm L}  {u} (\tau_{\rm p})$. 

%CHECK
%As $St$ increases, the difference between the PDFs of  $w_{\rm f}$ and $\Delta_{\rm L} {u} (\tau_{\rm p})$ 
%decreases and, at $St  \gsim 6.21$, the normalized PDFs of $w_{\rm f}$ and  $\Delta_{\rm L}{u} (\tau_{\rm p})$ 
%almost coincide. This perhaps suggests that, for $\tau_{\rm p} \gsim 6.21 \tau_\eta$, neither $w_{\rm f}$ nor $\Delta_{\rm L} {u}$ 
%at $\Delta \tau \simeq \tau_{\rm p}$ is significantly affected by the intense vortical structures at small 
%scales. 
%We note that the PDF shapes of $w^{\rm f}$, $\Delta_{\rm L} u (\tau_{\rm p})$ and 
%$\Delta_{\rm E} u (\tau_{\rm p})$ are close to each other for $\tau_{\rm p} \gsim 12.4 \tau_\eta$. 
%Ask Paolo for the acceleration PDF....

For all particles, the innermost part of the PDF of $w_{\rm f}$ shows a smooth Gaussian-like shape, 
suggesting that the PDF of  trajectory temporal velocity difference, $\Delta_{\rm T} {\bs u}$, also 
takes a similar shape in the inner part. This is supported  by the observation in Appendix A that the 
central parts of the PDFs of $\Delta_{\rm L} {u}$ and $\Delta_{\rm E} {u}$ have a more-or-less 
Gaussian shape. %We attempted to fit the tails with stretched exponentials, 
The fat non-Gaussian PDF tails of $w_{\rm f}$ can be fit by stretched exponential 
functions defined as, 
\begin{equation}
P_{\rm se} (x) =\frac{\alpha}{2 \beta \Gamma(1/\alpha)} \exp\left[- \left(\frac{|x|}{\beta} \right)^{\alpha}\right], 
\label{se}
\end{equation}
where $\Gamma$ is the Gamma function (Paper I). The variance of $P_{\rm se}$ is given by $\beta^2 \Gamma(3/\alpha)/\Gamma(1/\alpha)$. 
The parameter $\alpha$ controls the fatness of the PDF, and a smaller $\alpha$ corresponds to fatter tails. As expected, 
the $\alpha$ value for the tails of $P(w_{\rm f}, \tau_{\rm p})$ increases with increasing $\tau_{\rm p}$ or $St$. 
For the eight lines shown in the right panel of Fig.\ \ref{gaspartPDF}, the 
best-fit $\alpha$ for the PDF tails are 0.7, 0.78, 0.95, 1.23, 1.33, 1.7, 1.95 and 2 for 
$St=0.10$, 0.39, 1.55, 6.21, 12.4, 24.9, 99.4, and 795, respectively. 

The black dashed line in Fig.\ \ref{gaspartPDF} is the stretched exponential function 
with $\alpha =\frac{4}{3}$ that best fits the PDF tail of $St=12.4$ particles. The friction time of these particles is close 
to the Lagrangian correlation time $T_{\rm L}$ of the flow. In Paper I, we found that this stretched exponential can also well 
fit the relative velocity PDF of equal-size particles with $St=12.4$  and $St=24.7$ (see also \S 3.2). 
This similarity in the shape of the PDF tails for ${\bs w}_{\rm f}$ and the monodisperse 
relative velocity of $\tau_{\rm p} \simeq T_{\rm L}$ particles is of 
particular interest. Its implication will be discussed in \S 4.2.1. 

Finally, we point our that, if the flow velocity were exactly Gaussian, then the temporal flow velocity 
difference would be Gaussian, and our model picture indicates that the PDF of the 
particle-flow relative velocity would be Gaussian for particles of any sizes. In that case, 
the relative velocity PDF between different particles approaches Gaussian once the generalized acceleration 
contribution dominates. This suggests that using a Gaussian model flow velocity (see, e.g., Hubbard 2013) 
to study the particle relative velocity PDF would underestimate the degree of non-Gaussianity.

%The result will be explained based on 
%the PDF shapes of  $\Delta_{\rm L} {u} (\Delta \tau)$ and $\Delta_{\rm E} {u} (\Delta \tau)$ 
%The explains the fattening trend of the PDF, $P(w_{\rm f}, \tau_{\rm p})$, of the particle-flow relative velocity with decreasing $St$, as shown in Fig.\ \ref{gaspartPDF}.  
%(see also Appendix C). 
%where we show that ${\bs w}_{\rm f}$ is highly non-Gaussian for small
%particles with $St \lsim1$, and 
%the degree of non-Gaussianity decreases with increasing $St$. 
%The PDF finally approaches Gaussian at $\tau_{\rm p} \gsim 7T_{\rm L}$.

\subsection{Equal-size Particles}

%{\bf [Could you shorten significantly this section? I find it very long.]}

The relative velocity PDF of equal-size particles has been discussed in details in 
Paper I.  For convenience, we refer to this monodisperse case with 
$St_2 =St_1$ as Limit II. In Paper I, we found that the relative velocity 
PDF of identical particles is extremely non-Gaussian. %, and, using the physical picture of PP10, 
We identified two sources of non-Gaussianity, i.e., the imprint of the non-Gaussianity 
(or intermittency) of the turbulent flow and an intrinsic contribution arising from the particle 
dynamics. Our simulation data showed that the fatness of the monodisperse 
PDF first increases with $St$, reaches a peak at $St\simeq 1$, and then continuously decreases 
as $St$ increases above 1. 
%of the monodisperse PDF 
In the following, we briefly review the explanation for this behavior based on the PP10 picture. 

%We briefly summarize the  
%Consider a component, $w$, either radial or tangential, of the relative velocity 
%of an identical particle pair. 
For equal-size particles, only the shear term contributes, and the relative velocity, 
$w$, can be estimated by eq.\  (\ref{approxshear}) for $w_{\rm s}$. For a particle 
pair with $\tau_{\rm p} \lsim T_{\rm L}$, the factor $t_{\rm p}/(t_{\rm p} + \tau_{\rm p})$ in eq.\ (\ref{approxshear}) 
is of oder unity, and, qualitatively, $w$ can be approximated by the spatial 
flow velocity difference, $\Delta u (r_{\rm p})$, across the primary distance, $r_{\rm p}$ (Paper I).  
At a given $r$, the primary distance, $r_{\rm p} \simeq (r^2+w^2 \tau_{\rm p}^2)^{1/2}$, 
of particle pairs with $|w| \lsim r/\tau_{\rm p}$ is close to $r$. Therefore, the inner part 
of the PDF with $|w| \lsim r/\tau_{\rm p}$ follows  $\Delta u$ at a distance of $r$.  
The distribution of $\Delta u$ at small scales is already  non-Gaussian
% corresponding to the intermittency of spatial velocity structures 
(e.g., Frisch 1995), and it leaves a non-Gaussian imprint on the PDF of $w$. Particle pairs lying at higher PDF tails separate faster backward in time and have larger $r_{\rm p}$, 
so their relative velocity ``remembers"  $\Delta u(\ell)$ at larger scales $\ell$.  
Here, $\Delta u(\ell)$ denotes a component of  $\Delta {\bs u}({\bs \ell}) \equiv  {\bs u} ({\bs x} + {\bs \ell} ,t) - {\bs u} ({\bs x}, t)$ at a 
separation ${\bs \ell}$. Since the PDF of $\Delta u$ is wider at larger $\ell$, this leads to an enhancement at the tail part of the 
PDF. The effect, named as a self-amplification in Paper I, corresponds to the sling effect or caustic formation that occurs in flow 
regions with large velocity gradients, $\partial_j u_i$ (Falkovich et al.\ 2002, Wilkinson \&  Mehlig 2005, 
Wilkinson et al.\  2006, Falkovich \& Pumir 2007, Bewley et al.\ 2013). 
As $St$ increases, the self-amplification proceeds from 
the tails toward smaller $|w|$, while the innermost part of the PDF remains unchanged at first.  
As a result,  the overall shape of the PDF becomes fatter. 

The fattening trend ends at $St \simeq 1$. As $St $ increases above 1, %$r_{\rm p}$ increases, and 
the amplification moves deeper toward the central part, and  % For particles with larger $\tau_{\rm p}$, 
the PDF at the same value of $w$  samples $\Delta u$ at larger scales. Since the PDF shape of $\Delta u (\ell)$ 
is progressively thinner with increasing $\ell$ (see Appendix B of Paper I), 
this causes the overall fatness of the PDF of $w$ to decrease continuously at $St \gsim 1$. 
The PDF shape is thus the fattest at $St \simeq 1$. 
%a  phenomenon known intermittency in turbulence studies.  

For particles with $\tau_{\rm p}$ close to $T_{\rm L}$, the PDF tails of $w$ were found to be well described 
by a 4/3 stretched exponential distribution (see eq.\ (\ref{se})). Such a distribution were obtained in 
Paper I from a phenomenological argument based on the PP10 picture.  Assuming 
exactly Gaussian flow velocity and Kolmogorov scaling for $\Delta u (\ell)^2 \propto C (\bar{\epsilon} \ell)^{2/3}$ with $C$ 
the Kolmogorov constant,  the probability of finding a particle pair with a relative velocity of $w$ is estimated as 
$\propto \exp[-w^2/(2 C \bar{\epsilon}^{2/3} (r^2 + w^2 \tau_{\rm p}^2)^{1/3})]$, which is a 4/3 stretched exponential at 
$w \gg r/\tau_{\rm p}$ (see also Gustavsson et al.\ 2008).  
Using the same argument,  one can show that non-Gaussianity is an intrinsic feature of the particle 
dynamics: Even if the flow velocity were exactly Gaussian, the particle relative velocity would be 
non-Gaussian for any scaling behavior of $\Delta u({\ell})$ with $\ell$ (unless it is constant 
with $\ell$ as in the case of a white noise). 

%(Paper I). 
%The two assumptions used in the derivation are both valid only for  $\tau_{\rm p} \sim T_{\rm L}$,  and the 4/3 stretched exponential PDF 
%only applies to these particles.
%The prediction was based on the facts that the PDF of $\Delta u$ is 
%close to Gaussian at the typical primary distance of these particles, $r_{\rm p}$, and that  $r_{\rm p}$ is 
%still within the inertial range such that the Kolmogorov scaling, $\Delta u\propto r_{\rm p}^{1/3}$, applies. 

Finally, for $\tau_{\rm p} \gg T_{\rm L}$, the flow correlation or memory time is typically 
shorter than the particle memory, and the last factor in eq.\ (\ref{approxshear}) reduces the 
PDF width by a factor of $(T_{\rm L}/\tau_{\rm p})^{1/2}$. In the limit $\tau_{\rm p} \to \infty$,  
the particle motions are similar to Brownian motion, because even the largest eddies 
in the flow act as random noise when viewed over the friction time of these large particles. 
Therefore, the relative velocity PDF is  expected to approach 
Gaussian as $\tau_{\rm p} \to \infty$.
%Also, as the typical $r_{\rm p}$ exceeds the flow integral scale, $\Delta u(\ell)$ becomes independent of $\ell$, and 
A Gaussian shape is almost reached for the largest particles 
with $\tau_{\rm p} = 54 T_{\rm L}$ in the simulation of Paper I. 
%Since the generalized shear term is the only contribution for  
%The PDF trend described above for equal-size particles  provides useful hints 
%for the distribution of the shear term for particles of different sizes. 

The comparison between Limits I and II at equal $\tau_{\rm p}$ is useful 
to understand the PDF trend in the general bidisperse case (see \S 4). 
For a small particle with $\tau_{\rm p} \to 0$, the PDFs 
of its relative velocity with respect to the local flow velocity 
and to a nearby particle of the same size are determined by the 
distributions of the flow acceleration, ${\bs a}$,   
%In the case of equal-size particles with $\tau_{\rm p} \to 0$, the distribution of their relative velocity, ${\bs w}$, 
and the velocity gradients, $\partial_j u_{i}$, respectively (see \S 2).
In Appendix B, we compute the normalized PDFs of ${\bs a}$  and $\partial_j u_{i}$ 
in our simulated flow, and show that the distribution of ${\bs a}$ is generally 
fatter, consistent with previous studies (e.g., Ishihara et al.\ 2007). This suggests that, in the $St \to 0$ 
limit, the PDF of ${\bs w}_{\rm f}$ has a fatter shape than the relative velocity of 
equal-size particles.  %As $\tau_{\rm p}$ increases, the trends of the PDFs 
%for ${\bs w}_{\rm f}$ and for the relative velocity ${\bs w}$ of equal-size particles 
%are different.  
As $St$ increases toward $\simeq 1$, the PDF for equal-size particles fattens significantly due to 
the tail amplification (see discussion above),  while the PDF for  the particle-flow relative velocity becomes thinner 
continuously  (see Fig.\ \ref{gaspartPDF}). 
%On the other hand, $\simeq1$, and then becomes thinner for $St$ above $1$. 
The opposite trends of the two PDFs for $St \lsim 1$ have an 
important consequence. The PDF for equal-size particles is found to be already fatter 
than the distribution  of $w_{\rm f}$ at the smallest Stokes number, $St =0.1$, in 
our simulation, and, at $St \simeq 1$, the former has a much stronger non-Gaussianity 
than the latter. For $St$ above $1$, the PDF fatness for both cases decreases with increasing $St$, 
and we will show in \S 4.1.1 that the relative velocity PDF of equal-size particles remains to be significantly fatter than 
${\bs w}_{\rm f}$ at all $St \gsim 1$.  Thus, for all particle species  in our simulation, 
the PDF for Limit II has a higher degree of non-Gaussianity than for Limit I.   
%The  at the same $\tau_{\rm p}$, the PDF for equal-size particles (or for Limit II) is 
%always fatter than the PDF of $w_{\rm f}$, i.e., Limit (see \S 4.1).

\subsection{The $\tau_{\rm p2}\to \infty$ Limit and the One-particle Velocity}

If one of the two particles, say particle (2), is extremely large with $\tau_{\rm p2} \gg T_{\rm L}$, 
its velocity is tiny, and the relative velocity with respect to the smaller particle (1) is essentially the 
one-particle velocity, ${\bs v}^{(1)}$, of particle (1). We refer to this case as Limit III, corresponding to $St_2 \to \infty$.  
Paper I found that the PDF of the one-particle velocity, ${\bs v}$, for particles of any size is Gaussian.  
As discussed in Paper I, this is easy to see in the small and large particle limits. 
Small particles closely follow the local flow velocity, and the PDF of ${\bs v}$ is approximately given by the 1-point 
PDF of the flow velocity, ${\bs u}$, which is known to be approximately 
Gaussian. On the other hand, the motions of large particles with $\tau_{\rm p}\gg T_{\rm L}$ are 
similar to Brownian motion, because, for these particles, turbulent eddies of 
all sizes act as random noise when viewed over a timescale of $\simeq \tau_{\rm p}$. 
Therefore, the one-particle velocity PDF in the large particle limit  is also Gaussian.  
The Gaussianity of the one-particle velocity ${\bs v}$ for particles of any size suggests that, for any 
given $St_1$, the bidisperse relative velocity PDF approaches Gaussian as $St_2 \to \infty$.

%The dynamics of all particles in the first 12 species ($\tau_{\rm p} \lsim 14 T_{\rm L}$) 
%is thus fully relaxed by the end of the simulation, and we measured the statistics of the 
%particle velocity using the last few snapshots covering $5-6$ $T_{\rm L}$. 
%On the other hand, it is uncertain whether the largest particles from the last two 
%species are well relaxed. From the analysis of our simulation data, we find that the 
%velocity statistics of the largest particles do reach a steady state toward the end of the run, suggesting that 
%these particles may also be dynamically relaxed. 

\section{The PDF of the Particle Relative Velocity }

We analyze the probability distribution function of the particle relative velocity for 
all Stokes number pairs in our simulation. To compute the relative velocity statistics, we search pairs  
of particles from all species at given distances, $r$. We will consider $r=1$, $\frac{1}{2}$,  and $\frac{1}{4} \eta$. 
It would be useful to measure the PDF at even smaller $r$ because the size of dust particles in 
protoplanetary disks is much smaller than the Kolmogorov scale of protoplanetary turbulence. 
However, the number of particle pairs available in the simulation decreases with decreasing 
$r$, and, at $r \ll 1\eta$, the number is too limited to provide sufficient statistics. 
We will thus show simulation results for $\frac{1}{4} \eta \le r\le1\eta$, and 
discuss the $r-$dependence (or convergence) of the PDF in this range. 

For each particle (1), we locate particles (2) in a distance shell $[r-dr/2, r+dr/2]$. 
The shell thickness, $dr$, is set to $0.08r$, $0.08r$, and $0.16 r$ for 
$r=1$, $\frac{1}{2}$, and $\frac{1}{4}\eta$, respectively. For $r=\frac{1}{4}\eta$ with $dr= 0.16 r$,
the number of particle pairs for any two species is $\gsim 10^4$, 
about sufficient for the PDF measurement.  
%Smaller $r$ values are desirable, but that would require including a larger 
%number of particles to ensure accurate statistics.  
%We find that the number of pairs at $r=\frac{1}{8}\eta$ is already limited and is too small to study the high-order statistics such as the PDF tails. 
%We will only measure the first-order moment of the relative velocity (and the collision kernel) 
%at this distance. To increase the number of pairs,  we set $\delta r = 0.32 r$ for $r=\frac{1}{8}\eta$.   
%
%{\bf [Somewhere in the paragraph below it is important to say why we study all three cases: radial, tangential and total amplitude. What does one learn from each of them.]}
%
We examine the PDFs of the radial and tangential components, 
as well as the 3D amplitude, of the relative velocity. %The computation of the 3D amplitude 
%is straightforward. 
The 3D amplitude is of practical interest as it determines the total 
collision energy and hence the collision outcome. The PDFs of  
the radial and tangential components, on the other hand, are helpful for a 
thorough understanding of the underlying physics. In particular, the features 
of the radial and tangential PDFs provide very useful tests for the PP10 picture. 
To obtain the radial and tangential components, we set up a local coordinate system for each selected pair. 
In terms of the grid base vectors, ${\bs e}_1$, ${\bs e}_2$, and ${\bs e}_3$, the local system is taken to be ${\bs e}'_1 
= \cos\theta\cos\phi{\bs e}_1 +  \cos\theta\sin\phi {\bs e}_2  +  \sin\theta {\bs e}_3$, 
${\bs e}'_2= -\sin\phi {\bs e}_1 + \cos\phi{\bs e}_2$ and ${\bs e}'_3= -\sin\theta \cos\phi{\bs e}_1- \sin\theta \sin\phi {\bs e}_2 + 
\cos\theta {\bs e}_3$. The Euler angles are defined as $\sin\theta = r_3/r$, $\cos\theta = (r_1^2 +r_2^2)^{1/2}/r$, 
$\cos\phi= r_1/(r_1^2 +r^2_2)^{1/2}$, and $\sin\phi= r_2/(r_1^2 +r_2^2)^{1/2}$, with $r_1$, $r_2$ and $r_3$ 
the components of ${\bs r}$ in the grid coordinate. We decompose ${\bs w}$ as 
$w_{\rm r} = {\bs w} \cdot {\bs e}'_1$, $w_{\rm t2} = {\bs w} \cdot {\bs e}'_2$, 
and $w_{\rm t3} = {\bs w} \cdot {\bs e}'_3$. %andmeasure the PDF of each component. 
The tangential PDF, $P(w_{\rm t}; St_1, St_2)$, is set to the average of 
$P(w_{\rm t2}; St_1, St_2)$ and  of $P(w_{\rm t3}; St_1, St_2)$. 
%The rms or variances of the relative velocity can be 
%easily computed from the measured PDFs. 
The analysis is the same as in Paper I for equal-size particles. In \S 4.1, we 
fix the Stokes number, $St_1$, of particles (1), and examine the trend of the relative velocity PDF 
as a function of $St_2$. In \S 4.2, we will show the PDFs at fixed Stokes number ratios, $f \equiv St_{\ell} /St_{ h}$. 
We will pay particular attention to the non-Gaussianity of the PDF, and discuss how the PDF shape changes with 
the Stokes numbers. As a reminder, in our terminology, a ``fat" (or ``thin" ) PDF shape refers to higher (or lower) probabilities 
specifically at the \emph{tail} part of the PDF.    

%to which we refer the reader for details. 
%In this section, we  consider the root-mean-squares or variances of the relative velocity PDFs.

\subsection{The PDF at Fixed $St_1$}

If one of the Stokes numbers is fixed at $St_1$, the three interesting limits, 
i.e., $St_2 \to 0$, $St_2 =St_1$, and $St_2 \to \infty$, discussed in \S 3 are useful to 
confine the trend of the PDF shape as a function of  $St_2$. These cases were 
named as limits I, II, and III, respectively. The shape of the PDF for $0 \le St_2 \le St_1$ 
is expected to lie in between Limit I and Limit II, while $St_1 \le St_2 < \infty$ corresponds to a range 
between Limits II and III. 

%In other words, the three limits are important delimiters for the behavior of the relative velocity 
%PDF in the general bidisperse case. 
% does the monodisperse PDF reach a Gaussian shape.}
%0.63 for St=0.1
%0.64  St=0.19    
%0.73  St=0.39    
%0.85  St=0.78    
%0.91  St=1.55   
%1 St =3.11   1.05 
%1.2     St = 6.21
%1.45   St =12.4
%1.72     St = 24.9
%1.8     St =49.7
%1.95     St =99.7 
%2        Sr =199
%2         St =398
%2         St =795

\subsubsection{The PDFs of the radial and tangential relative speeds}

%As summarized in the previous section, the rms relative speed of two particles at separation of $r$ is estimated 
%by $w \simeq \delta v \left( R_{\rm p} \right)$ where $\delta v (\ell)$ represents the flow velocity difference across 
%a distance $\ell$ and $R_{\rm p}$ is the primary particle separation, $R_{\rm p}$, backward in time.  The picture is 
%illustrated in Fig.\ 1. This estimate essentially assumes that the particle relative speed is determined by their memory of the 
%flow velocity difference  at the primary  separation, $R_{\rm p}$. As discussed earlier, the primary separation, 
%$R_{\rm p}$,  is the particle distance at time $-\tau_{\rm p}$. Assuming a ballistic separation behavior within 
%a friction timescale,  we have $R_{\rm p}^2= r^2  + w^2 \tau_{\rm p}^2$.  Therefore  $w \simeq \delta v \left( (r^2 + w^2 \tau_{\rm p}^2)^{1/2} \right)$, 
%which can be used to derive the scaling behavior of the rms relative speed with the Stokes number.    

\begin{figure*}[t]
\includegraphics[height=2.9in]{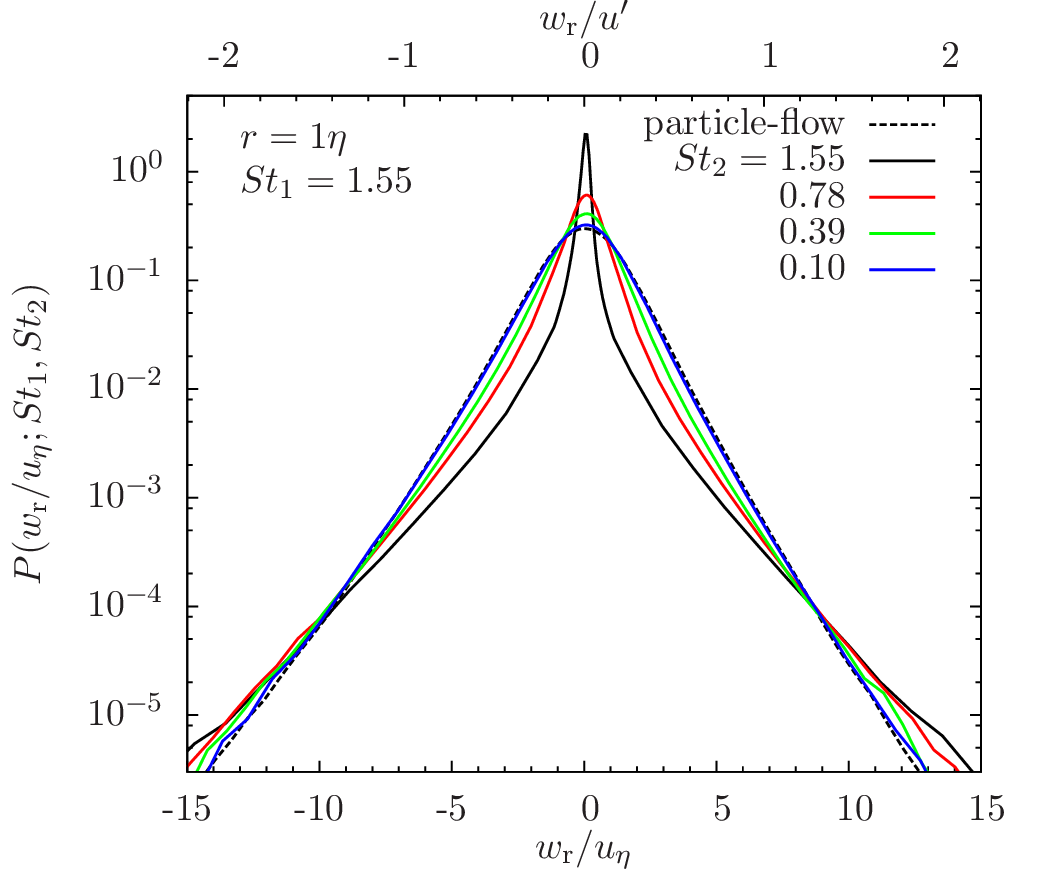}
\includegraphics[height=2.9in]{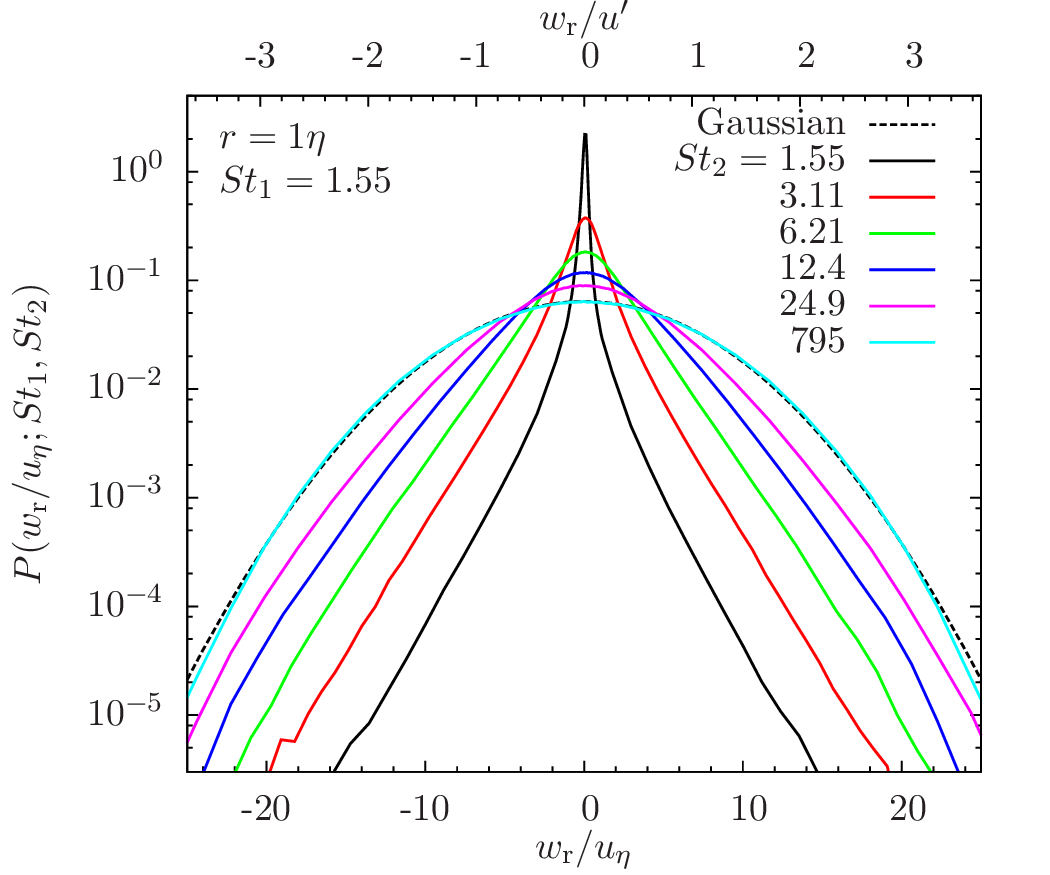}
\caption{PDF of the radial relative speed at $r=1\eta$ as a function of $St_2$ for $St_1 =1.55$. The 
left and right panels show results for $St_2 \le St_1$ and $St_2 \ge St_1$, respectively. 
The solid black curves in both panels correspond to the same monodisperse PDF at $St =1.55$. 
Note that, for clarity of the PDF shapes for $St_2 \le St_1$, the left panel shows a narrower range of $w_{\rm r}$ than in the right panel.
The dashed black line in the left panel is the PDF of the particle-flow relative velocity of $St =1.55$ particles,  
while the dashed line in the right panel shows the best-fit Gaussian PDF for $St_1=1.55$ and 
$St_2 = 795$.} %All the PDFs are measured at $r=1\eta$.}
%{\bf  The normalizations of the particle friction time can be converted 
%using $\Omega = St/14.4$ and $\Omega_{\rm eddy} =St/19.2$}.}
\label{radialpdf4} 
\end{figure*}

In Fig.\ \ref{radialpdf4}, we plot the PDFs, $P(w_{\rm r}; St_1, St_2)$, of the 
radial relative velocity for $St_1$ fixed at $1.55$.  The relative velocity is measured 
at $r=1\eta$. The left and right panels show results for $St_2 \le St_1$ and 
$St_2 \ge St_1$, respectively. In both panels, the solid black line is 
the PDF for identical particles with $St=1.55$, corresponding to Limit II. 
The monodisperse PDF at $St=1.55$ is highly non-Gaussian, and the physical 
origin of the non-Gaussianity has been explained in \S 3.2 
using the PP10 picture. This PDF is also negatively skewed. 
The left and right wings of the PDF correspond to approaching and separating particle pairs. 
The asymmetry of the two wings indicates faster relative velocity for approaching pairs. 
As discussed in Paper I, there are two reasons for this asymmetry. 
%and the physical origin have been discussed in \S 6.5.
First, the radial relative speed PDF for small equal-size particles inherits a 
negative skewness from the PDF of the longitudinal flow velocity difference 
or gradients of the flow (see Appendix B and Paper I). Second, approaching particle pairs have a larger separation 
in the near past than separating pairs. This suggests that, for small particles, 
the primary distance, $r_{\rm p}$, of approaching pairs is larger than separating ones, 
which enhances the asymmetry as $St$ increases at small $St \lsim 1$.  %As shown in Paper I, for $St$ above $\simeq 0.4$,  
As $St$ keeps increasing, the particle distance ($r_{\rm p}$) at a friction time 
ago is less dependent of the particle separation in the near past, and the 
asymmetry decreases (see Figs.\ \ref{radialpdf6} and \ref{radialpdf9}). 
%to the asymmetry in the radial PDF of small particles ($St\lsim1$), especially at 
%small positive values of $w_{\rm r}$ (see Fig. X in Paper I). 

\begin{figure*}[t]
\includegraphics[height=2.9in]{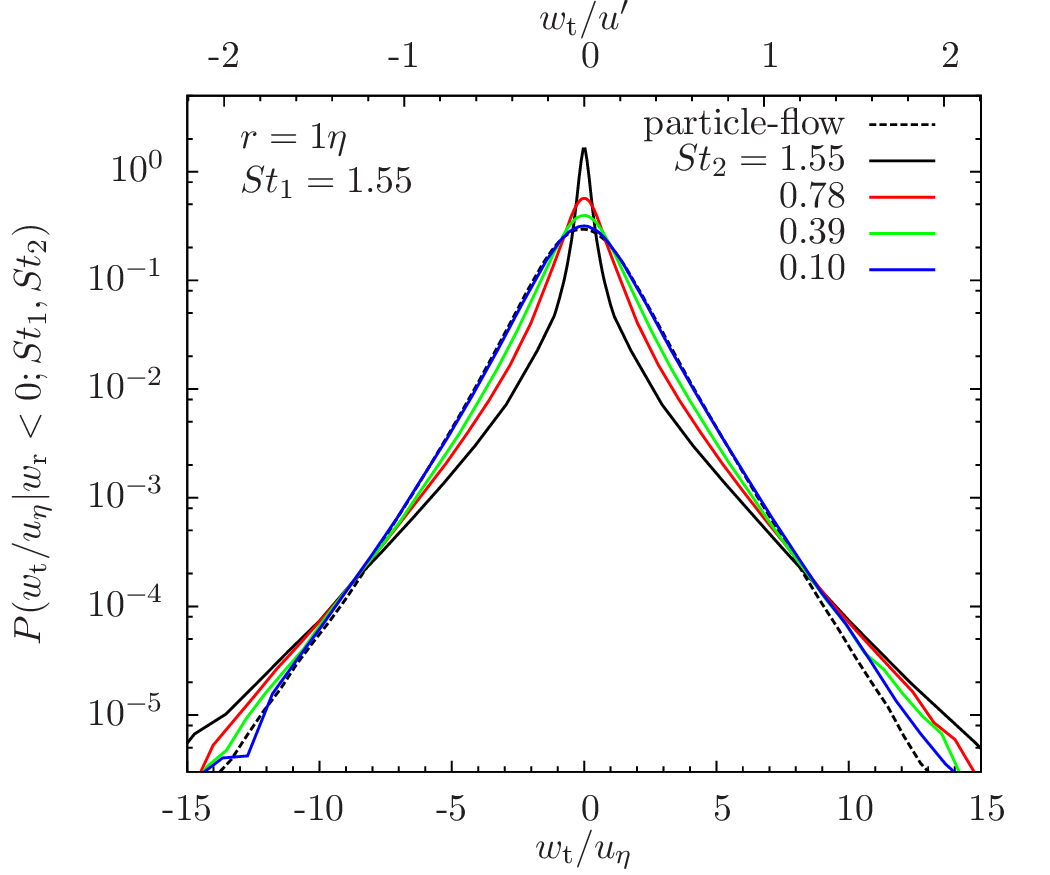}
\includegraphics[height=2.9in]{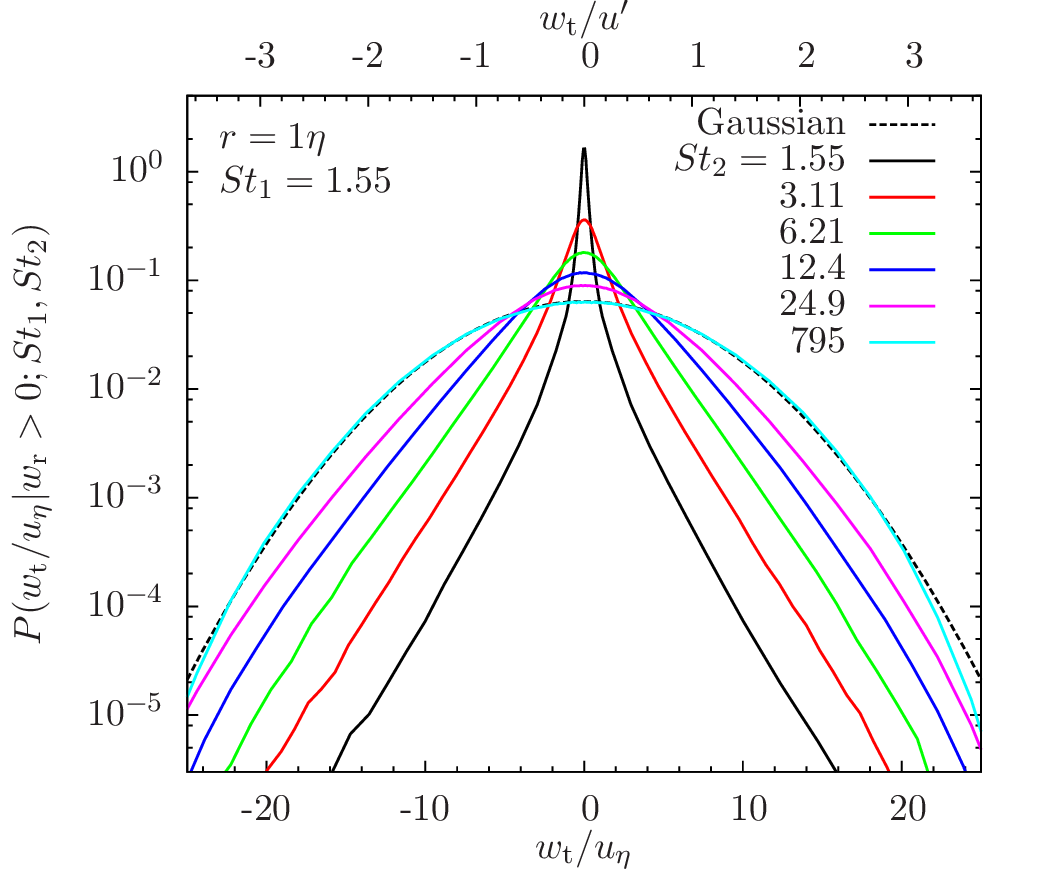}
\caption{PDF of the tangential relative velocity for approaching particle pairs ($w_{\rm r} <0$) 
with one of the Stokes number $St_1$ fixed at $1.55$. 
The figure is plot in the same way as Figure \ref{radialpdf4}. }
%{\bf  The normalizations of the particle friction time can be converted 
%using $\Omega = St/14.4$ and $\Omega_{\rm eddy} =St/19.2$}.}
\label{tangentialpdf4} 
\end{figure*}

The black dashed line in the left panel plots the PDF of the radial 
component,  $w_{\rm f,r}$, of the relative velocity between $St=1.55$ particles 
and the flow element, corresponding to Limit I with $St_1 =1.55$ and $St_2 \to 0$.  
For a consistent comparison with the particle-particle case, here the PDF of  $w_{\rm f,r}$ 
is measured at the same distance, $r=1\eta$.  It is thus not the same as the 
particle-flow relative velocity PDFs shown in Fig.\ \ref{gaspartPDF} at zero distance. 
To compute $w_{\rm f,r}$ at a finite distance,  we used the TSC interpolation to 
obtain the flow velocities at a separation of $r$ from the position of each particle in the 
three base directions of the simulation grid. % the particle-flow relative velocity at $r$ is computed as ${\bs w}_{\rm f} = {\bs u}({\bs X}(t)+r{\bs e}_{i}, t)-{\bs v}(t)$. 
We then averaged the PDFs of $w_{\rm f,r}$ measured from the three directions. 
%The method used to measure this radial PDF is given Appendix C. 
Unlike the particle-flow relative velocity at zero distance, here the PDF of  $w_{\rm f,r}$ 
at $r=1\eta$ for $St =1.55$ is asymmetric. This is because, at a finite distance, the 
generalized shear term has a nonzero contribution (see eq.\ \ref{approxshear}) to the relative 
velocity, which gives rise to %the shear term provides a contribution, and, for $St \lsim 3.11$ particles, 
a noticeable negative skewness in the PDF of $w_{\rm f,r}$ for small particles with $St \lsim 3.11$. 
%This is different from the PDF of $w_{\rm f}$ at zero distance, which is symmetric for all 
%particles (see Fig.\ \ref{gaspartPDF}).   
%At exactly zero distance,  the particle-flow relative velocity is essentially 1-point statistics, and its PDF 
%is expected to symmetric from statistical isotropy. Another perspective is that, at $r=0$, only the generalized 
%acceleration contributes to the particle-flow relative velocity, and its PDF is symmetric about $w_{\rm f} =0$. 
%is inherited from the longitudinal flow velocity difference (see the dashed black curve).
%To compute the particle-flow relative velocity, we interpolated the flow velocity at a distance of $1\eta$ 
%from position of each particle using the TSC method. 
 %${\bs w}_{\rm s}$
 
As expected, the bidisperse PDFs in the left panel of Fig.\ \ref{radialpdf4} lie in between 
Limit II (black solid line) and Limit I (dashed solid line). Since a negative  skewness exists in 
both limits, the PDF is asymmetric for all $St_2 \le St_1$. At the smallest $St_2$ ($=0.1$; the blue line) 
shown in the figure, the radial PDF, $P(w_{\rm r}; St_1, St_2)$, approaches Limit I. 
As $St_2$ decreases below $St_1$, the central part of the PDF widens, 
and the rms of the PDF increases (Paper II). This corresponds to the increase in the contribution 
of the generalized acceleration term, ${\bs w}_{\rm a}$. 
%The acceleration contribution depends on the difference in the particle friction times. 
As discussed in \S 2, ${\bs w}_{\rm a}$  is related to the temporal flow velocity difference, 
$ \Delta_{\rm T} {\bs u}$, on the particle trajectory. 
%$\propto |\tau_{\rm p2} -\tau_{\rm p1}|$
Applying eq.\ (\ref{approxaccel}) to the Stokes number pairs in the left panel of  Fig.\ \ref{radialpdf4}, we have 
${\bs w}_{\rm a} \simeq (1-f) \Delta_{\rm T} {\bs u}(\tau_{\rm p1})$, where $f=St_2/St_1$ for $St_2\le St_1$. 
Clearly, as $St_2$ decreases,  $f$ decreases and the contribution of ${\bs w}_{\rm a} $ increases.  
On the other hand, it can be seen from eq.\ (\ref{approxshear}) that the shear contribution decreases with 
decreasing $St_2$ (see Paper II).  

The black solid line for Limit II  has a fatter overall shape than  the black dashed line 
for Limit II (see the discussion at the end of \S 3.2),  suggesting that  the distribution of the 
generalized acceleration contribution is thinner than the shear contribution. As ${\bs w}_{\rm a}$ 
increases with decreasing $St_2$, the overall fatness of the relative velocity 
PDF decreases. Interestingly, despite the increase in the rms width, the 
probability at the far tails of the relative velocity PDF becomes smaller at smaller $St_2$, 
indicating a lower probability of finding particle pairs that collide with extremely 
high velocity. In general, whether the probability at the far tails increases or 
decreases with decreasing $St_2$ is determined by the competition of two effects. 
First, if the PDF shape is given, an increase in the rms  tends to give a larger probability 
at the far tails. On the other hand, if the rms is fixed, the thinning of the PDF shape would 
lead to a lower probability at high tails. It appears that the effect of thinning PDF shape 
wins in the case of $St_1=1.55$ and $St_2 \le St_1$.

The right panel of Fig.\ \ref{radialpdf4} shows the radial relative speed PDF 
for $St_2  \ge St_1 = 1.55$.  As $St_2$ increases, the PDF becomes wider, 
and its shape becomes thinner. This is again due to the increase in the 
contribution from the acceleration term. Using eq.\ (\ref{approxaccel}) here, 
we have ${\bs w}_{\rm a} \simeq (1-f) \Delta_{\rm T} {\bs u}(\tau_{\rm p2})$, 
where $f = St_1/St_2$. As $St_2$ increases, both 
the time lag in $\Delta_{\rm T} {\bs u}$ and the factor $1-f$ increase, leading to the increase 
of the rms width of the PDF.  
%and can be explained using the same reasoning for the $St_2 \le St_1$ case 
%in the left panel. 
The PDF shape of $\Delta_{\rm T} {\bs u}(\tau_{\rm p2})$ is expected 
to become thinner with increasing $ \tau_{\rm p2}$.  This is based on the observation of the 
thinning trend of the Eulerian and Lagrangian temporal velocity differences with 
increasing time lag $\Delta \tau$ (see Appendix A). The argument was used 
earlier  to explain the thinning trend of the particle-flow relative velocity with particle 
inertia (see \S 3.1). The thinning of the  distribution 
of $\Delta_{\rm T} {\bs u}(\tau_{\rm p2})$ with increasing $St_2$ makes 
the PDF of the relative velocity thinner.  
At $St_2 \ge 49.7$, the PDF  is  close to Gaussian, consistent with the expectation for Limit III (\S 2.3).  The black dashed line in the 
figure is the Gaussian fit to the $St_2 = 795$ case. 
We note that the PDF becomes symmetric for $St_2 \gsim 6.21$.

Since the PDF shape becomes thinner for $St_2$ both above or below $St_1$, 
the fatness  of the PDF peaks at equal-size particles. The PDF shapes in both limits I and III are thinner than Limit II. 
A computation of the skewness of the radial PDFs shows 
that it decreases as the Stokes number difference increases.  
This is because the generalized acceleration is independent of the relative motions of the two particles 
and thus provides a symmetric contribution to the two wings of the radial relative velocity PDF (see discussion in \S 2). 

\begin{figure*}[t]
\includegraphics[height=2.9in]{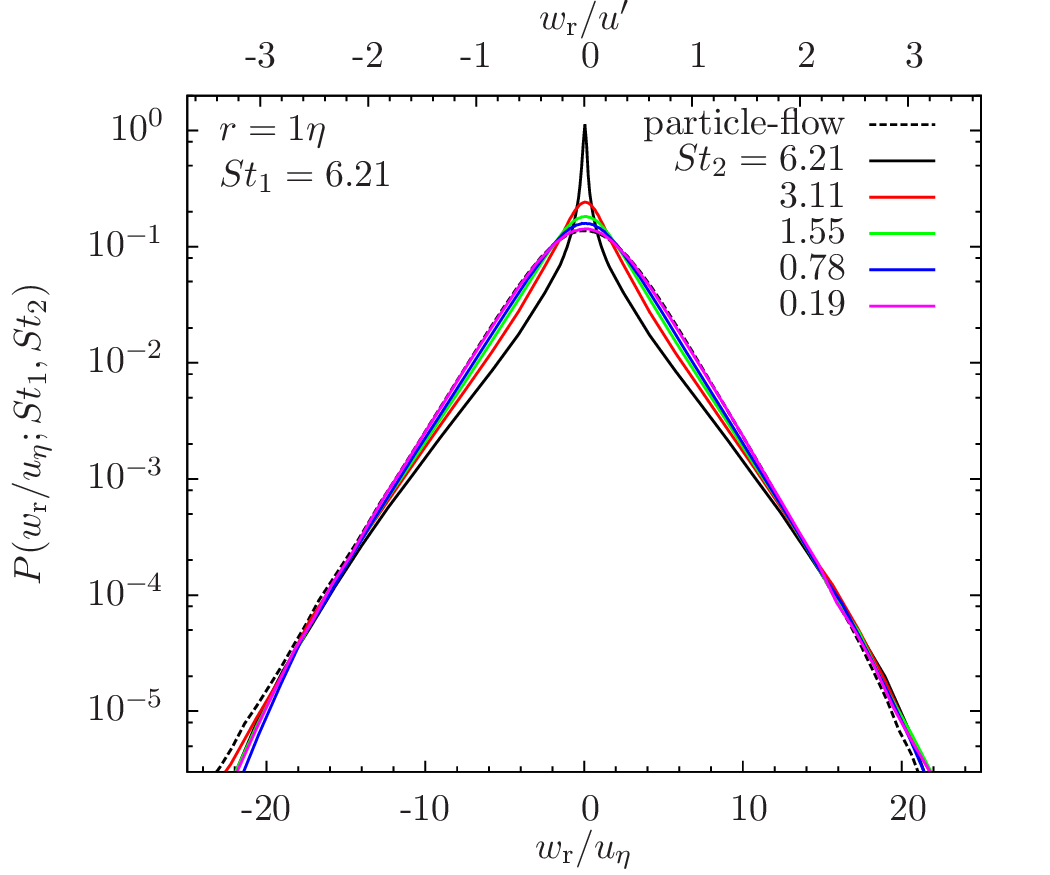}
\includegraphics[height=2.9in]{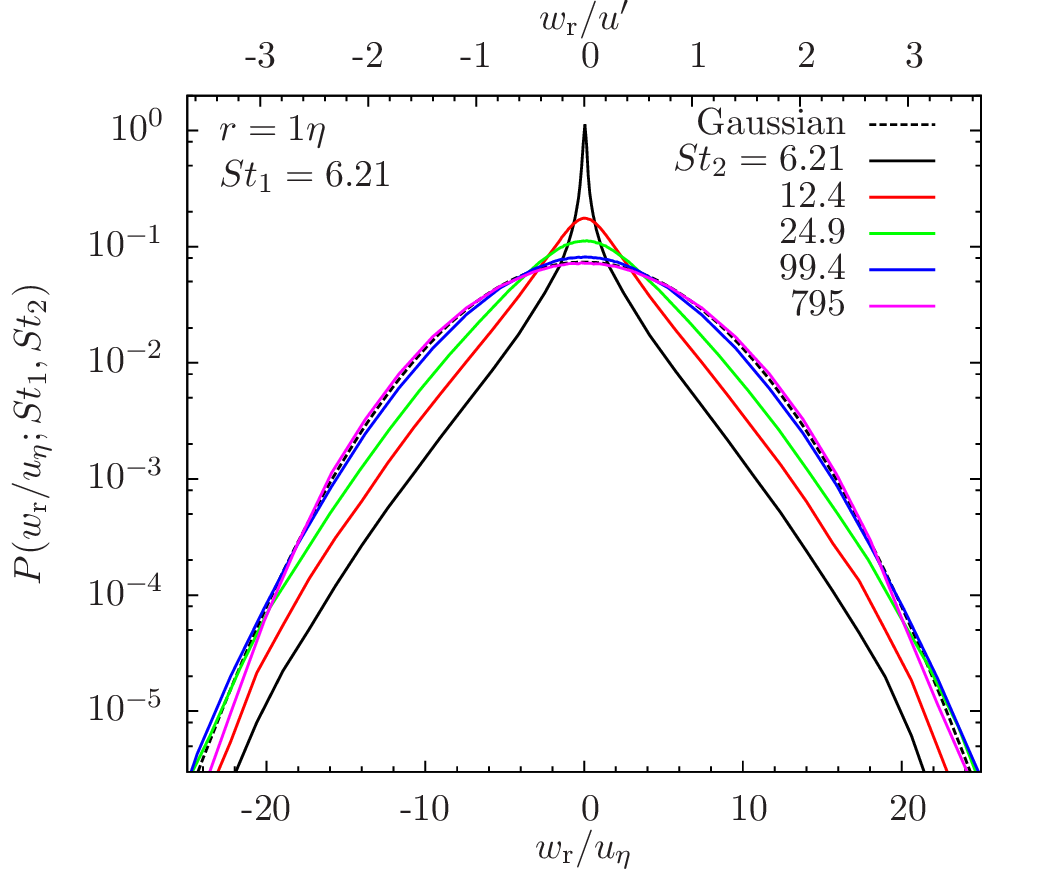}
\caption{PDF of the radial relative speed as a function of $St_2$ for $St_1 =6.21$. 
The figure is plot in the same way as Fig.\ \ref{radialpdf4}.
%The left and right panels showthe results for $St_1 \le St_2$ and  $St_1 \ge St_2$, respectively. 
%The solid black curves in both panels correspond the monodisperse PDF at $St =6.21$. 
%The dashed black line in the left panel is the radial PDF of the particle-relative 
%velocity at $r=1\eta$ for $St =6.21$ particles. 
The dashed black line in the right panel  shows the 
best Gaussian fit to the bidisperse PDF for $St_2 = 795$. %All the PDFs are measured at $r=1\eta$.
}
%{\bf  The normalizations of the particle friction time can be converted 
%using $\Omega = St/14.4$ and $\Omega_{\rm eddy} =St/19.2$}.}
\label{radialpdf6} 
\end{figure*}

\begin{figure*}[t]
\includegraphics[height=2.9in]{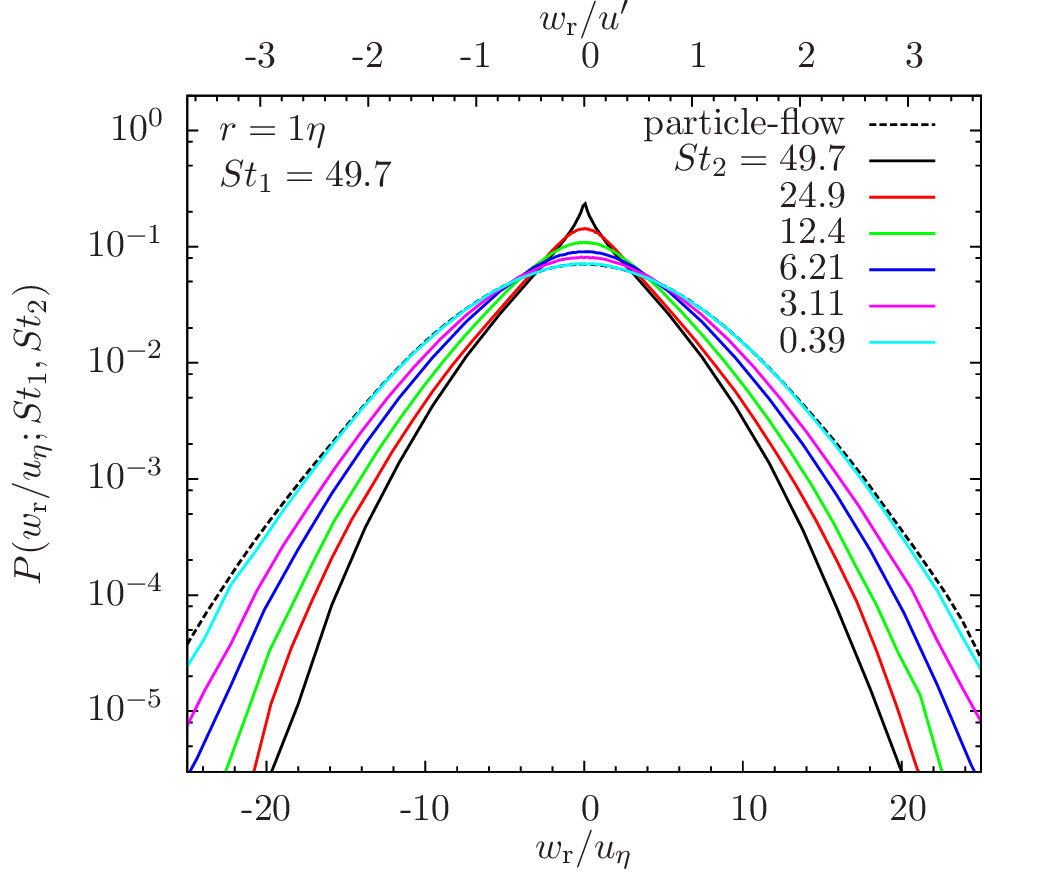}
\includegraphics[height=2.9in]{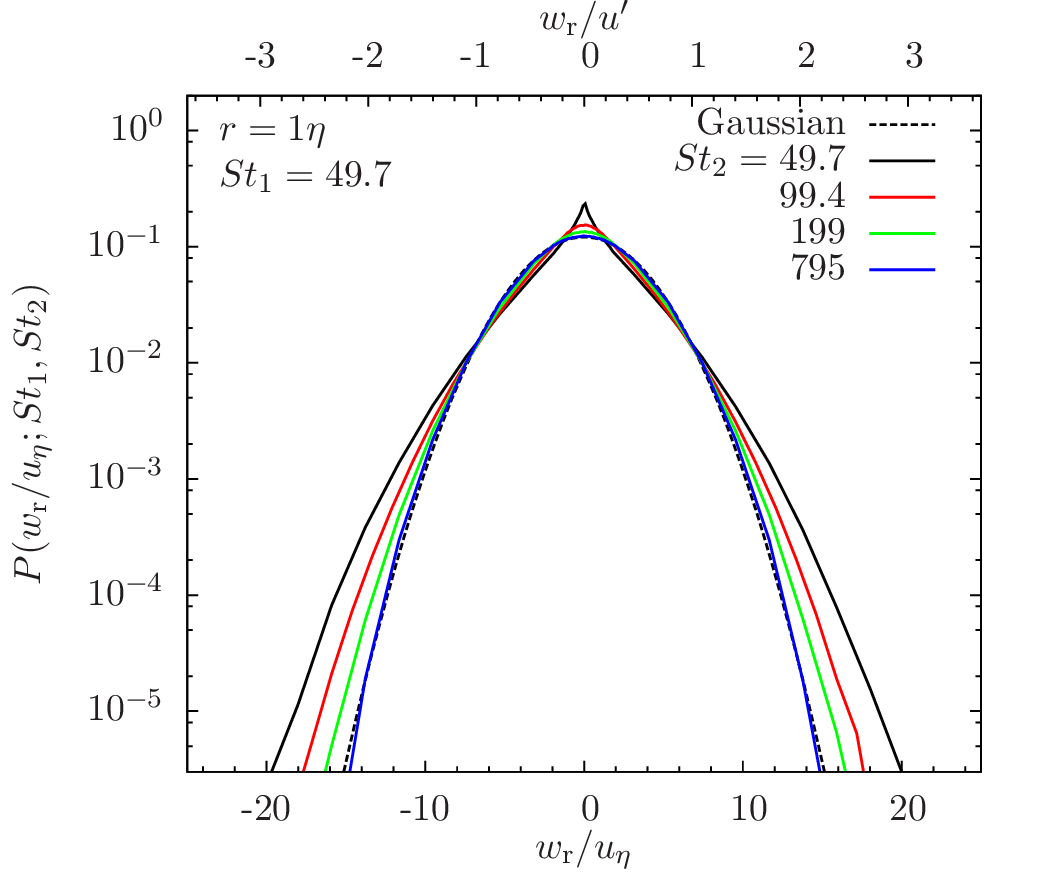}
\caption{Same as Figure \ref{radialpdf6}, but for $St_1 =49.7$.}
%{\bf  The normalizations of the particle friction time can be converted using $\Omega = St/14.4$ and $\Omega_{\rm eddy} =St/19.2$}.}
\label{radialpdf9} 
\end{figure*}

In Fig.\ \ref{tangentialpdf4}, we show the PDF of the tangential relative speed, 
$P(w_{\rm t}|w_{\rm r} \le 0; St_1, St_2)$ at a distance of $r=1\eta$ for $St_1 =1.55$. 
The conditioning on a negative radial relative speed ($w_{\rm r} \le 0$) indicates that only particle pairs 
approaching each other are counted. %
Since only approaching particles may lead to collisions, 
it is of practical interest to separate them out from the pairs moving away from each other. 
Fig.\ \ref{tangentialpdf4} is plot in the same way as Fig.\ \ref{radialpdf4}. 
The trend of $P(w_{\rm t}|w_{\rm r} \le 0; St_1, St_2)$ 
as a function of $St_2$ is similar to that of $P(w_{\rm r}; St_1, St_2)$. 
Unlike the radial component, the two wings of the tangential PDF are symmetric, as expected from 
statistical isotropy. We find that, for any $St_2$, the left wing of $P(w_{\rm r}; St_1, St_2)$ 
coincides with that of $P(w_{\rm t}|w_{\rm r} \le 0; St_1, St_2)$. 
The coincidence in the monodisperse case was found in Paper I  for all particles with 
$St\gsim 0.1$ in our simulation.\footnote{%As discussed in footnote 8, 
%the equalization of the radial and tangential components for separating 
%identical particle pairs is less efficient than the approaching ones. The right 
%wing of $P(w_{\rm r}; St)$  coincides with that of $P(w_{\rm t}|w_{\rm r} \ge 0; St)$ 
%only for $St \gsim 1$.
Because the longitudinal ($\Delta u_{\rm r}$) and transverse ($\Delta u_{\rm t}$) flow velocity differences in a turbulent 
flow are not equal, as $\Delta u_{\rm t} > \Delta u_{\rm r}$, the coincidence of   
the radial and tangential PDFs for approaching particle pairs  of equal size are not trivial. As discussed in Paper I, the 
equalization of  $w_{\rm r}$ and $w_{\rm t}$  
is essentially caused by the randomization of the direction of the primary separation, ${\bs r}_{\rm p}$, with respect to ${\bs r}$, due to 
the deviation of the particle trajectories from the flow elements and the turbulent separation of particle pairs backward in time (see \S 3.2).  
} %For $St$ below $\simeq 1$, the right wing of $P(w_{\rm r}; St)$ 
%is narrower than $P(w_{\rm t}|w_{\rm r} \ge 0; St)$.}. 
%A physical discussion on the mechanism that tends to equalize $w_{\rm r}$ and $w_{\rm t}$ in the monodispserse case was given in footnote7.
%Note that, for tracer particles with $St=0$, the left wing of $P(w_{\rm r}; St)$ is narrower than that of $P(w_{\rm t}|w_{\rm r} \le 0; St)$. 
%As discussed inPaper I, there are reasons responsible for approximately equal left wings of $P(w_{\rm r}; St)$  and 
%$P(w_{\rm t}|w_{\rm r} \le 0; St)$ for inertial particles in the monodisperse case. First,  the direction of ${\bs w}$ is randomized relative to ${\bs r}$ due to the derivation of particle 
%trajectories from flow elements. Second, the particle memory of the spatial flow velocity and the stochastic separation of particle pairs 
%backward in time also tend to make the radial and tangential components equal.  
For the bidisperse case, the generalized acceleration term gives equal contributions to $w_{\rm r}$ and $w_{\rm t}$ (\S 2), 
and it thus enhances the equalization of the radial and tangential PDF. 

%The coincidence  of the left  wings of $P(w_{\rm r}; St_1, St_2)$ and $P(w_{\rm t}|w_{\rm r} \le 0; St_1, St_2)$ 
%suggests that each radial or tangential component provides equal amount 
%of collision energy for any two approaching particles.  

The central part of the monodipserse PDF is very sharp with a cusp-like shape\footnote{The formation 
of the sharp cusp in the monodisperse PDF can be explained by the picture described in \S 3.2. 
For small particles, the faster backward separation of particle pairs at higher PDF tails causes a
 self-amplification of the tails, while the innermost part remains unchanged. 
As the amplification proceeds toward 
the center of the PDF with increasing $St$, the range of the unaffected central part is narrower, 
and  the inner part would appear sharper relative to the outer parts.  
As $St$ increases further, the innermost part of the PDF would eventually be affected, and the cusp 
then shrinks and finally disappears for $\tau_{\rm p} \gsim T_{\rm L}$.}
(see also Fig.\ \ref{radialpdf6} for the $St_1=6.21$ case, where the cusp in 
the monodisperse PDF is even sharper). 
%Note a sharp cusp forms at the central part of the PDF (also see the solid black lines in Figs.\ \ref{tangentialpdf4}, and \ref{radialpdf6})). 
In the bidisperse case, the central cusp smoothes rapidly as the Stokes number 
difference increases.  %In fact,  the central PDF quickly takes a Gaussian-like shape, 
%as $St_2$ decreases below or increase above $St_1$. 
This is caused by the contribution of the acceleration term.  As argued  
in \S 3.1, the PDF of the temporal trajectory velocity difference, $\Delta_{\rm T} {\bs u}$, 
is expected to take a smooth, Gaussian-like shape in the 
innermost  part.  Then, since $ {\bs w}_{\rm a} \propto \Delta_{\rm T} {\bs u}$ (eq.\ (\ref{approxaccel})), 
the bidisperse PDF would have a smoother central part, 
when the acceleration contribution increases. 
%with increasing Stokes number difference.

Fig.\ \ref{radialpdf6} plots the PDF of the radial relative velocity for $St_1 =6.21$. 
%The left and right panels correspond to $St_2 \le St_1$ and $St_2 \ge St_1$, respectively. 
The figure is plot in the same way as Fig.\ \ref{radialpdf4} for the radial PDF 
for $St_1 =1.55$.  As found in Paper I,  the fatness of the PDF shape for 
identical particles decreases with $St$ for $St \gsim 1$, and the tails of the 
monodisperse PDF (the black solid lines) for $St \ge 6.21$ are thinner than the $St_1 =1.55$ 
case shown in Fig.\ \ref{radialpdf4} 
(see explanation in \S 3.2). At $St \ge 6.21$, the skewness for the monodisperse 
radial PDF disappears. % and the left and right PDF wings are approximately symmetric. 
An explanation for the recovery of the symmetry at large $St$ 
was discussed earlier in this section.
%is because, for $St \ge 6.21$ particles, the 
%particle separation at a friction time ago is significant and insensitive to the initial condition 
%at $\tau \simeq 0$, and its direction is also completely random with respect to ${\bs r}$.  Since the contribution of 
%therefore, for $St_1 \gsim 6.21$, $P(w_{\rm r}, St_1,St_2)$ is symmetric for all $St_2$. 

Similar to the $St_1 =1.55$ case, the rms width of the PDF increases as $St_2$ decreases 
below or increases above $St_1$, while the overall shape of the PDF becomes thinner. 
Interestingly, in the left panel, the high PDF tails almost coincide for  all $St_2$  below $St_1$. 
In particular, the far tails in the dashed black line for Limit I (i.e., the particle-flow relative velocity) 
happen to be close to those in the black solid line for equal-size particles. 
As mentioned earlier, two effects, i.e., the increase of the rms and the thinning of the 
overall PDF shape with decreasing $St_2$, determine the trend of the 
probability at the far tails. The coincidence of the far tails suggests that, 
at $St_1 =6.21$, these two effects roughly cancel out. In the right panel, as $St_2$ 
increases above $St_1 =6.21$, the PDF keeps widening, and the PDF shape 
approaches Gaussian (the black dashed line) at $St_2 \gsim 99.4$. 

%Further check
%We also attempted to fit the PDF tails for $St_1 =6.21$ with stretched exponentials.  For $St_2$ below $St_1$, the far 
%tails of the PDF barely changes, and the best-fit $\alpha$ for the tails increase from 1.1 for 
%$St_2 =St_1$ to 1.3 for $St_2 =0.1$. For $St_2 \ge 6.21$, $\alpha$ 
%increases from 1.1 for the monodispers case to 1.4, 1.8 and 2 for 
%$St_2=24.9$, $99.7$ and $397$, respectively. 

%08   1.4
%09   1.6
%10   1.8
%11   1.9
%12   2 

Fig.\ \ref{radialpdf9} plots the PDF of the radial relative velocity for $St_1 =49.7$. 
In the left panel, we see that the PDF width increases with decreasing $St_2$. 
The widening of the PDF width wins over the thinning trends of the overall 
PDF shape with decreasing $St_2$, and thus the probability at the far tails 
keeps increasing with decreasing $St_2$. In the right panel, the PDF becomes thinner as 
$St_2$ increases above $St_1$, and finally approaches Gaussian, as expected for 
Limit III. 
%Note, however, that 
The rms width of all PDFs shown in this panel is almost the same. 
% Reconsider
%(see the right panel of Fig.\ \ref{3drms79} for the roughly constant 3D rms at $St_2 \ge St_1$). 

To summarize, for a given $St_1$, the tail shape of the relative velocity 
PDF is the fattest for the monodisperse case with $St_2 = St_1$. 
As $St_2$ increase above or decreases below $St_1$, %the rms width of the PDF broadens, but 
the generalized acceleration contribution increases, 
and the overall PDF shape becomes thinner. %as the distribution of the acceleration term is thinner than the shear term. 

\begin{figure*}[t]
\includegraphics[height=2.9in]{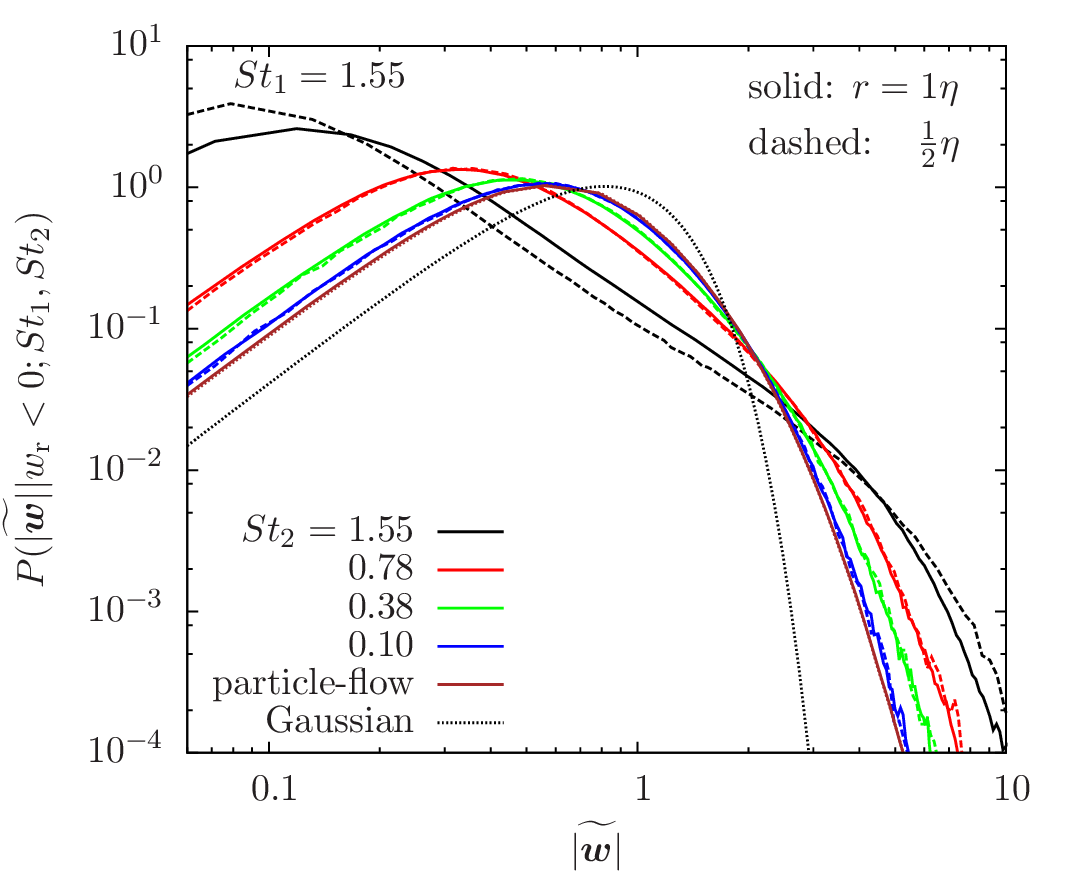}
\includegraphics[height=2.9in]{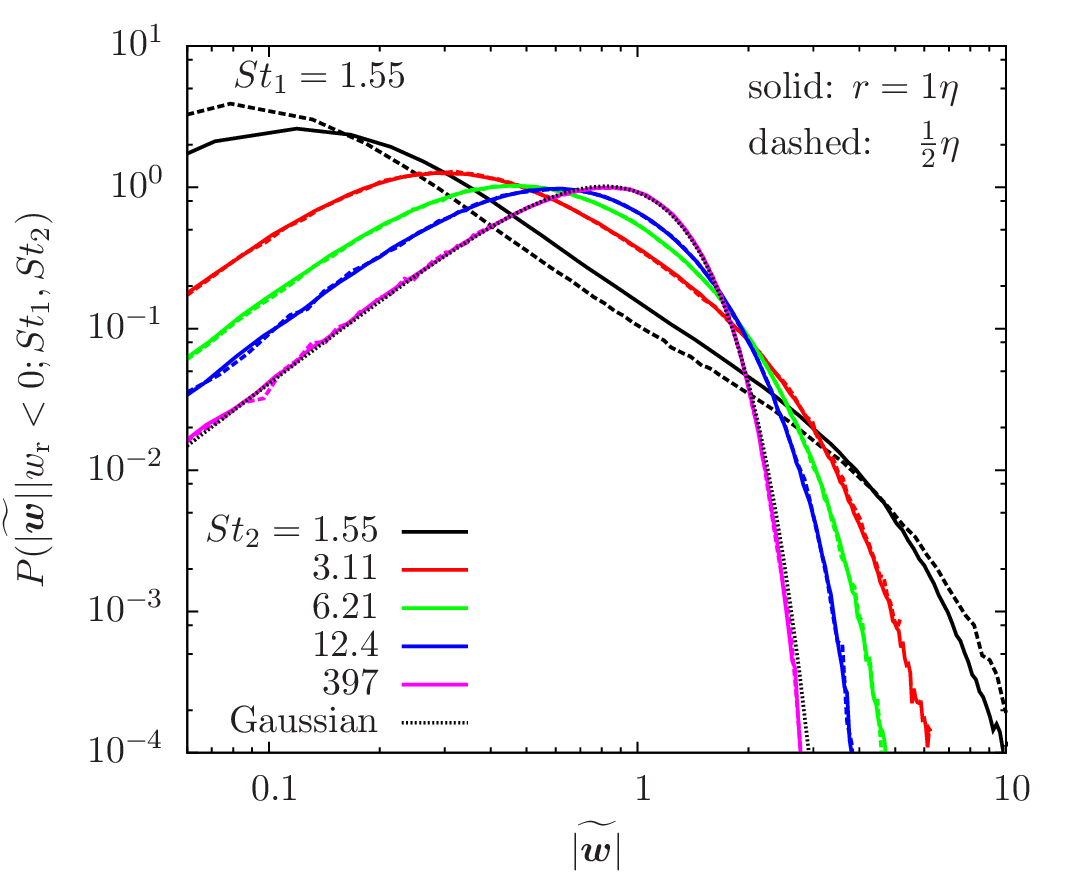}
\caption{Normalized PDF of the 3D amplitude, $|{\bs w}|$, of the 
relative velocity of approaching particle pairs ($w_{\rm r} <0$) 
at $r=1\eta$ (solid lines) and $\frac{1}{2}\eta$ (dashed lines). 
In both panels, black lines correspond to identical particles with $St_2 =St_1=1.55$. 
For each $St_2$, the amplitude, $|{\bs w}|$, is normalized to its rms value. %and the normalized PDF has unit variance. 
Left and right panels show results for $St_2 \le St_1$ and
%and the black dotted line corresponds to the particle-flow relative velocity 
%(i.e., $St_2=0$) at $r=0$.    
%3d assuming the PDF of each component is independent and has the same PDF $\propto \exp(-|x/0.19|^{0.6})$. 
$St_2  \ge St_1$, respectively.  In the left panel, the brown lines are the normalized 
PDFs of the particle-flow relative velocity at $r= 1\eta$ and $\frac{1}{2}\eta$.
Black dotted lines in both panels are the normalized PDF, $\sqrt{\frac{54}{\pi}} \widetilde{|{\bs w}|}^2\exp(-3\widetilde{|{\bs w}|}^2/2)$, 
for the amplitude of a 3D Gaussian vector.}
%{\bf  The normalizations of the particle friction time can be converted 
%using $\Omega = St/14.4$ and $\Omega_{\rm eddy} =St/19.2$}.}
\label{3dpdf4} 
\end{figure*}

\begin{figure*}[t]
\includegraphics[height=2.9in]{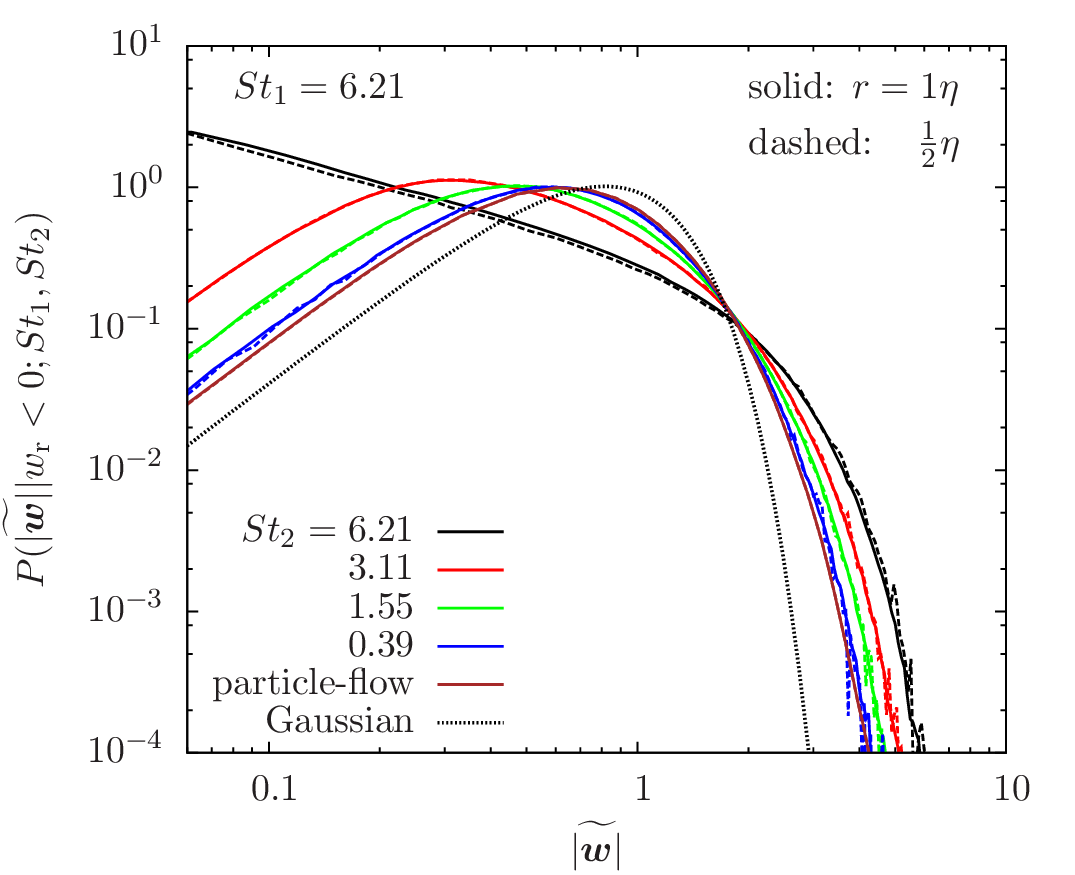}
\includegraphics[height=2.9in]{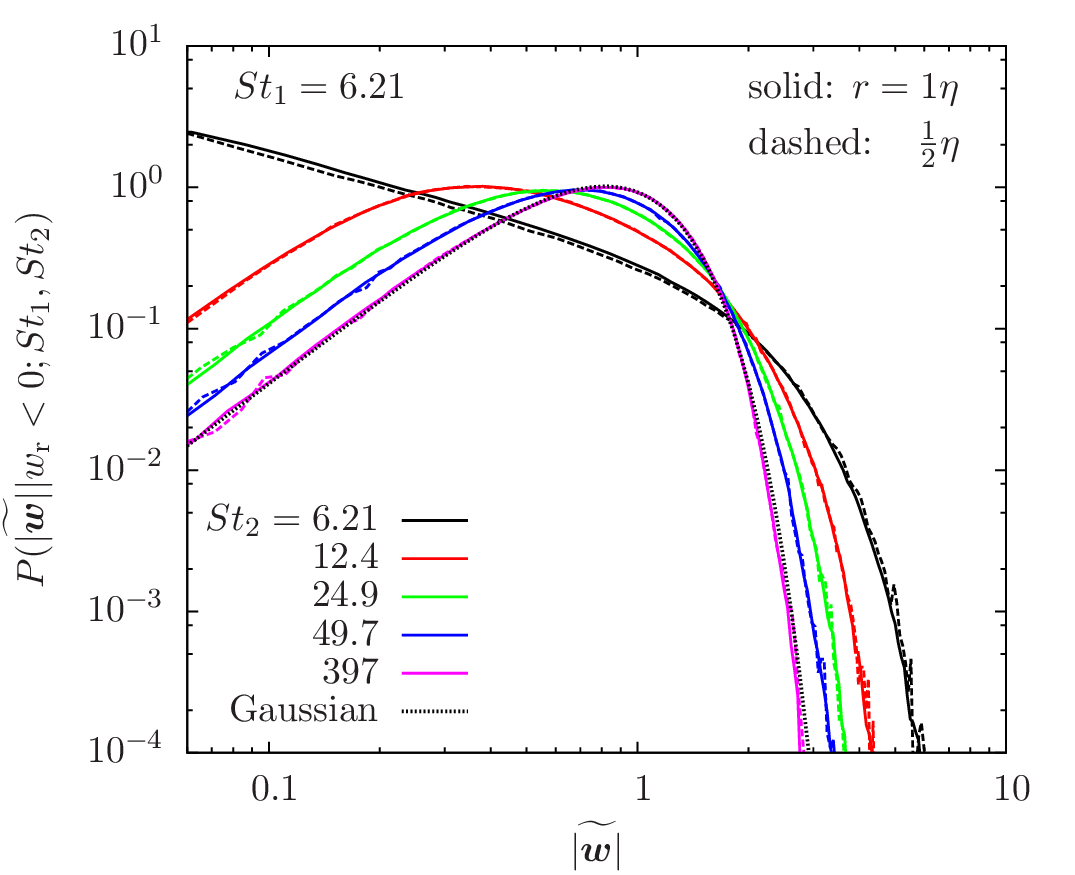}
\caption{Similar to Fig.\ \ref{3dpdf4}, but for $St_1 =6.21$. %Note that 
%the black dotted for the particle-flow relative speed at $r=\frac{1}{2}\eta$ 
%coincides with the magenta lines for $St_1 = 0.1$.
}
%{\bf  The normalizations of the particle friction time can be converted 
%using $\Omega = St/14.4$ and $\Omega_{\rm eddy} =St/19.2$}.}
\label{3dpdf6} 
\end{figure*}

%{\bf [Please move all the long paragraph below to the Appendix, and leave here only a brief sentence referring to the appendix.]}

An important conclusion of theoretical interest is that the distribution of the generalized shear
contribution, ${\bs w}_{\rm s}$, is fatter than the acceleration term, ${\bs w}_{\rm a}$. 
This is seen from a comparison of Limits I (black dashed lines) and II  (black solid lines) 
in Figs.\  \ref{radialpdf4}, \ref{radialpdf6} and \ref{radialpdf9}, which are dominated by the generalized 
acceleration and shear terms, respectively. Since ${\bs w}_{\rm a}$ and ${\bs w}_{\rm s}$ are related to the 
temporal ($\Delta_{\rm T} {\bs u}$) and spatial ($\Delta {\bs u}$) flow velocity differences, 
respectively, one may understand their distributions by considering the statistics of 
spatial and temporal velocity structures in turbulence.  We first note that the spatial velocity structures in turbulent flows do not  have a higher degree 
of non-Gaussianity than the temporal structures. %This, however, is not true.  
In fact, the Lagrangian temporal structures are known to be more intermittent than the Eulerian spatial structures (see Appendix A). 
Therefore, the finding that the distribution of ${\bs w}_{\rm s}$ is fatter than ${\bs w}_{\rm a}$ % that the fatter distribution of ${\bs w}_{\rm s}$ 
cannot be interpreted by a simple comparison of the temporal and spatial intermittency.
A key to understand the fatter distribution of ${\bs w}_{\rm s}$ is %why the distribution of ${\bs w}_{\rm s}$ is fatter than ${\bs w}_{\rm a}$ is 
that ${\bs w}_{\rm s}$ and ${\bs w}_{\rm a}$ sample the temporal and spatial flow velocity differences in different ways. 
Unlike the PDF of ${\bs w}_{\rm a}$, which roughly samples $\Delta_{\rm T} {\bs u} (\Delta \tau) $ at a single time lag $\Delta \tau \simeq \tau_{\rm p,h}$, 
${\bs w}_{\rm s}$ is not controlled by the spatial flow velocity difference, $\Delta {\bs u}$, at a single length 
scale. In fact, different parts of the distribution of ${\bs w}_{\rm s}$ 
sample the flow velocity at different scales. For example, in the case of equal-size particles 
with a friction time of $\tau_{\rm p}$, the PDF at relative velocities below/above the rms 
value, i.e., at $|w| \lessgtr \langle w^2 \rangle^{1/2}$, depends on 
the flow velocity at scales $\ell \lessgtr (r^2 + \langle w^2 \rangle \tau_{\rm p}^2)^{1/2}$, 
respectively. Considering that the PDF width of  $\Delta {\bs u}$ increases with increasing $\ell$, 
this implies that the PDF shape of ${\bs w}_{\rm s}$ is significantly fatter than the PDF of $\Delta {\bs u}$ 
at a single scale $\ell \simeq (r^2 + \langle w^2 \rangle \tau_{\rm p}^2)^{1/2}$. 
This fattening effect is responsible for why the distribution of ${\bs w}_{\rm s}$ 
has a higher degree of non-Gaussianity than ${\bs w}_{\rm a}$, even though 
the spatial flow velocity structures are not more intermittent than the temporal ones.

\subsubsection{The PDF of the 3D amplitude}

\begin{figure*}[t]
\includegraphics[height=2.9in]{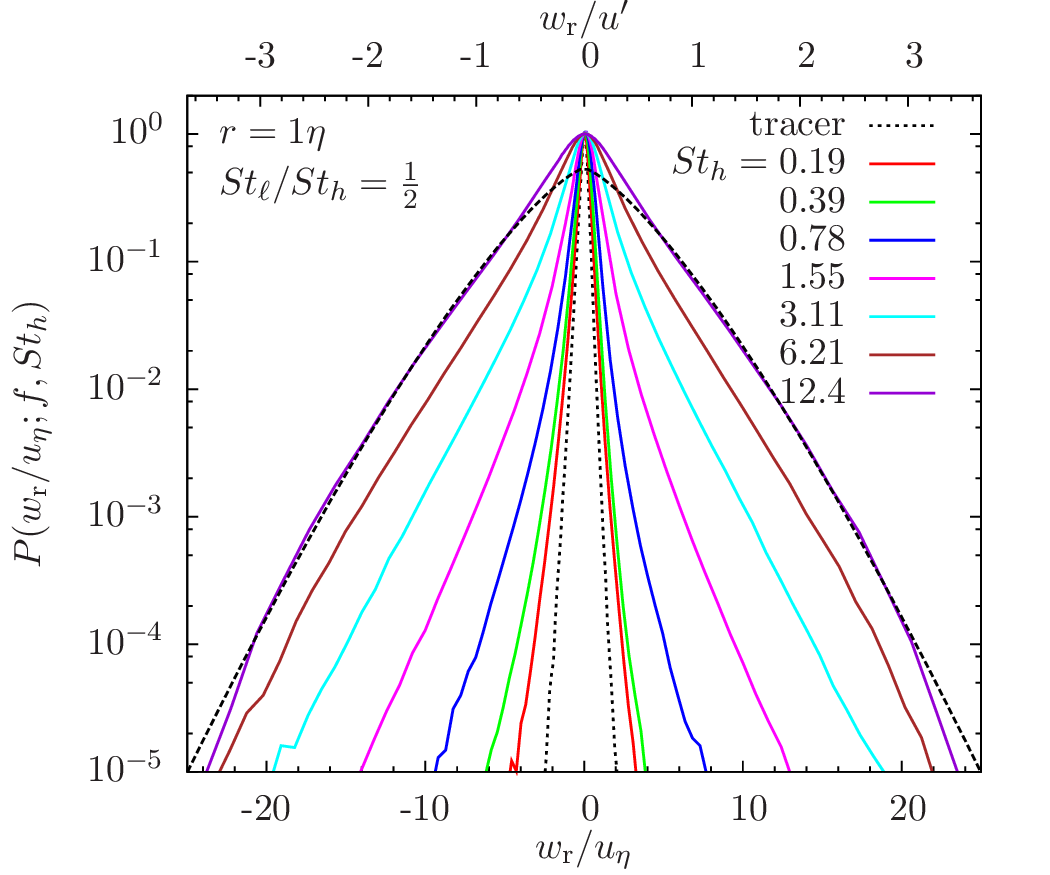}
\includegraphics[height=2.9in]{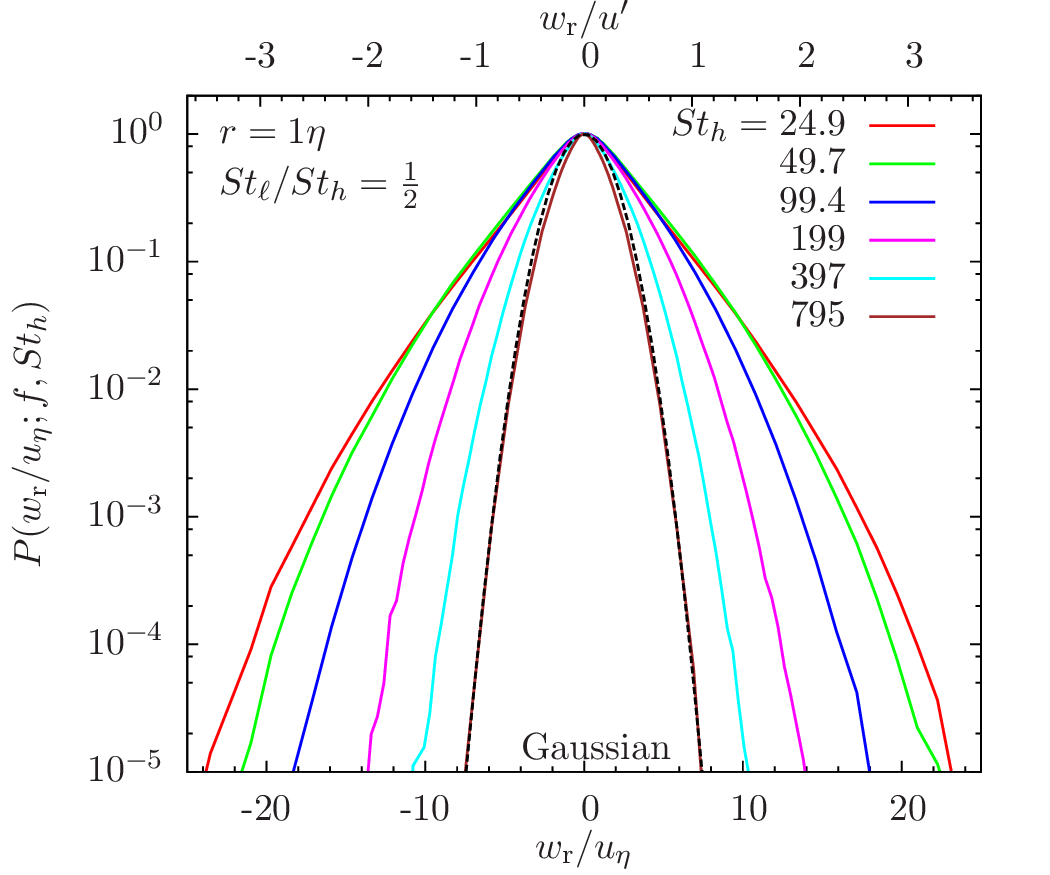}
\caption{PDF of the radial relative velocity at $r=1\eta$ for 
a  Stokes ratio $f=\frac{1}{2}$. Left  and right panels show $St_{ h} \le 12.4$ and $St_{ h} > 12.4$, 
respectively. All PDFs are normalized to the central peak values. 
The black dashed line in the left panel corresponds to the PDF of tracer particles, 
while the long dashed line is the stretched exponential fit with $\alpha =4/3$ 
to the PDF tails for $St_{ h} =12.4$. 
The black dashed line in the right panel is the Gaussian fit to the PDF of 
%the relative velocity 
the two largest particles in our simulation.  
}
\label{rfixedratio2} 
\end{figure*}

\begin{figure*}[t]
\includegraphics[height=2.9in]{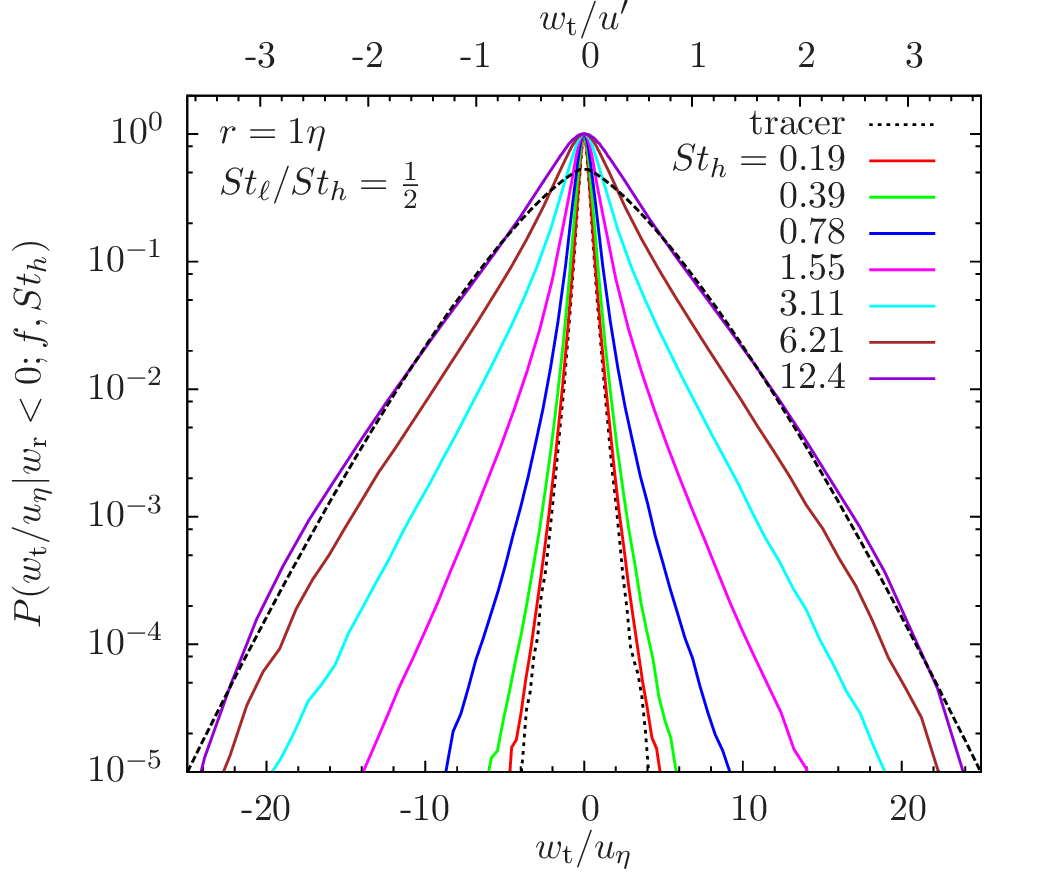}
\includegraphics[height=2.9in]{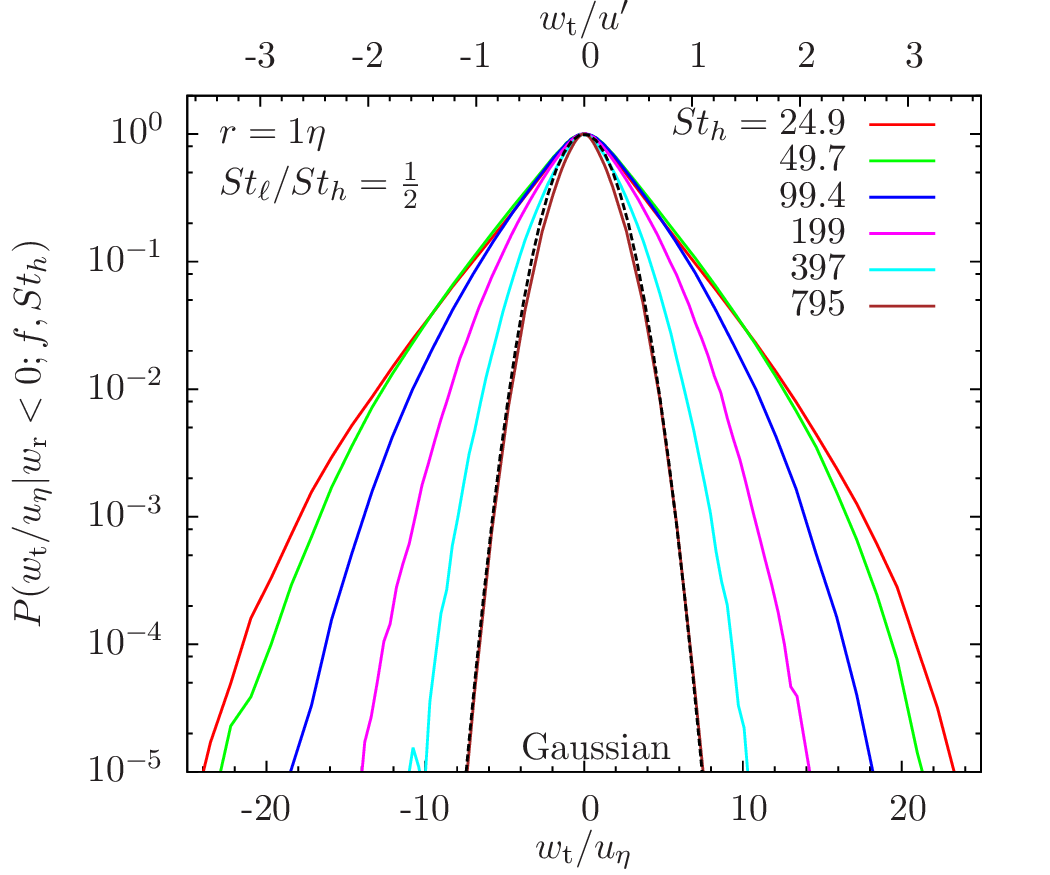}
\caption{The PDF of the tangential relative velocity at $r=1\eta$ 
for a  Stokes ratio of $f=\frac{1}{2}$. The figure is plot in the same way as Fig.\ \ref{rfixedratio2} 
for the radial PDF.}
\label{tfixedratio2} 
\end{figure*}

In Fig.\ \ref{3dpdf4}, we show the PDF of the 3D amplitude, $|{\bs w}|$, 
of the relative velocity of approaching particles with $w_{\rm r} <0$.  
In both panels, $St_1$ is fixed at $1.55$, and the left and right 
panels plot the PDFs for $St_2 \le St_1$ and $St_2 \ge St_1$, respectively. 
We normalized $|{\bs w}|$ to its rms value, i.e., $\widetilde{|{\bs w}|} \equiv |{\bs w}|/\langle w^2 \rangle_{-}^{1/2}$,
where $\langle w^2 \rangle_{-}^{1/2}$ is the 3D rms relative velocity of approaching particle pairs 
with negative $w_{\rm r}$. %Each normalized PDF has a unit variance. 
As discussed in Paper I,  $|{\bs w}|$ is related to the particle 
collisional energy and its PDF plays an important role in determining 
the collision outcome.  

In both panels of Fig.\ \ref{3dpdf4}, the solid and dashed black curves 
show the PDFs of equal-size particles with $St =1.55$ at $r=1$ and $\frac{1}{2}\eta$, 
respectively. The black dotted lines correspond 
to the normalized PDF for the amplitude of a 3D Gaussian vector. 
As shown in Paper I, the monodipserse PDF is extremely non-Gaussian, with 
large probabilities distributed at both very small and large relative speeds.
The large probabilities at  small $\widetilde{|{\bs w}|}$($\ll 1$) correspond 
to the sharp cusps at the central parts of the radial and tangential PDFs (see Figs.\ \ref{radialpdf4} 
and \ref{tangentialpdf4}). The solid and dashed color curves are bidisperse PDFs 
at $r=1$ and $\frac{1}{2}\eta$, respectively. As $St_2$ moves away from $St_1$, 
the degree of non-Gaussianity in the PDF decreases, consistent with the results in \S 4.1.1.  
When the Stokes numbers differ by a factor of 2, there is a rapid decrease in the probability at the left tail. 
This is because, due to the acceleration contribution, the central parts of the radial and tangential PDFs become smooth and 
the sharp central cusps disappear (see \S 4.1.1). 
As the Stokes number difference increases, the right PDF tail of $\widetilde {|{\bs w}|}$ become thinner, and  
%and a peak forms in the central part. % with $\widetilde{|{\bs w}|} \simeq 1$  
we also see that the peak of the PDF moves to the right  
toward $\widetilde {|\bs{w}|} \simeq 1$, meaning that more probabilities are distributed around the rms relative velocity.  %As the Stokes number ratio increases, the 
%PDF becomes closer to Gaussian (black dashed lines). 
In the left panel, as $St_2$ decreases to $0.1$, 
the PDF shape approaches that of the particle-flow relative velocity (brown lines), 
which is still fatter than Gaussian (the black dotted line). In the right panel, the PDF of  $|{\bs w}|$ approaches 
Gaussian in the limit of large $St_2$. At $St_2 \gsim 398$, the PDF coincides with the Gaussian 
distribution. The trend of the PDF shape as a function of $St_2$ is 
again consistent with the expectation that it lies in between the three limits, 
$St_2 \to 0$, $St_2 =St_1$ and $St_2 \to \infty$.  
Fig.\ \ref{3dpdf6} shows the simulation result  for $St_1 =6.21$, which is similar to  Fig.\ \ref{3dpdf4} for the $St_1 =1.55$ case.

In the monodisperse case, the PDF converges with decreasing $r$ already at $r\simeq 1\eta$, for particles with $St \gsim 6.21$ 
(see black solid and dashed lines in Fig.\ \ref{3dpdf6}). However, for smaller particles of equal-size with $St \lsim 3.11$, 
the PDF shape has an $r-$dependence at $r \simeq \frac{1}{2} \eta$ (see Fig.\ \ref{3dpdf4}). 
The convergence for the PDF of identical particles of small size is very challenging to reach, 
e.g., in the study of Lanotte et al.\ (2011), the convergence is barely reached at 
$r \simeq 0.086 \eta$ for  particles with $St =3.3$.  
In the bidisperse case, the convergence is easier to achieve because the contribution from the 
generalized acceleration term is $r-$independent (see \S 2 and Paper II). All the bidisperse PDFs for different particles 
shown in Figs.\ \ref{3dpdf4} and Fig.\ \ref{3dpdf6} already converge at $\frac{1}{2}\eta$. %and thus directly applicable to dust particle collisions at $r \to 0$.   
In fact, we find that, except for the two smallest particles with $St_1=0.1$ and $St_2 =0.19$ in our simulation (see Fig.\ \ref{3dpdfratio2} in \S 4.2.4), 
the convergence of the PDF shape is reached at $r \simeq \frac{1}{2} \eta$ for all non-equal Stokes pairs.

\subsection{The Relative Velocity PDF at Fixed Stokes Ratios}

In this subsection, we examine the relative velocity PDF for particle pairs with fixed Stokes number ratios, $f \equiv St_{\ell}/St_{h}$, and 
show how the PDF changes with $St_{ h}$.  As a reminder, $St_{\ell}$ and $St_{h}$ are the Stokes numbers of the smaller and larger particles, respectively. 
With a fixed $f$, 
it is easier to compare the bidisperse PDF with that of equal-size 
particles ($f =1$) discussed in Paper I. At a given $f$, there are two interesting limits, 
$St_{ h} \to 0$ and $St_{ h} \to \infty$. For $St_{ h} \to 0$, both particles become 
tracers, and the particle relative velocity PDF approaches the PDF,  $P_{\rm u} (r)$, of the flow 
velocity difference, $\Delta u$, across the particle distance, $r$. In the opposite limit $St_{ h} \to \infty$, 
the relative velocity is essentially the 1-particle velocity of the smaller particle, and 
its PDF approaches Gaussian (\S 3.3).  A particularly interesting result we find is 
that, if the friction time $\tau_{\rm ph}$ of the larger particle is close to the Lagrangian correlation time $T_{\rm L}$ of the flow, 
the PDF tails can be approximately described by a 4/3 stretched exponential 
for any value of $f$.   

\subsubsection{The  PDFs of the radial and tangential relative speeds}

In Fig.\ \ref{rfixedratio2}, we show the PDF, $P(w_{\rm r}; f, St_{ h})$, 
of the radial relative velocity at $r=1\eta$ for a Stokes ratio of $f=\frac{1}{2}$. The left 
and right panels plot  results for $St_{ h} \le 12.4$ and $St_{ h} \ge 24.9$, respectively. 
Each PDF is normalized to its value at the central peak (i.e., at $w_{\rm } =0$).  
In the left panel, the black short-dashed line is the PDF of the radial 
relative velocity between tracer particles (i.e., $St_{ h} =0$) 
at $r=1\eta$, and the long-dashed line is the stretched exponential 
function with $\alpha = 4/3$  (see eq.\ (\ref{se})) that best fits the 
PDF tails for $St_{ h} =12.4$. The dashed line in the right panel is the Gaussian 
fit for $St_{\ell} =397$ and $St_{ h} =795$. The figure is plot in the same way 
as Fig.\ 10 of Paper I for the monodisperse case (i.e., $f=1$).  
The PDF width first increases with $St_{ h}$, corresponding 
to the increase of $\Delta_{\rm T} {\bs u} (\tau_{\rm p, h})$ with $\tau_{\rm p,h}$ in the generalized 
acceleration term  (eq.\ (\ref{approxaccel})) 
and the increase of $r_{\rm p}$ with $\tau_{\rm p,l}$ in the shear term  (eq.\ (\ref{approxshear})).  
But for large $\tau_{\rm p, l}$ and $\tau_{\rm p, h}$, 
the $(1+\Omega_{\rm l})^{-1/2}$ and $[t_{\rm p}/(t_{\rm p} +\tau_{\rm p, h})]^{1/2}$ 
factors in eqs.\ (\ref{approxaccel}) and (\ref{approxshear}) take effect,  leading to the 
decrease of the PDF width at $\tau_{\rm p,h} \gg T_{\rm L}\simeq 14.4 \tau_\eta$ (the right panel). 
The behavior of the width or rms of the PDF at fixed $f$ as a function of 
$St_{ h}$  has been studied and explained in Paper II in the context of the PP10 picture.
%right panel of Fig.\ \ref{radialpm}. 
The left and right wings are asymmetric due to the shear contribution. 
The asymmetry first increases with $St_{ h}$ as $St_{ h}$ increases to 0.39, then 
decreases at larger $St_{ h}$. This is similar to the case of equal-size particles 
($f=1$; see Paper I) and consistent with the expected behavior of the 
shear contribution (see a physical discussion in \S 4.1.1). The PDF reaches 
symmetry at $St_{ h} \gsim 3.11$.

\begin{figure}[t]
\centerline{\includegraphics[width=1.1\columnwidth]{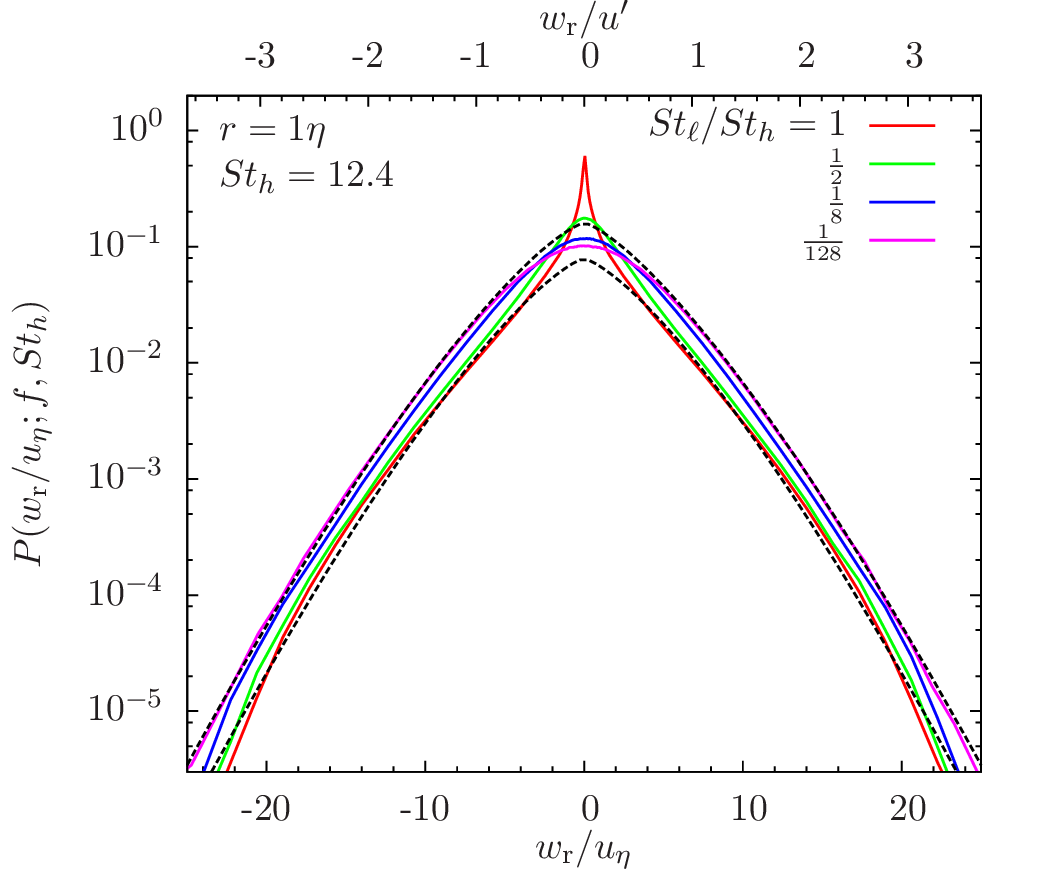}}
\caption{Radial relative velocity PDF for $St_{ h} =12.4$ and $f=1$, $\frac{1}{2}$, $\frac{1}{8}$, 
and $f\frac{1}{128}$. The friction time of $St_{ h} =12.4$ particles is 
close to the Lagrangian correlation time, $T_{\rm L}$, of the flow. Black dashed lines 
are stretched exponential fits 
with $\alpha =4/3$ to the PDF tails for $f=1$ and $\frac{1}{128}$, respectively. %For $\tau_{\rm p2} \simeq T_{\rm L}$, 
%the shape of the PDF tails appears to be invariant with $f$.
}
\label{stretch} 
\end{figure}

%\begin{figure}[h]
%\includegraphics[height=2.5in]{rlowr4.eps}
%\includegraphics[height=2.5in]{rhighr4.eps}
%\caption{The PDF of the radial relative velocity at $r=1\eta$ 
%for a  Stokes ratio of $f=\frac{1}{4}$. The figure is plot in the same way as Fig.\ \ref{rfixedratio2}.}
%\label{rfixedratio4} 
%\end{figure}
  
A comparison of  the left  panel of Fig.\ \ref{rfixedratio2} with that of Fig.\ 10 in  Paper I for identical 
particles  reveals an interesting difference. As explained in \S 3.2 (see Paper I for details), in the 
monodisperse case, the innermost part of the PDF closely follows the flow velocity difference PDF 
(the short-dashed line for tracers in Fig.\ 10 of Paper I). On the other hand, the central part of the 
bidisperse PDF at any $St_{ h}$ in the left panel of Fig.\ \ref{rfixedratio2} is wider than the short-dashed 
line for tracers. This is because the generalized acceleration term, which is absent for equal-size 
particles, contributes to the central part of the bidisperse PDF. Also, at the same $St_{ h}$, the PDF 
shape for $f = \frac{1}{2}$ is thinner than for the case of equal-size particles with 
$f=1$ (see \S 4.1.1 and also \S 4.2.3 below).  

In Fig.\ \ref{tfixedratio2}, we show the PDF of the tangential relative velocity, 
$P(w_{\rm t}|w_{\rm r} \le 0; f, St_{ h})$, for approaching particle pairs 
with $f=\frac{1}{2}$. The figure is plot in the same way as Fig.\ \ref{rfixedratio2} for 
the radial case. The two wings of the tangential PDF are symmetric for any $St_{ h}$. 
%as expected from statistical isotropy. 
The tangential PDF as a function of $St_{ h}$ shows a similar trend 
as  the case of the radial relative speed (Fig.\ \ref{rfixedratio2}). We find again that the left 
wings of the radial PDF and the tangential PDF, $P(w_{\rm t}|w_{\rm r} \le 0; f, St_{ h})$, 
of approaching pairs coincide (see \S 4.1.1).  
In Appendix C, we consider the tangential PDF, $P(w_{\rm t}|w_{\rm r} > 0; f, St_{ h})$, 
of separating particle pairs (with $w_{\rm r} > 0$), and compare it with 
$P(w_{\rm t}|w_{\rm r} \le 0; f, St_{ h})$. 
Similar to the asymmetric wings of the radial PDF,  for small particles of similar sizes  
there is a difference  between  $P(w_{\rm t}|w_{\rm r} \le 0; f, St_{ h})$ and  
$P(w_{\rm t}|w_{\rm r} > 0; f, St_{ h})$ for approaching and separating pairs.     

\begin{figure*}[t]
\includegraphics[height=2.9in]{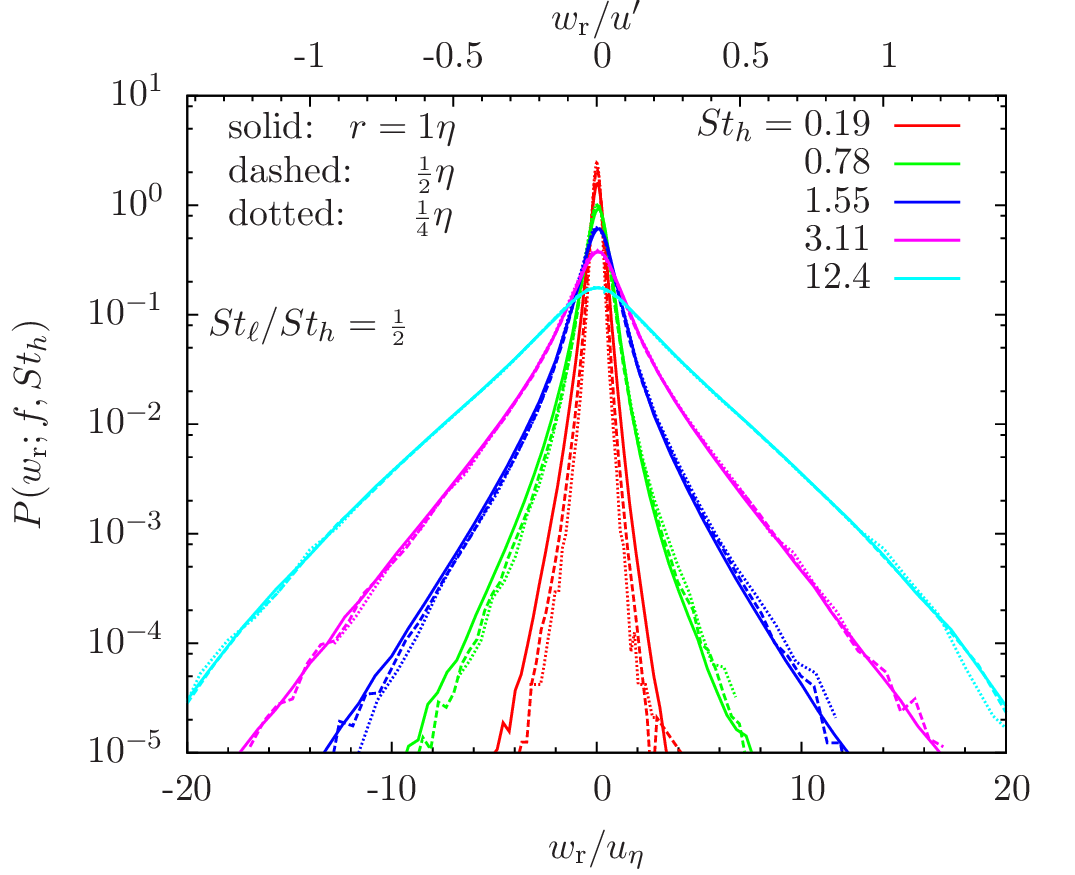}
\includegraphics[height=2.9in]{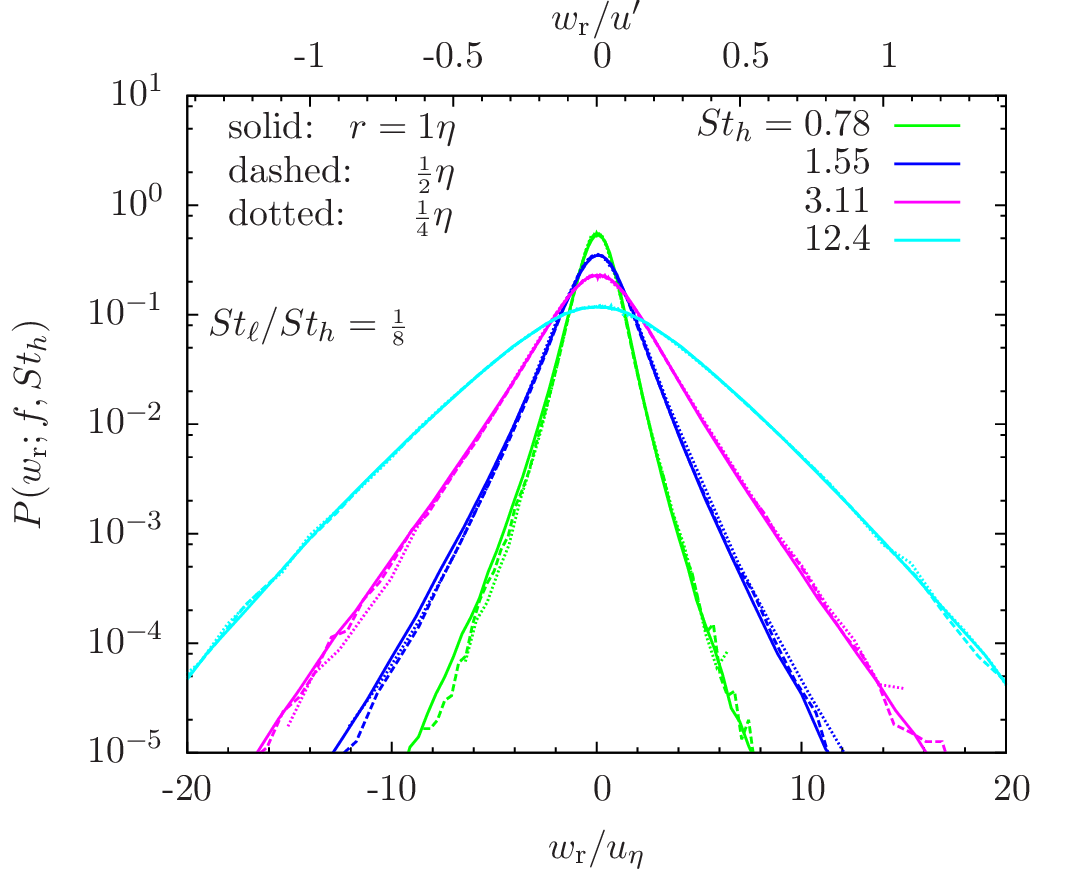}
\caption{The radial relative velocity PDF for particle pairs with $f=\frac{1}{2}$ (left panel) and $\frac{1}{8}$ 
(right panel) at $r=1\eta$ (solid), $\frac{1}{2}\eta$ (dashed), and $\frac{1}{4} \eta$ (dotted). 
}
\label{Rrdepend} 
\end{figure*}

We also examined the PDFs for other values of $f$.  %Fig.\ \ref{rfixedratio4} shows 
%the results for the radial PDF at $f= \frac{1}{4}$.  
The qualitative behavior of the PDF for different $f$ with 
increasing $St_{ h}$ is similar to the case of $f= \frac{1}{2}$. In \S 4.2.3, 
we will carry out a detailed quantitative analysis of the fatness of the PDF shape 
as a function of $St_{ h}$ at a few values of $f$.

In Paper I, we showed that the PDF tails for identical particles with $St = 12.4$ and $24.9$ ($\tau_{\rm p} \sim T_{\rm L}$) 
can be well described by a stretched exponential function with $\alpha =4/3$. 
A phenomenological argument for this stretched exponential based on the PP10 model was given in \S 3.2 (see also Paper I). 
From Figs.\ \ref{rfixedratio2} and \ref{tfixedratio2},  we see that such a stretched exponential fits well the PDF tails also 
in the bidisperse case with $f=\frac{1}{2}$ and $St_{ h} = 12.4$. 
The applicability of the $4/3$ stretched exponential is actually even more general. 
In Fig.\  \ref{stretch}, we show the radial PDFs for $St_{ h} = 12.4$ at 
4 values of $f$ in the range $ \frac{1}{128}\le f \le 1$.   
The shape of the PDF tails  is more or less invariant with $f$, and can be generally 
approximated by the 4/3 stretched exponential. In other words, a 4/3 stretched exponential 
applies as long as the friction time of the larger particle is close to $T_{\rm L}$.  However, 
note that, despite the invariance of the tail shape,  the central part and hence the 
overall shape do vary with $f$.

%One perspective to understand this shape invariance at the tail parts is the 
In \S 3.1, we observed that the particle-flow relative velocity, ${\bs w}_{\rm f}$, 
(corresponding to  $f=0$) for $St=12.4$ particles (see Fig.\ \ref{gaspartPDF}) 
shows 4/3 stretched exponential tails.  Since ${\bs w}_{\rm f}$ is largely controlled  
by the temporal flow velocity difference, $\Delta {\bs u}_{\rm T}$, along 
the particle trajectory, this suggests that, for particles with $\tau_{\rm p} \simeq T_{\rm L}$,
the PDF tails of $\Delta {\bs u}_{\rm T}(\Delta \tau)$ at $\Delta \tau \simeq T_{\rm L}$ 
are close to a 4/3 stretched exponential. Then, one may infer  from 
eq.\ (\ref{approxaccel}) that the acceleration term ${\bs w}_{\rm a}$ ($\propto \Delta {\bs u}_{\rm T}$)
would have 4/3 stretched exponential tails if the friction time 
of the larger particle, $\tau_{\rm p,h}$,  is close to $T_{\rm L}$.  
Considering that it fits well also the PDF tails of the generalized shear 
term, ${\bs w}_{\rm s}$, in the monodisperse case ($f=1$) with $\tau_{\rm p} \simeq T_{\rm L}$,  
%for particles of slightly different sizes also has 4/3 stretched exponential tails 
%if the fictions times of the two particles are both close to $T_{\rm L}$.
we would expect that the 4/3 stretched exponential  
applies to any $f$ as long as $\tau_{\rm p,h} \simeq T_{\rm L}$.
%This is actually seen in Fig.\ \ref{stretch}. 
%It is likely that 
The general validity of the 4/3 stretched exponential for $\tau_{\rm p, h} \simeq T_{\rm L}$ may 
have a profound physical origin, which, however,  is currently not clear to us.

\subsubsection{The $r-$dependence of the radial relative velocity PDF}

Fig.\ \ref{Rrdepend} shows the $r-$dependence of the radial PDF for 
$f= \frac{1}{2}$ (left panel) and $f= \frac{1}{8}$ (right panel). As discussed in \S 2, 
the generalized acceleration contribution only depends on individual trajectories of the 
two particles and is thus $r-$independent. The $r-$dependence of the relative 
velocity PDF comes only from the shear contribution.
In the left panel for $f= \frac{1}{2}$, the PDF width decreases with 
decreasing $r$ for the smallest $St_{ h}$ ($=0.19$). 
For small particles, the shear term depends on the local flow velocity 
difference, and thus decreases with decreasing $r$.
%This corresponds to the $r-$dependence of the radial rms relative speed shown in 
%Fig.\ \ref{rt} for the same Stokes pair.  
As $St_{ h}$ increases, the $r-$dependence of the PDF width becomes weaker, 
and in both panels the PDF is almost $r-$independent for $St_{ h} \gsim 3.11$.  
For larger $St_{\ell}$ and $St_{ h}$, the particle memory time is longer, 
and the particle distance at a friction/memory time ago is less sensitive to $r$. 
Therefore, the shear contribution becomes less dependent 
on $r$. At the same time, the increase of the acceleration contribution with 
$St_{ h}$ also tends to reduce the $r-$dependence. A comparison of the left and right panels shows that, at the same $St_{ h}$, the $r-$dependence 
of the PDF is weaker for smaller $f$, again due to the relatively larger acceleration 
contribution. The $r-$dependence for $f=\frac{1}{2}$ and $\frac{1}{8}$ shown here  is much weaker 
than the equal-size case ($f=1$).
It appears that, for $f=\frac{1}{8}$, the PDF already converges at 
$r=\frac{1}{4} \eta$ for all $St_{ h} \gsim 0.78$. 

\begin{figure*}[t]
\includegraphics[height=2.9in]{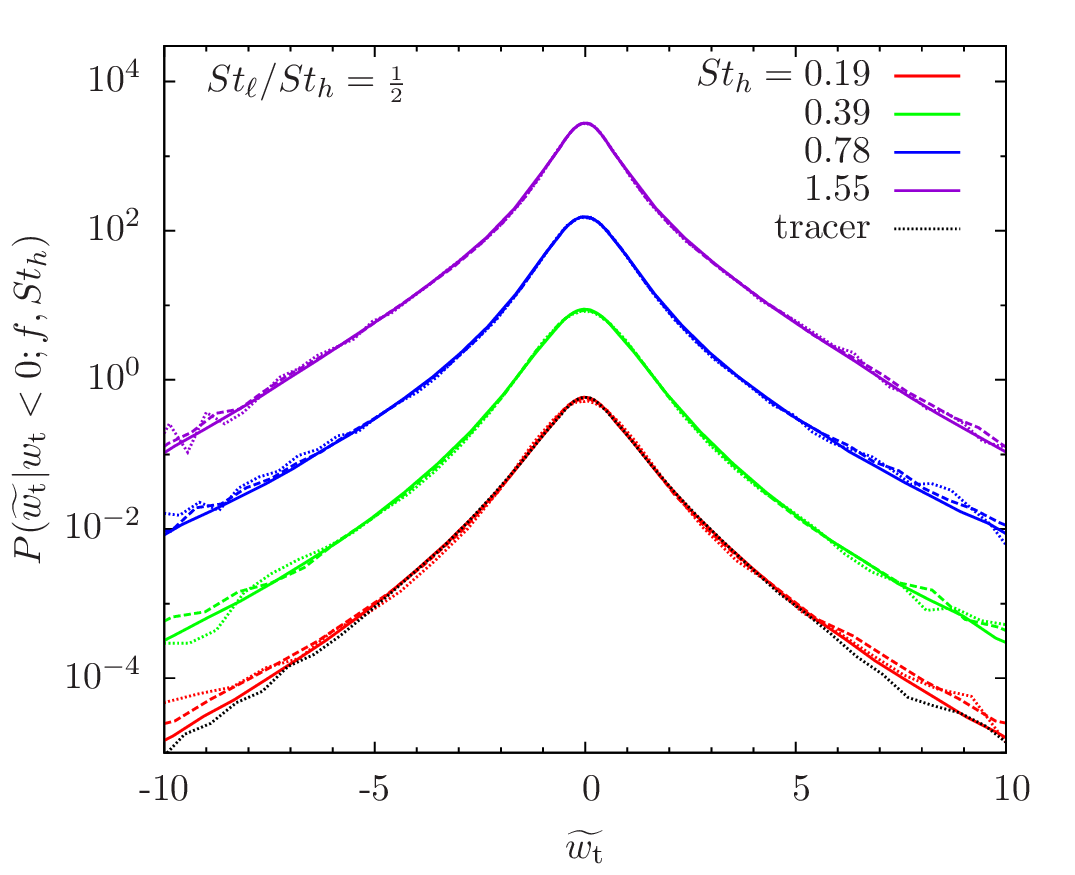}
\includegraphics[height=2.9in]{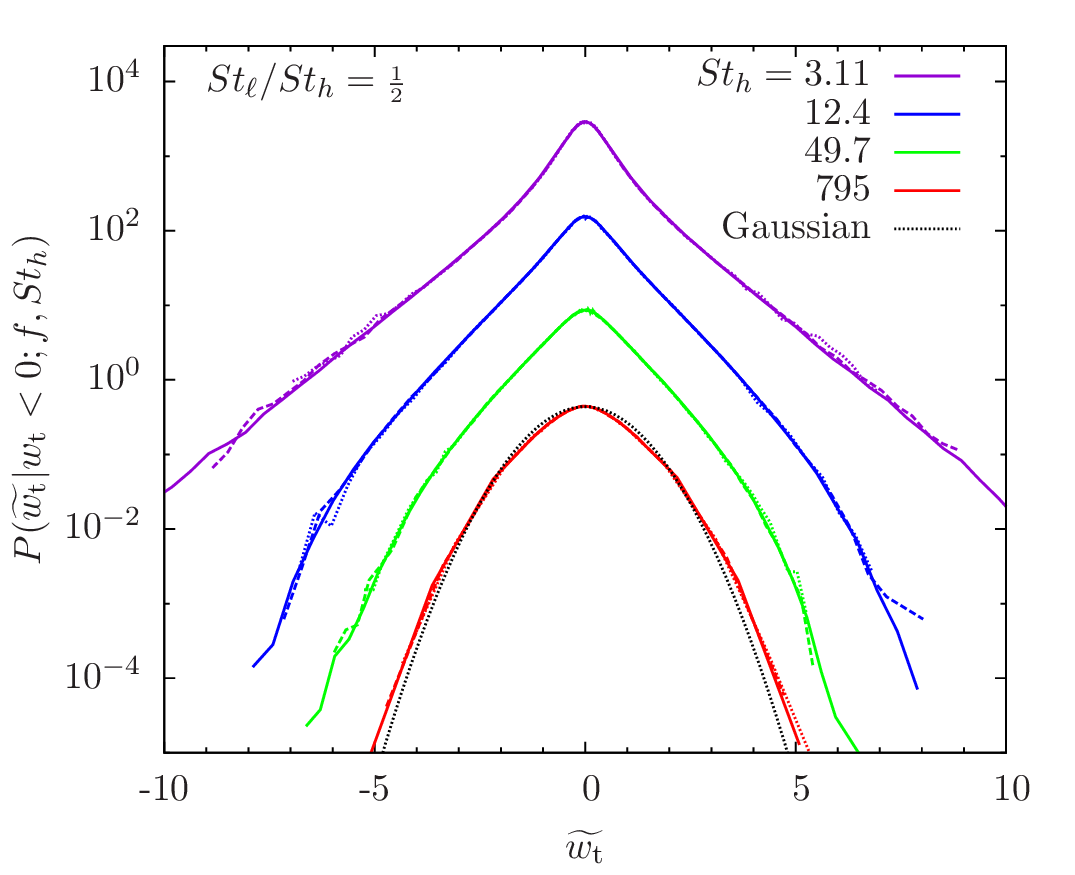}
\caption{The normalized PDF of the tangential relative velocity, $w_{\rm t}$, for  approaching 
pairs with $f=\frac{1}{2}$ at $r=1$ (solid), $\frac{1}{2}$ (dashed) and $\frac{1}{4}\eta$
(dotted). For each PDF,  $w_{\rm t}$ is normalized to its rms value, i.e.,  $\widetilde{w_{\rm t}} \equiv  w_{\rm t}/\langle w_{\rm t}^2 \rangle_{-}^{1/2}$. 
Left and right panels show $St_{ h} \le 0.78$ and $St_{ h} \ge 1.55$, respectively. In both panels, 
the bottom lines show the actual PDF values, while upper lines for each larger (left panel) or smaller (right panel) 
$St_{ h}$ are shifted upward by a factor of 16 for clarity. The black dotted lines in the left and right panels are the PDF 
of approaching tracer particles at $r=1\eta$ and the Gaussian fit for $St_{ h} =795$, respectively.}
%{\bf  The normalizations of the particle friction time can be converted using $\Omega = St/14.4$ and $\Omega_{\rm eddy} =St/19.2$}.}
\label{tangnormpdf2} 
\end{figure*}

\begin{figure*}[t]
%\centerline{\includegraphics[width=0.5\columnwidth]{scaleTmratio8.eps}}
\includegraphics[height=2.9in]{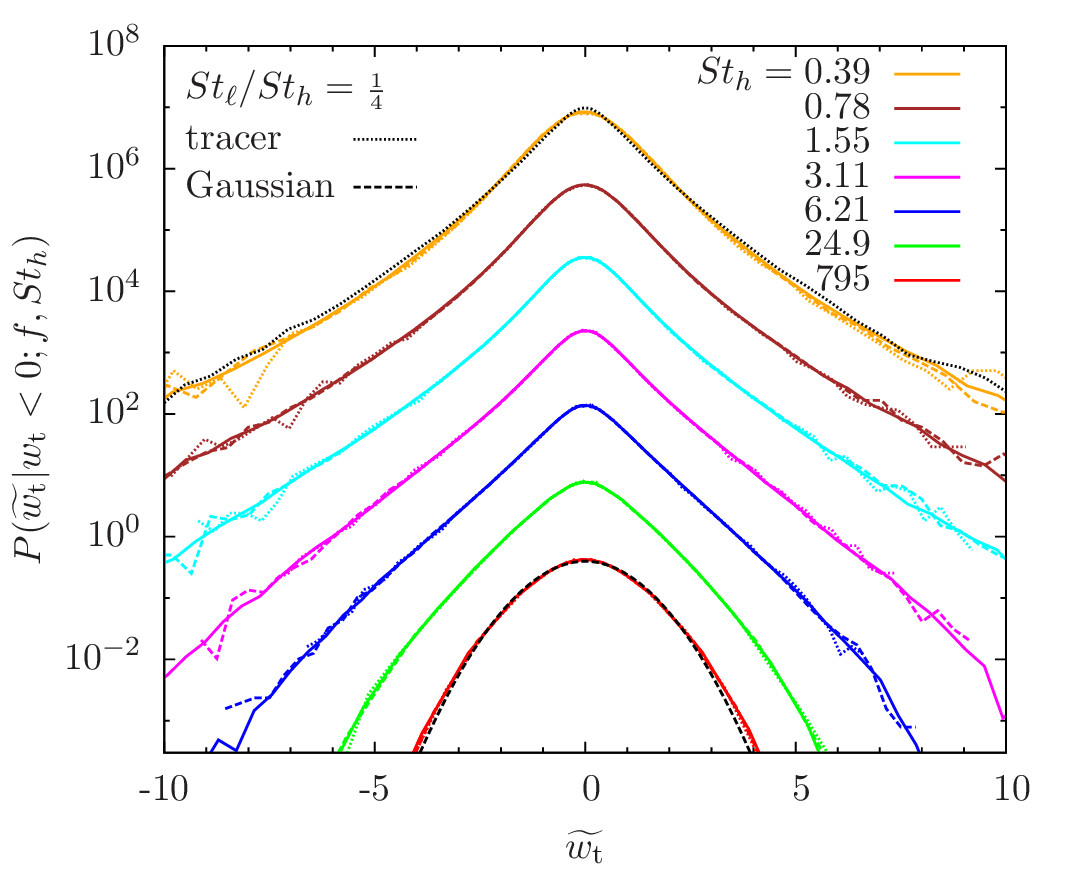}
\includegraphics[height=2.9in]{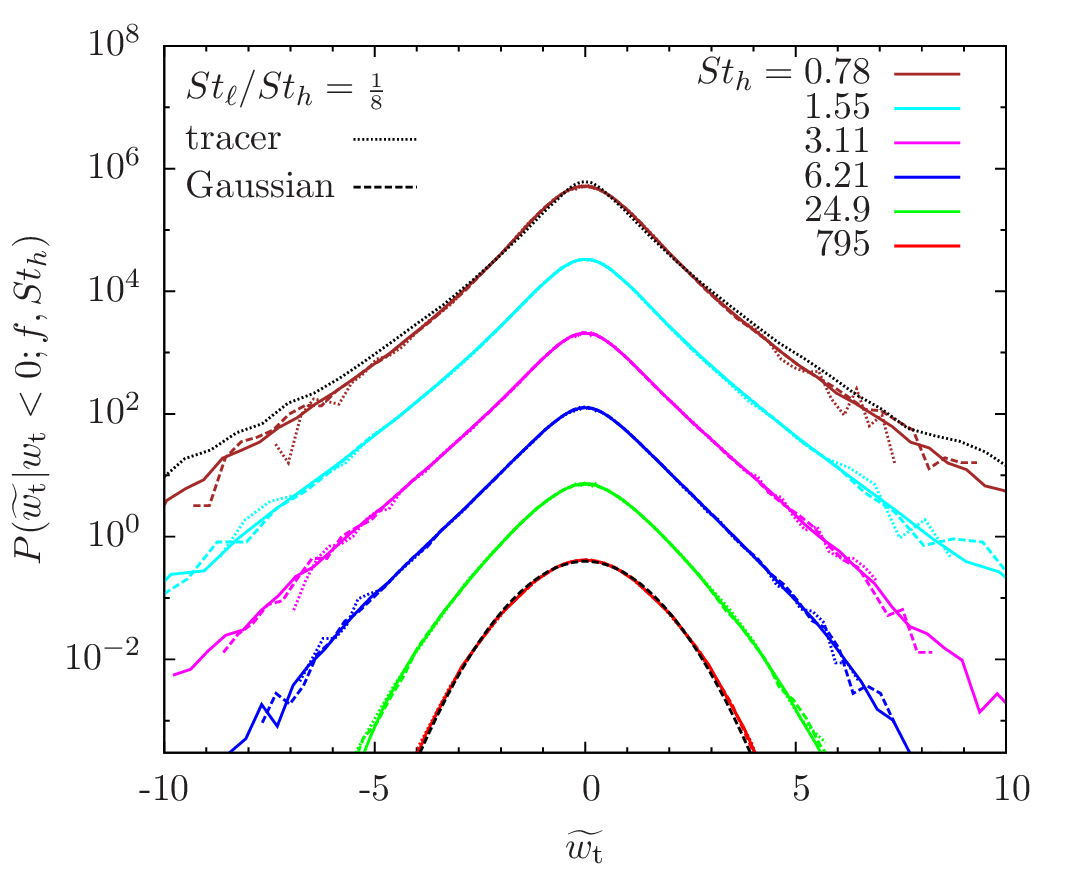}
\caption{The normalized tangential relative velocity PDF for  approaching 
pairs with $f=\frac{1}{4}$ (left panel) and $f=\frac{1}{8}$ (right panel) at $r=1$ (solid), $\frac{1}{2}$ (dashed) and $\frac{1}{4}\eta$ (dotted). 
The bottom lines for $St_{ h} =795$ show the actual values of the PDF, while 
the PDFs for each smaller  $St_{ h}$ is shifted upward by a 
factor of 16 for clarity. The black dotted and dashed lines in both panels are the PDF 
of tracer particles at $r=1\eta$ and the Gaussian 
fit for $St_{ h} =795$, respectively.}
%The fatness of the PDF decreases with increasing $St_{ h}$.}
\label{tangnormpdf48}
\end{figure*}

%Reconsider:
%The left wing of the PDF represents approaching pairs and is of practical interest for particles 
%collisions. At $St_{ h} \lsim 3.11$, the width of the left wing shows a slight $r-$dependence.  
%%This wing is $r-$independent for $St_{ h} \gsim 3.11$. But
%By normalizing the left wing to its own rms, we attempted to check whether 
%and how the shape of the left wing changes with $r$. It turns out that, except for $f=\frac{1}{2}$ 
%and $St_{ h} =0.19$, 
%the shape of the left wing is essentially $r-$independent, i.e., it  converges at $r\gsim \eta/4$. 
%The case with $f=\frac{1}{2}$ and $St_{ h} =0.19$ corresponds to the smallest two 
%particles in our simulation, and, for this Stokes pair, the left PDF tail slightly fattens 
%with decreasing $r$ (see also \S 4.2.3 for the normalized tangential PDF of approaching pairs).
%For $f= \frac{1}{8}$, the minimum $St_{ h}$ available in our simulation is 0.78.  
%The right of tail of the radial PDF, corresponding to separating particle pairs 
%has a more complicated trend with $St_{ h}$. It first becomes narrower 
%as $St_{ h}$ increases to $0.39$. This is due to the shear contribution: the  

At a given $f$ and $St_{ h}$, the left and right tails of the 
radial PDF tend to be more symmetric as $r$ decreases. In particular, for $St_{ h}= 0.78$ 
and $1.55$ in both panels, the right tail slightly increases with decreasing $r$, while 
the left wing becomes narrower. The opposite trends of the two tails both reduce the 
negative skewness of the PDF.  The increase in the right tail for these  values of 
$St_{ h}$ with decreasing $r$ is due to the shear contribution. 
%as due to the particle memory of the flow velocity 
%difference in the past. 
The right wing corresponds to separating pairs, and, backward in time, 
the distance of these pairs first decreases in near past  and then starts to increase 
as the two particles move past each other. At smaller $r$, it takes a shorter time for 
the distance of these pairs to switch from decreasing to increasing. Therefore, 
the primary distance, $r_{\rm p}$, of separating pairs at a friction time ago 
could increase with decreasing $r$, leading to a larger shear contribution 
and the increase in the right wing. On the other hand, for approaching pairs, $r_{\rm p}$ always 
tends to be smaller at smaller $r$, and the width of the left wing decreases with 
decreasing $r$. Overall, as $r$ decreases, the difference of $r_{\rm p}$ between 
approaching and separating pairs decreases, reducing the asymmetry of the two wings.  

We attempted to check how the shape of the left wing for approaching particles changes with 
decreasing $r$ by normalizing the wing to its own rms.  It turns out that, except for 
$f=\frac{1}{2}$ and $St_{ h} =0.19$, the shape of the left wing is almost invariant 
with $r$, indicating that the $r-$dependence of the shape is weaker than that of the width or rms.  
The trend of the shape of the left wing with $f$ and $St_{ h}$ 
is the same as the tangential PDF of approaching pairs, which we discuss in the next section.

\subsubsection{The normalized PDF of  the tangential relative velocity}

In Fig.\ \ref{tangnormpdf2}, we show the normalized PDF, $P(\widetilde{w_{\rm t}}|w_{\rm r} <0; f, St_{ h})$, 
of the tangential relative speed for approaching particle pairs ($w_{\rm r}<0$) with a Stokes ratio of $f=\frac{1}{2}$. 
%at different particle distances. 
We normalized $w_{\rm t}$ to its rms value, i.e., $\widetilde{w_{\rm t}} \equiv  w_{\rm t}/\langle w_{\rm t}^2 \rangle_{-}^{1/2}$, 
where the subscript ``-" indicates that only approaching pairs are counted. All the normalized PDFs have 
a unit variance. The normalization gives a clearer comparison of the 
PDF shape. The black dotted line in the left panel is the PDF of approaching tracer particles\footnote{The PDF {\it shape} of the relative velocity 
of trace particles already converges at $r=1\eta$, even though its width decreases with decreasing $r$. The tracer particle follows the local 
flow velocity, and, as discussed in Appendix B, the PDF shape of the flow velocity difference approaches the velocity 
gradient distribution at sufficiently small $r$.} at $r=1\eta$, corresponding to 
$St_{ h} =0$. As $St_{ h}$ increases to $0.78$, the tails of $P(\widetilde{w_{\rm t}}|w_{\rm r} <0; f, St_{ h})$ 
show a slow fattening trend. The fattening appears to be weak, but is verified by a computation of the kurtosis.   
Then, starting from $St_{ h} =1.55$, the PDF shape becomes continuously 
thinner (the right panel). Similar to the monodisperse case, the PDF fatness peaks at $St_{ h} \simeq 1$. 
%The figure is plot in the same way as Fig.\  12 of Paper I for the monodisperse case.  
%Consistent with previous observations, a companion with Fig.\  12 of 
%Paper I shows that the radial PDF for $f=\frac{1}{2}$ is significantly thinner than in 
%the monodisperse case.  

We tried to obtain a quantitative estimate of the PDF shape for $f=\frac{1}{2}$ 
by fitting the tails of $P(w_{\rm t}|w_{\rm r} \le 0; f, St_{ h})$ with stretched exponential functions (eq.\ (\ref{se})). 
%We give a more quantitative description for the fatness of the PDF by fitting the tails of 
%the tangential PDF,  $P(w_{\rm t}|w_{\rm r} \le 0; St_1, St_2)$, with stretched exponential 
%functions, eq.\ (\ref{se}). 
%The fits also apply to the left wing of the radial PDF $P(w_{\rm r}; f, St_{ h})$, as it coincides with 
%that of $P(w_{\rm t}|w_{\rm r} \le 0; f, St_{ h2})$. 
We find that, as $St_{ h}$ increases from 0.19, to 0.39 and 0.78, the best-fit value of 
$\alpha$ decreases from 0.7($r=1\eta$), to 0.67 and 0.6, indicating a slight tail fattening of the PDF.  %increases toward $1$. 
As $St_{ h}$ increases further, $\alpha$ starts to increase. The best-fit $\alpha$ is 
0.65,  0.8,  1.15, 1.33,  1.45,  1.55,1.65, 1.7, 1.75  and 1.9 for $St_{ h} =1.55$,  3.11, 6.21, 12.4,   24.9, 49.7, 99.4, 199, 397 and 795, 
respectively. For $St_{ h} = 795$, the PDF is close to Gaussian 
(the black dotted line in the right panel), but the best-fit $\alpha$ is $1.9$, slightly smaller than 2.  
The measured $\alpha$ values also apply to the left wing of the radial PDF for $f=\frac{1}{2}$, 
which coincides with the left wing of $P(w_{\rm t}|w_{\rm r} <0; f, St_{ h})$ (see \S4.2.1). These values are generally larger 
than those obtained in Paper I  for the monodisperse PDF with $St$ equal to 
$St_{ h}$ listed here. This is due to the effect of the acceleration contribution which tends to make the PDF 
tails thinner. 
%Similar to the monodisperse case, the PDF fatness peaks at $St_{ h} \lsim 1$. 

%{\bf [I wonder if the two paragraphs below are strictly necessary. It gets a bit tiring and boring, and somewhat repetitive. Could they be cut?
%The paper is very long anyway.....]}

To understand the trend of the PDF shape at $f=\frac{1}{2}$, 
we consider again the behavior of the generalized acceleration (${\bs w}_{\rm a}$) and shear (${\bs w}_{\rm s}$) 
contributions to the PDF. %as a function of $St_{ h}$. 
%As discussed in \S 2.2, ${\bs w}_{\rm a}$ can be approximately estimated as $\propto \Delta {\bs u}_{\rm T} (\tau_{\rm ph})$. 
%Since ${\bs w}_{\rm a} \propto \Delta {\bs u}_{\rm T} (\tau_{\rm p,h})$ and the distribution of $\Delta {\bs u}_{\rm T}$ 
%is thinner at larger time lag, 
As discussed earlier, the PDF shape of ${\bs w}_{\rm a}$ would become thinner continuously with increasing $\tau_{\rm p,h}$ (see \S 3.1 and \S 4.1.1). 
%Therefore,  the contribution of ${\bs w}_{\rm a}$ 
%tends to make the relative velocity PDF thinner at larger $St_{ h}$. 
Using the monodisperse case as a guideline for the generalized shear contribution, 
%${\bs w}_{\rm s}$, 
%The shear term, ${\bs w}_{\rm s}$, is determined by the particles' 
%memory of the spatial flow velocity difference, $\Delta u (\ell)$, in the past (\S 2.2 
%and in Paper I).  Particle pairs lying at higher PDF tails separate faster 
%backward in time, and thus ``remember"  $\Delta u (\ell)$ at larger scales $\ell$, giving rise to an amplification at the tail part of the PDF.  
%Based on the discussion in \S 3.2,  the effect of tail amplification fattens 
the distribution of ${\bs w}_{\rm s}$ would first fatten as $St_{ h}$ increases toward 1, and then become thinner
%due to the amplification at the tail part.     
%the reason for the tail amplification is that
as the Stokes number increase further above 1. 
%the shear contribution samples 
%the PDF, $P_{\rm u}(\Delta u, \ell)$, of the flow velocity difference, $\Delta u (\ell)$, 
%at larger scales, and its distribution becomes thinner due to the thinning trend 
%of $P_{\rm u}(\Delta u, \ell)$ with increasing $\ell$. 
% the thinning trend of the PDF of the the shear term starts from $St \gsim 1$. 
%Therefore, the PDF of ${\bs w}_{\rm s}$ first fattens with $St_{ h}$ and then becomes thinner as $St_{ h}$ keeps increasing.  
%The discussion above suggests that, 
Therefore, at small $St_{ h} \lsim 1$, the distributions of ${\bs w}_{\rm a}$ and ${\bs w}_{\rm s}$ have 
opposite trends with increasing $St_{ h}$, and their competition determines the fatness 
of the relative velocity PDF.  For $f=\frac{1}{2}$, it appears the fattening trend by ${\bs w}_{\rm s}$ 
wins, and the PDF of ${\bs w}$ becomes fatter as $St_{ h}$ increases to $0.78$. 
Since the acceleration contribution tends to counteract the fattening trend, the 
increase of the tail fatness in the range $0.19 \le St_{ h} \le 0.78$ 
is significantly weaker than in the monodisperse case.  As $St_{ h}$ increases above 1, 
the PDF of ${\bs w}_{\rm s}$ starts the thinning trend, and, 
together with the thinning effect by the acceleration term, ${\bs w}_{\rm a}$, 
and the fatness of the PDF 
of ${\bs w}$ decreases, %for $St_{ h}$ above $\simeq 1$, 
as observed in Fig.\ \ref{tangnormpdf2}.

%In an attempt to fit the left wing of the radial 
%PDF and the tangential PDF (not shown) of approaching pairs for $f= \frac{1}{4}$, 

% f=1, 0.67
% 0.67, 0.52,  0.48,   0.49,    1,    1.1,    1.3,   1,33, 1.45, 1.5, 1.65, 1.75, 1.9
%f=1/2
%0.7,    0.67,    0.6,   0.65,  0.8,   1.15,  1.33,  1.45,  1.55   1.65, 1,7, 1,75, 1.9
% f= 1/4
% XX,   0.78,   0.78,   0.8,   0.93, 1.15,  1.35,  1.5,   1.65,   1.8,  1.9,    2,   2
% f=1/8
% XX,   XX,      0.85,   0.9,     1,    1.18,   1.35,   1.6,   1. 8,   1.9 ,  2,   2  ,   2, 

Fig.\ \ref{tangnormpdf48} plots the normalized tangential PDF for $f=\frac{1}{4}$ (left panel) 
and $f=\frac{1}{8}$ (right panel). In both panels, the black dotted line is the PDF for tracer 
particles ($St_{ h} =0$) at $r=1\eta$, and the black dashed line is the Gaussian fit to 
$St_{ h} =795$. For $f=\frac{1}{4}$, the minimum $St_{ h}$ available in our simulation is 
$0.39$, and, for this  $St_{ h}$, the PDF shape is close to that of the tracers. It turns out that, for 
$f=\frac{1}{4}$, the PDF shape remains roughly unchanged as $St_{ h}$ increases 
to $0.78$, and fitting the PDFs for $St_{ h} \le 0.78$ with stretched exponentials 
gives $\alpha =0.78$. This invariance of the PDF shape is probably due to the fact that
the fattening effect of the shear term cancels out the thinning trend of 
the acceleration term in this range of $St_{ h}$.  For larger $St_{ h}$ ($\gsim 1$), 
the PDF becomes thinner with increasing  $St_{ h}$, and %as both the shear and acceleration terms have thinner distributions at larger $St_{ h}$. 
the best-fit $\alpha$ increases from $\simeq 0.8$ at $St_{ h} =1.55$ to 2 at $St_{ h} \gsim 397$.
%Note that the PDF tails for $St_{ h}=12.4$ are again fit a 4/3 stretched exponential (the black dashed lines).  
%We also examined the PDF behavior for other values of $f$. 

For $f = \frac{1}{8}$, the minimum $St_{ h}$ we have is 0.78, %The PDF shape shows a continuous thinning trend as $St_{ h}$ increases  above $0.78$. 
%A more detailed discussion on the behavior 
%of the PDF fatness as a function of $St_{ h}$ and $f$ will be given in \S 4.2.3.
%The minimum $St_{ h}$ available in our simulation for this case is 
%$0.78$. 
and, as $St_{ h}$ increases above 0.78, the PDF shape becomes continuously thinner (right panel).  
For $f=\frac{1}{8}$ and $St_{ h} \gsim 0.78$, the acceleration term, 
${\bs w}_{\rm a}$, dominates the contribution to the relative velocity, and the PDF of 
${\bs w}_{\rm a}$ is thinner at larger $\tau_{\rm p,h}$.

%At $St_{ h}$ below $0.78$, the shear term would be more 
%important, and an interesting question is whether the PDF tails has a flattening trend with increasing 
%$St_{ h}$ for $St_{ h} \lsim 0.78$. The answer depends on whether the fattening effect of 
%the shear contribution, ${\bs w}_{\rm s}$, at small $St_{ h}$ due to the tail amplification 
%is stronger than the thinning trend of the acceleration term in the $f =\frac{1}{8}$ case.
%A definite answer  would  require future simulations that include smaller particles and 
%allow smaller values of $St_{ h}$ for $f=\frac{1}{8}$. 

From  Fig.\ \ref{tangnormpdf2} and  Fig.\ \ref{tangnormpdf48}, we see that the PDF shape 
is almost independent of $r$.  Except for $f=\frac{1}{2}$ and $St_{ h} = 0.19$, 
the shape of all the PDFs in these figures  already converges at $r \simeq \frac{1}{2}\eta$. 
%the PDF still has a weak $r-$dependence at $\widetilde{|{\bs w}|} <1$ for $r=\frac{1}{4} \eta$. 
The $r-$dependence of the PDF shape is much weaker than 
the monodisperse case, and the dependence decreases  with decreasing $f$. 
As explained before, this is due to the acceleration contribution in the bidisperse case, which is independent of $r$.  
%As seen Fig.\ \ref{tangnormpdf48},. 
In the case with $f=\frac{1}{2}$ and $St_{ h} = 0.19$, corresponding to the two smallest particles in our simulation, 
the PDF tails become slightly fatter with decreasing $r$. For this Stokes pair, 
the relative velocity at  $r \gsim \frac{1}{4}\eta$ still has a significant $r-$dependent shear contribution. 
To achieve convergence for this case, a resolution below $r \lsim \frac{1}{4}\eta$ is needed.

In Appendix C, we compare the tangential PDFs for approaching and separating pairs, which is of 
theoretical interest. We show that, because the generalized acceleration term is independent 
of the relative motions of the two particles, the difference between the PDFs of 
approaching and separating pairs decreases when ${\bs w}_{\rm a}$
increases with decreasing $f$.   

\subsubsection{The PDF of the 3D amplitude}

\begin{figure*}[t]
\includegraphics[height=2.9in]{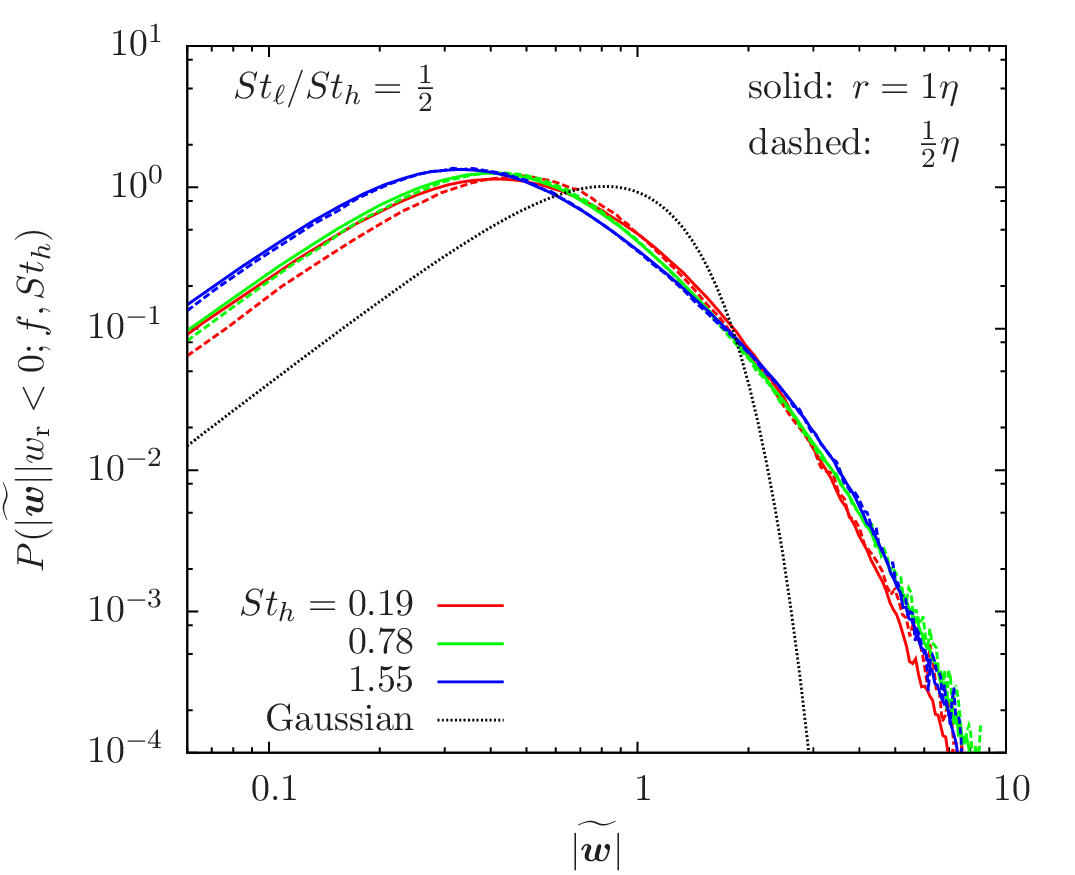}
\includegraphics[height=2.9in]{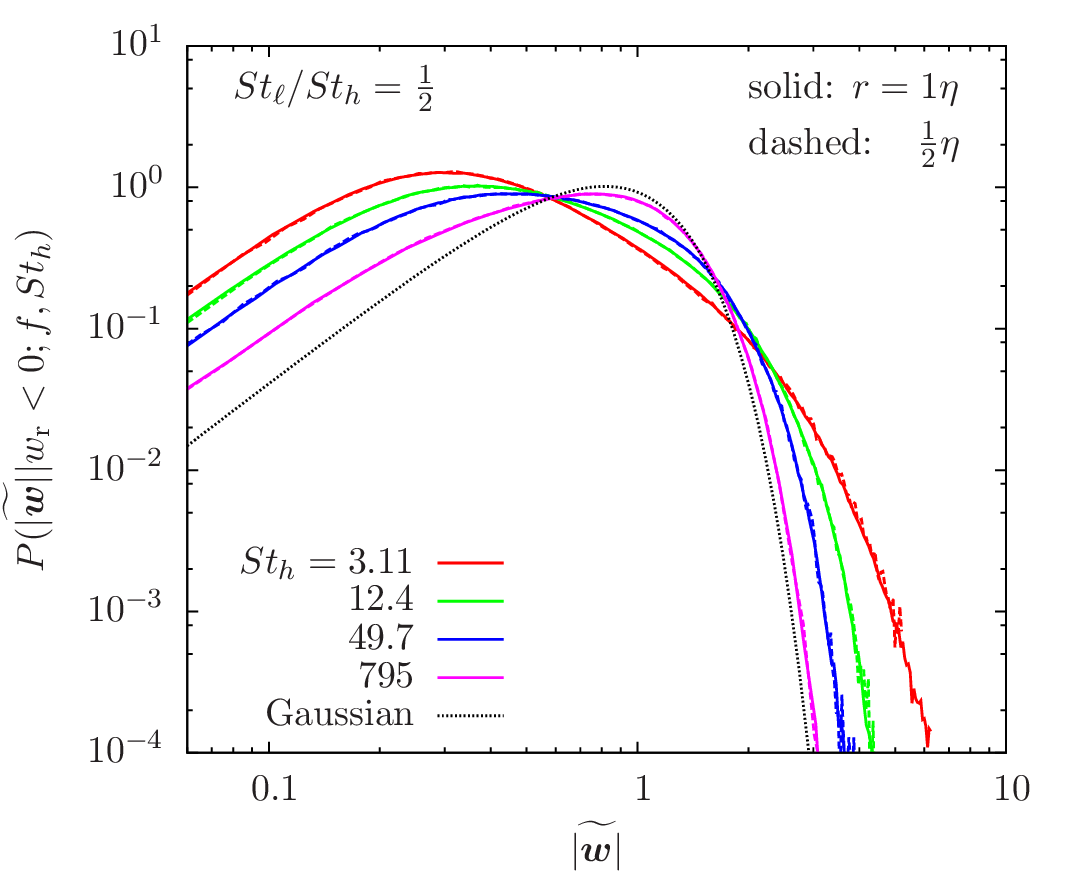}
\caption{The PDF of the 3D relative velocity  amplitude for approaching particle pairs 
at $r=1\eta$ (solid) and $ \frac{1}{2} \eta$ (dashed). The Stokes ratio is fixed at $f=\frac{1}{2}$. 
The 3D amplitude, $|{\bs w}|$, is normalized to the rms value, 
i.e., $\widetilde{|{\bs w}|} = |{\bs w}|/\langle w^2\rangle_{-}^{1/2}$, 
so that each PDF in the figure has unit variance. 
The left and right panels show $St_{ h} \le 1.55$ and $St_{ h} \ge 3.11$, respectively. 
In both panels, the black dotted line is the normalized PDF of the amplitude of a 3D 
Gaussian vector.}
\label{3dpdfratio2} 
\end{figure*}

In Fig.\ \ref{3dpdfratio2}, we show the normalized PDF, $P(\widetilde{|{\bs w}}||w_{\rm r}<0; f, 
St_{ h})$, of the 3D amplitude, $|{\bs w}|$, of the relative velocity for approaching particle 
pairs with a Stokes ratio $f=\frac{1}{2}$. For each PDF, we normalized $|{\bs w}|$ to its rms, 
$\langle w^2\rangle_{-}^{1/2} $. The figure is plot in a similar way as Fig.\ 14 of Paper I.  
But unlike that figure, which shows  the PDF only at one distance ($r=1\eta$), here we plot the 
results for  $r=1\eta$ and $\frac{1}{2}\eta$.  
%The figure is plot in a  similar way as Fig.\ 10 in Paper I for identical particles ($f=1$).
In the left panel, we see that, as $St_{ h}$ increases from $0.19$ (red) to 1.55 (blue), 
the PDF around the rms value (i.e., $\widetilde{|{\bs w}|} \simeq 1$) slightly 
decreases, and more probabilities are transferred to the left and right parts of the PDF 
at $\widetilde{|{\bs w}|} \ll 1$ and $\widetilde{|{\bs w}|} \gg 1$, respectively. 
This trend corresponds to the fattening of the PDFs of the radial and tangential 
relative speeds with increasing $St_{ h}$ at small $St_{ h}$ 
(see Fig.\ \ref{tangnormpdf2} and discussions in \S 4.2.3). 
Note that the far right PDF tail  for $St_{ h} =1.55$ 
already becomes thinner than that for $St_{ h} =0.78$, consistent with the 
observation in Fig.\ \ref{tangnormpdf2} that the thinning of the tangential PDF tails starts 
at $St_{ h} \gsim 0.78$. %A detailed explanation for the PDF behavior 
%of the tangential relative speed  %for $f=\frac{1}{2}$ 
%was given in \S 4.2.3. 

\begin{figure*}[t]
\includegraphics[height=2.9in]{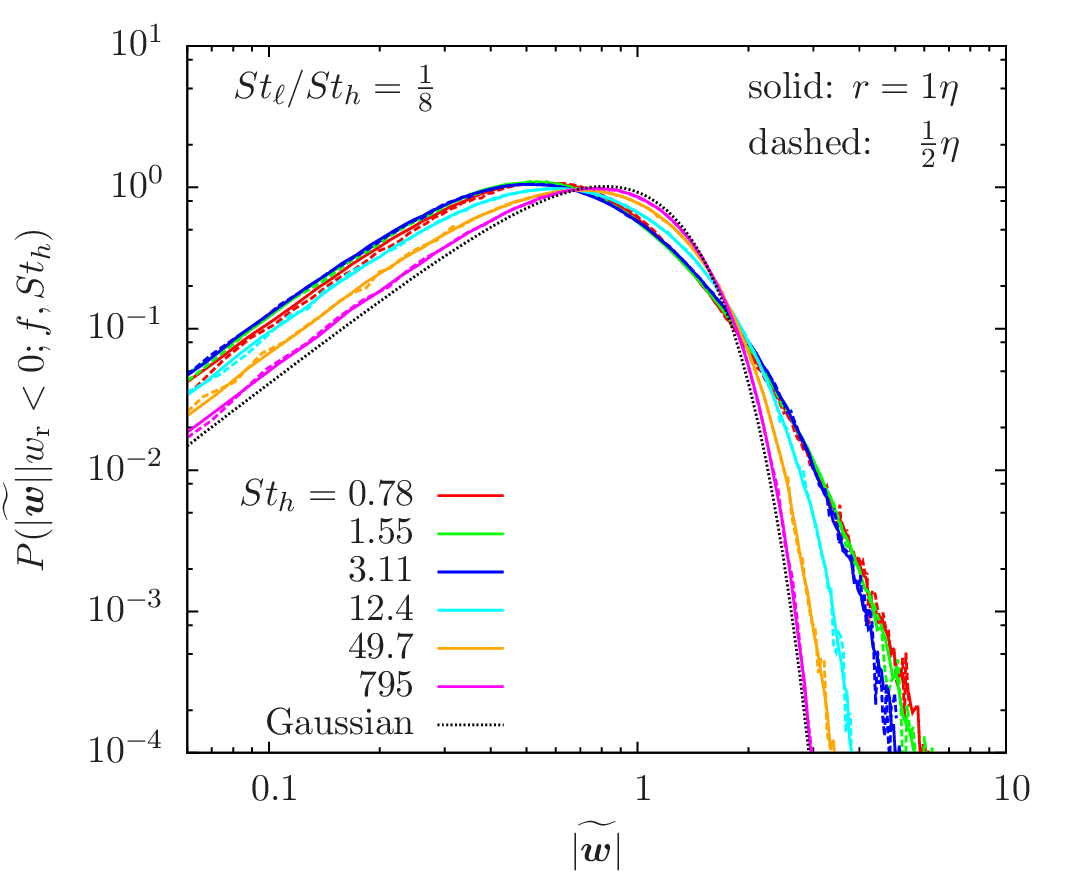}
\includegraphics[height=2.9in]{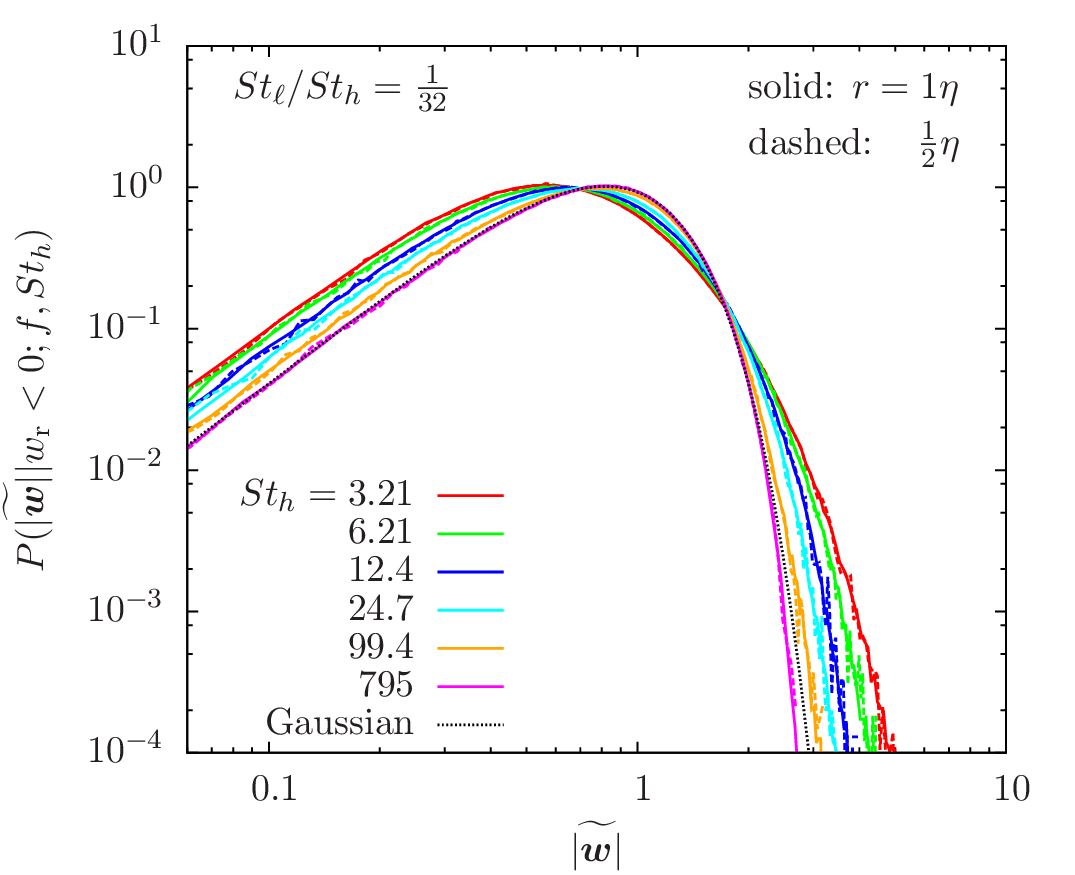}
\caption{The normalized 3D relative velocity PDFs for $f=\frac{1}{8}$ (left) and $f=\frac{1}{32}$ (right).
The solid and dashed lines correspond to $r=1$ and $\frac{1}{2}\eta$, respectively. 
The black dotted line is the normalized PDF for the 
amplitude of a 3D Gaussian vector.}
\label{3dpdfratio8} 
\end{figure*}

For $St_{ h}$ above 3.11, the PDF shape has an opposite trend. 
With increasing $St_{ h}$, the probability is more concentrated 
around the rms value, and the PDF at very small and large relative 
velocities decreases,  
%This trend of the PDF shape with $St_{ h}$ above 1.55 is again similar to that in the monodisperse case.  
as expected from the thinning of the radial and tangential 
PDFs for $St_{ h}$ in this range. In the limit $St_{ h}\to \infty$, 
the PDF is expected to approach Gaussian (black dotted line), but for the largest 
$St_{ h}$ ($=795$) in the simulation, the PDF is still slightly non-Gaussian, 
especially at $\widetilde{|{\bs w}|} \ll 1$. 
The general trend of the 3D amplitude PDF for $f=\frac{1} {2}$ as a 
function of $St_{ h}$ is similar to the monodisperse case shown in Fig.\ 14 of  
Paper I. Comparing with that figure, we see again that the PDF tails in the bidisperse 
case are significantly less fat than for equal-size particles.  

%The left part This implies that the shear term provides a PDF contribution with the largest fatness at $St \simeq 1$.  
%We found that a  similar trend holds for the bidisperse case with $f$ fixed at $\frac{1}{4}$, where the 
%PDF tails fatten slightly as $St_1$ increases to 0.78, and then shows a thinning trend for $St_1$ 
%above 0.78. 

The left panel Fig.\ \ref{3dpdfratio8} shows the normalized PDF of the 3D 
amplitude for $f=\frac{1}{8}$. The minimum $St_{ h} $ available
is $0.78$, and, consistent with the early results (see the right panel of Fig.\ \ref{tangnormpdf48}) 
for the tangential PDF, the right tail of  $P(\widetilde{|{\bs w}|}|w_{\rm r} <0; f, St_{ h})$ 
at large $\widetilde{|{\bs w}|}$ keeps thinning for all $St_{ h} \ge 0.78$. 
The PDF becomes close to Gaussian (dotted black line) for $St_{ h} \gsim 199$. 
The PDF for $f=\frac{1}{32}$ in the right panel has the same trend. 

%The PDF shape is found be $r$-dependence for all Stokes pairs in our simulation for $f=\frac{1}{8}$. 

Although the bidispserse PDF is thinner than the case of 
equal-size particles, significant non-Gaussianity still exists, especially if both particles 
are small. The PDF is close to Gaussian only if one or both particles are sufficiently 
large.  
%as seen in Figs.\  \ref{3dpdfratio2} and \ref{3dpdfratio8}. 
It is of practical interest to examine the particle size range
in which the relative velocity PDF can be approximated by Gaussian. 
For a given $St_{ h}$ or $\tau_{\rm ph}$, the PDF becomes thinner and closer to 
Gaussian as $f$ decreases. Interestingly, we find that, for a fixed $\tau_{\rm ph}$, 
the shape of the PDF barely changes as $f$ decreases from $\frac{1}{16}$ to 0. 
The reason is that,  for $f$ in this range, the particle relative velocity, ${\bs w}$, is 
dominated by the generalized acceleration term, whose distribution is controlled by $\Delta {\bs u}_{\rm T} (\tau_{\rm ph})$. 
In the $f \to 0$ ($St_{\ell} \to 0$) limit,  ${\bs w}$ approaches 
the particle-flow relative velocity, ${\bs w}_{\rm f}$, and, based on our 
results in \S 3.1, the PDF of ${\bs w}_{\rm f}$ is close to Gaussian if 
the particle friction time is larger than $3.5 T_{\rm L}$. It follows that,  
for any $f$ in the range $0 \le f \le \frac{1}{16}$, 
%in the $f\to 0$ limit, 
the particle relative velocity PDF is approximately Gaussian 
if  $\tau_{\rm ph} \gsim 3.5 T_{\rm L}$ (corresponding to $St_{ h} \gsim 49.7$ in our flow). 
%Therefore, 
%the PDF of  the particle relative velocity is approximately Gaussian if $\tau_{\rm ph} \simeq 3.5-7 T_{\rm L}$ 
The condition for the relative velocity PDF to have a Gaussian shape is stronger 
for $\frac{1}{16}  < f \le 1$. 
%At a given $f$, there exists a minimum value for $St_{ h}$, above which the relative velocity 
%PDF reaches Gaussian. This minimum value increases as $f$ increases.  
For example, at $f=\frac{1}{8}$ and $f=\frac{1}{4}$, Gaussianity is reached only 
for $\tau_{\rm ph} \gsim 14 T_{\rm L}$ and $\tau_{\rm ph} \gsim 27 T_{\rm L}$, 
respectively. For $f=\frac{1}{2}$ (Fig.\ \ref{3dpdfratio2}) and $f=1$ (Fig.\ 14 of paper I), the PDF 
still shows some (slight) degree of non-Gaussianity even for the largest 
particles ($\tau_{\rm ph} = 54 T_{\rm L}$) in our simulation.  

In Figs.\  \ref{3dpdfratio2} and \ref{3dpdfratio8}, we see that the shape of the 
normalized PDF for the 3D amplitude already converges at $r=\frac{1}{2}\eta$, 
except for the relative velocity between the two smallest particles in our simulation 
(the red lines in the left panel of Fig.\ \ref{3dpdfratio2}). 
%This is consistent with the results in \S 4.2.3 for the PDF of the tangential component. 
The convergence is also found for other $f$ values not shown here.  
The $r-$dependence of the relative velocity PDF is stronger in the monodisperse case, and it 
does not converge at $r\simeq 1\eta$ for equal-size particles with $St \lsim 3.11$ (see 
Fig.\ \ref{3dpdf4}). In the bidisperse case,  the generalized acceleration contribution 
is independent of $r$ (\S 2) and its presence helps reduce the $r-$dependence 
of the relative velocity statistics. This finding is of particular interest for applications 
to dust particle collisions in protoplanetary disks.  Dust particles should be viewed 
as nearly-point particles (e.g., Hubbard 2012), and one needs to examine the collision 
velocity PDF in the $r \to 0$ limit. The weaker $r-$dependence of the bidisperse PDF 
makes it easier to resolve the relative velocity or collision statistics at $r \to 0$.  

In comparison to a Gaussian distribution,  the PDFs measured in our simulation 
have more probabilities at both extremely small and large collision speeds. 
Clearly, the fat high tails of the PDF may lead to significantly higher probability of  
fragmentation than predicted under the Gaussian assumption (Windmark et al.\ 2012b, Garaud et al.\ 2013). 
On the other hand, the PDF at  low collision velocity is also considerably higher than a Gaussian distribution,  which would 
favor sticking of the particles. The competition of  these two  effects is important for understanding 
how the  non-Gaussianity of the collision velocity affects the growth of dust particles in  protoplanetary disks.  
In next section, we give some speculations using our simulation results, but a definite answer requires 
evolving the particle size distribution with a coagulation model that accounts for the non-Gaussian collision velocity.  

\section{Implications of Non-Gaussianity on Dust Particle Collisons}

\subsection{Parameters and Assumptions}

We discuss the effect of the non-Gaussianity of turbulence-induced collision velocity on 
dust particle collisions in protoplanetary disks. Using the probability distribution 
measured in our simulation, we will roughly estimate the fractions of collisions leading to 
sticking, bouncing and fragmentation, as the particle size grows. 
We adopt a minimum mass solar nebula. The profiles of the gas density, sound speed 
and scale height are set to $\rho = 1.7 \times 10^{-9} (R/{\rm AU})^{-2.75}$ g cm$^{-3}$, 
$C_{\rm s} = 1.0 (R/{\rm AU})^{-0.25}$ km s$^{-1}$, and $H= 5 \times 10^6 (R/{\rm AU})^{1.25}$ km, with $R$ 
the distance to the central star. As our main purpose is to give a simple illustrating example, 
we primarily consider $R=1$ AU.  At 1AU,  the mean free path, $\lambda$, of the gas is  $\simeq 1$ cm, 
and, using the Epstein and Stokes formulae, we find the friction time $\tau_{\rm p} = 6 \times 10^3  (a_{\rm p}/{\rm cm})$ s 
and $\tau_{\rm p} = 2.3 \times 10^3  (a_{\rm p}/{\rm cm})^2$ s for particle size, $a_{\rm p}$, below and above 
$\lambda$, respectively.  The two formulae connect at $a_{\rm p} \simeq 2.6$ cm. 

We use the prescription of Cuzzi et al.\ (2001) to specify turbulence conditions in 
protoplanetary disks with the Shakura-Sunyaev parameter $\alpha_{\rm t}$. The 
large-eddy turnover time is set to $T_{\rm eddy} = \Omega_{\rm K}^{-1} = 5 \times 10^6 (R/{\rm AU})^{3/2}$ s 
with $\Omega_{\rm K}$ the Keplerian rotation frequency. 
The turbulent rms velocity and integral scale are given by $U = \alpha_{\rm t} ^{1/2} C_{\rm s}$ 
and $L=  \alpha_{\rm t}^{1/2} H$, respectively.  Note that here 
$U$ is the 3D rms velocity, different from $u'$ given earlier for 
the 1D rms of our simulated flow. Taking a fiducial value of $10^{-4}$ for $\alpha_{\rm t}$, we have 
$U= 10 (\alpha_{\rm t}/10^{-4} )^{1/2}$  m s$^{-1}$, and $L = 5 \times 10^4 (\alpha_{\rm t}/10^{-4} ) ^{1/2} $ 
km at 1AU. Adopting a dynamical viscosity of $9\times 10^{-5}$ g cm$^{-1}$ s$^{-1}$ for molecular hydrogen 
at $\simeq$ 300 K, the Reynolds number is estimated to be $Re\simeq 10^8 (\alpha_{\rm t}/10^{-4} )$. 
The Kolmogorov time and length scales are given by  $\tau_\eta \simeq T_{\rm eddy} Re^{-1/2} = 500 (\alpha_{\rm t}/10^{-4} )^{-1/2} $ s, and 
$\eta \simeq L Re^{-3/4} = 0.05 (\alpha_{\rm t}/10^{-4} )^{-1/4}  $ km.  At 1AU,  particles of mm size have 
$\tau_{\rm p} \simeq \tau_{\eta}$ or $St \simeq1$. We also define a Stokes number based on the 
large-eddy time as $St_{\rm eddy} \equiv \tau_{\rm p}/T_{\rm eddy}$, and, at 1AU, $St_{\rm eddy} \simeq 1$ corresponds to $a_{\rm p} \simeq 50$ cm.

The collision outcome for particles over a wide size range is complicated, and its dependence on the collision velocity and 
the particle internal structure is subject of ongoing investigations. A summary of experimental results is given 
by G\"uttler et al.\ (2010).  Here we adopt the assumption of Windmark et al.\ (2012b) and Garaud et al.\  (2013) to 
determine the collision outcome of silicate particles. Particles are assumed to stick if the 3D amplitude 
of the collision velocity, $|\bs{w}|$, is below a bouncing threshold $w_{\rm b}$ of 5 cm s$^{-1}$, while 
fragmentation takes place for $|\bs{w}|$ above a threshold $w_{\rm f} \simeq 1$ m s$^{-1}$. 
Between $w_{\rm b}$ and $w_{\rm f}$, the colliding particles bounce off each other. 
Following Garaud et al.\  (2013), we compute the fractions of sticking, bouncing 
and fragmentation as $F_{\rm s} = \int_{0}^{w_{\rm b}} |\bs{w}| P(|\bs{w}|; f, St_{h}) d|\bs{w}|/\langle |\bs{w}| \rangle$,  $F_{\rm b} = \int_{w_{\rm b}}^{w_{\rm f}} 
|\bs{w}| P(|\bs {w}|; f, St_{h}) d|\bs{w}|/\langle |\bs{w}| \rangle$, and $F_{\rm f} = \int_{w_{\rm f}}^{\infty} |\bs{w}| P(|{\bs w}|; f, St_{h}) d|\bs{w}|/\langle |\bs{w}| \rangle$, 
respectively, where $P(|\bs{w}|; f, St_{h})$ is the PDF of $|\bs{w}|$, and 
$\langle |\bs{w}| \rangle \equiv \int_{0}^{\infty}| \bs{w}|P(|\bs{w}|; f, St_{h}) d|\bs{w}|$ is the mean of $|\bs{w}|$. 
Unlike the fractions defined in Windmark et al.\ (2012b), a weighting factor $\propto | \bs{w}|$ is included in these equations, 
accounting for higher collision frequency for particle pairs 
with larger relative velocity (Garaud et al.\  2013).  The weighting factor used here is based on the cylindrical 
formulation of  the collision kernel\footnote{There is a different and perhaps more accurate kernel formulation, named the 
spherical formulation (Wang et al.\ 2000, Paper I), which depends on the radial component of the 
relative velocity and only counts approaching particle pairs. The collision-rate 
weighted fractions $F_{\rm s}, F_{\rm b} $ and $F_{\rm f}$ in the spherical  
formulation will be studied in a later work.}, where the collision rate $\propto |\bs{w}|$ (see Paper I).     
 
%for approaching particle pairs (with $w_{\rm r} <0$).  
%and fragmentation as $F_{\rm s} = \int_{0}^{w_{\rm b}} |\bs{w}| P(|\bs{w}||w_{\rm r} <0; f, St_{h}) d|\bs{w}|/\langle |\bs{w}| \rangle$,  $F_{\rm b} = \int_{w_{\rm b}}^{w_{\rm f}} 
%|\bs{w}| P(|\bs {w}||w_{\rm r} <0; f, St_{h}) d|\bs{w}|/\langle |\bs{w}| \rangle$, and $F_{\rm f} = \int_{w_{\rm f}}^{\infty} |\bs{w}| P(|{\bs w}|w_{\rm r} <0; f, St_{h}) d|\bs{w}|/\langle |\bs{w}| %\rangle$, respectively, where $P(|\bs{w}||w_{\rm r} <0; f, St_{h})$ is the PDF of the 3D amplitude of the relative velocity
%for approaching particle pairs (with $w_{\rm r} <0$).  

In our calculation, we adopt PDFs  measured at $r=\frac{1}{2}\eta$.  It is desirable to use 
PDFs at smaller $r$ because the relative velocity statistics of small particles  ($St \lsim 1$) 
of similar sizes ($f=1$ or $\frac{1}{2}$) have not converged at $r=\frac{1}{2}\eta$. However, the
measured PDFs at smaller $r$($\le \frac{1}{4}\eta$) are very noisy because the number of particle 
pairs available in our simulation becomes too limited at $r\le \frac{1}{4}\eta$. 
The choice of using the PDF data at $r=\frac{1}{2} \eta$ is sufficient for an illustration purpose, 
especially as we are more concerned with the collisions of inertial-range particles with 
$St$ considerably above 1 (corresponding to $a_{\rm p} \gg 1$ mm at 1AU). The 
PDFs for these relatively large particles already converge at $r=\frac{1}{2} \eta$. 
For small, similar-size particles, our estimates for the fractions need 
to be improved by future simulations that can accurately measure the PDFs at smaller $r$.  
%In order to see the effects of non-Gaussianity, we also compute the fractions based on the 
%assumption of Maxmellian distribution for $|\bs{w}|$.  

The  inertial range of our simulated flow is short, and one needs to be careful when applying 
the measured statistics to protoplanetary turbulence. Since the realistic $Re(\simeq10^8)$ in the 
disk cannot be reached by currently available computation power, the best solution 
is to extrapolate the simulation results to high $Re$. The extrapolation requires 
 the $Re-$dependence of the PDF, which is 
%A 1024$^3$ simulation with a wide range of particle sizes is computationally intensive, 
%and we are planning such a run in future work. 
currently unknown. We are thus forced to use rather strong assumptions. % in the application. %Again this is not a problem for our purpose of illustration.  
For a clear description of our assumptions, we distinguish the Stokes numbers in 
our simulated flow and in the real disk. We use $St^{\rm s}$ and $St_{\rm eddy}^{\rm s}$ to 
denote the Stokes numbers based on $\tau_{\eta}$ and $T_{\rm eddy}$ 
in our simulation, while $St^{\rm d}$  and $St_{\rm eddy}^{\rm d}$ are the corresponding numbers in the disk. 
In our flow, $T_{\rm eddy} \simeq 20 \tau_\eta$ and thus $St^{\rm s} \simeq 20 St_{\rm eddy}^{\rm s}$. 
Due to a much broader inertial range in the disk, 
we have $St^{\rm d} \simeq 10^4 St_{\rm eddy}^{\rm d}$ at 1AU. 
  
The only species of particles that definitely lie in the inertial range of our simulated 
flow are those with $St^{\rm s}= 6.21$ (or equivalently $St_{\rm eddy}^{\rm s}=0.31$). 
Larger particles with $St^{\rm s} \ge 12.4$ ($St_{\rm eddy}^{\rm s} \ge 0.62$) and smaller ones with $St^{\rm s} \le 3.11$ 
($St^{\rm s}_{\rm eddy} \le 0.16$) are affected by the flow structures at driving and dissipation scales, respectively. Accordingly, we divide dust particles in the 
disk into three ranges, i.e, the driving range with $St_{\rm eddy}^{\rm d} \ge 0.62$ ($St^{\rm d} \ge 6.2 \times 10^3$), 
the dissipation range with  $St^{\rm d} \le 3.11$ ($St_{\rm eddy}^{\rm d} \le 3.1 \times 10^{-4}$), 
and the inertial range with $3.11 \times 10^{-4} < St_{\rm eddy}^{\rm d} < 0.62$ ($3.11 < St^{\rm d} < 6.2\times 10^3$). 
%Below we describe our procedure to fix or approximate the the collision velocity statistics 
%for the three ranges, respectively. 
%$St_{\rm eddy}^{\rm s} = 0.31$ ($St^{\rm s}= 6.21$), 
%and  $St^{\rm s} \le 3.21$
Our measurement for $St_{\rm eddy}^{\rm s} \ge 0.62$ particles can 
be directly applied to corresponding particles with $St_{\rm eddy}^{\rm d} \ge 0.62$ 
in the disk because collisions of these particles are determined by large-scale structures 
that are resolved in our  simulation. For particles in the inertial range, we make a strong 
simplification. We use the PDF shape of the only inertial-range particle 
($St^{\rm s}_{\rm eddy} = 0.31$) in our flow for all inertial-range particles ($3.11 \times 10^{-4} < St_{\rm eddy}^{\rm d} < 0.62$) at 1AU
in the disk. %(i.e., all the inertial-range particles are assumed to have the same PDF shape).  
%In the realistic disk, for $St_{\rm eddy} \le 0.31$, there is still a wide range of particle size 
%belonging to the inertial range. 
%This is different from  in our simulation  $St_{\rm eddy} \le 0.16$ marks a transition 
%to the dissipation range.  We 
Finally, for dissipation-range particles with $St^{\rm d} \le 3.11$ in the disk, we assume 
the PDF shape at each $St^{\rm d}$ is the same as the PDF of particles with 
 $St^{\rm s} = St^{\rm d}$ in our flow.  

For the rms width of the PDF, we make use of the prediction of the PP10 model (Papers I \& II) with 
a Reynolds number ($Re \simeq 10^8$) appropriate for the disk. In particular, the model predicts 
a $(St^{\rm d})^{1/2}$ scaling for equal-size particles in the inertial range ($3.11 < St^{\rm d} < 6.2 \times 10^3$). 
%This latter assumption is based on the prediction of our model (see Paper I) for the rms relative velocity 
%of equal-size particles in the inertial range. 
%The trend of the rms for these particles with decreasing $St^{\rm d}$
%is set to be the same as  the measured $\langle w^2 \rangle^{1/2}$ for $St^{\rm s} \le 3.11$ in our flow. The amplitude of the 
%rms is properly scaled to smoothly connect with the inertial range.  
Similar approximations are used for different particles with a fixed Stokes ratio, $f\equiv{St_{\ell}}/St_{h}$. 
We divide particle pairs at a given $f$ into the same three ranges based on the Stokes number, 
$St_{h}^{\rm d}$ (or $St_{{\rm eddy}, h}^{\rm d}$), of the larger particle. We then take the same procedure for 
the PDF shape as a function of $St_{h}^{\rm d}$ using the simulation data for the three ranges. 
Again, the rms relative velocity at fixed $f$ is taken from our model, 
which predicts a $(St_{h}^{\rm d})^{1/2}$ scaling for $St_{h}^{\rm d}$ in the inertial range for any value of $f$. 
%The uncertainty of our simplifying assumptions will be
%discussed later. 
%with the Stokes number of the larger particle as $St_{h}^{1/2}$ for $St_{h}^{1/2}$ in the inertial range.  
%the measured PDF for $St \lsim 3.11$ particles in our simulation.  
%Note that this assumption underestimates the 
%The only particle species that definitely lies in the inertial range of our flow is particles with 
%$St= 6.21$ or $St_{\rm e}=0.31$. 
% we examine
% the ...  Interpolating the inertial range....
%(hence increasing non-Gassianity with increasing $Re$ ) 
%With the above assumptions, we calculated the fractions of sticking, bouncing and fragmentation with increasing particle sizes. 

\subsection{Fractions of Sticking, Bouncing and Fragmentation}

\begin{figure}[t]
\centerline{\includegraphics[width=3.9in]{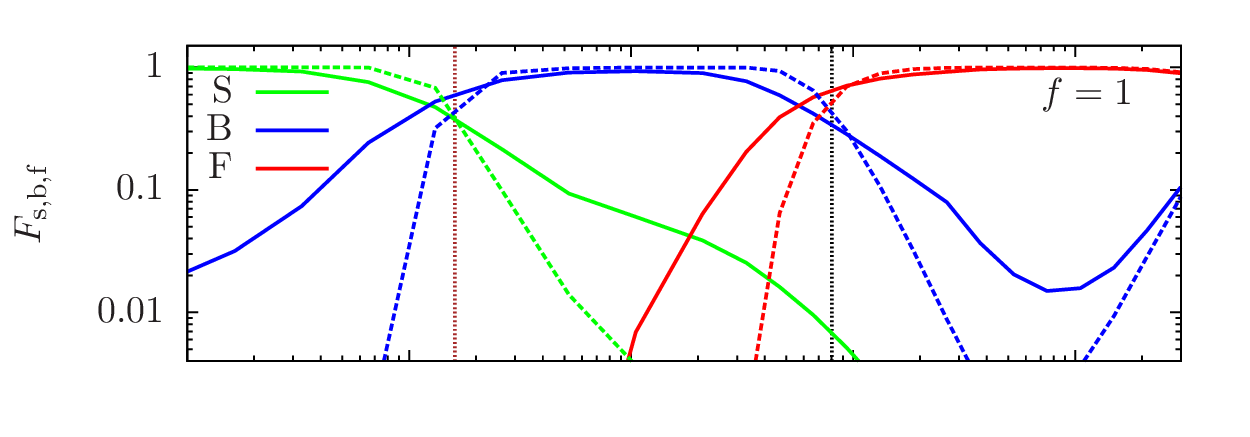}}
\vspace{-1cm}
\centerline{\includegraphics[width=3.9in]{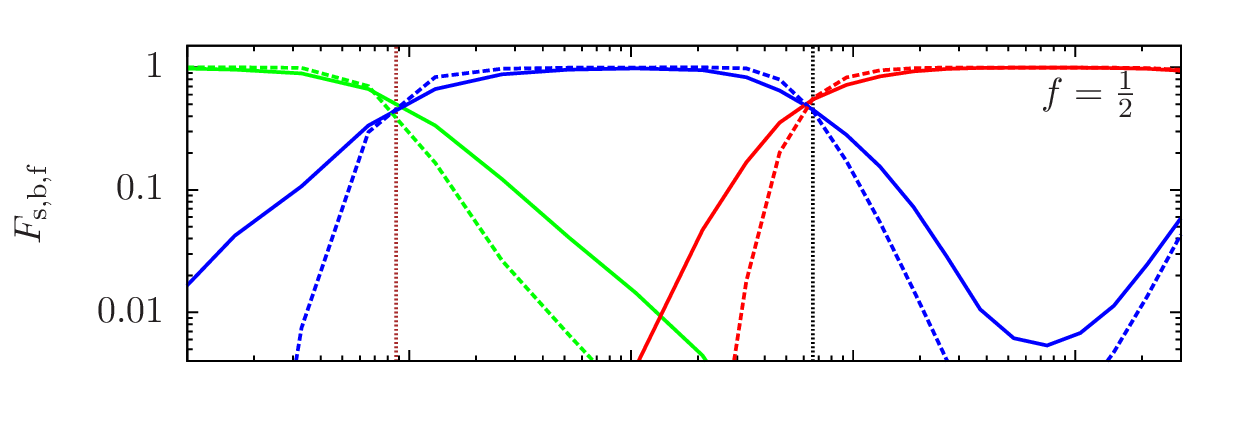}}
\vspace{-1.0cm}
\centerline{\includegraphics[width=3.9in]{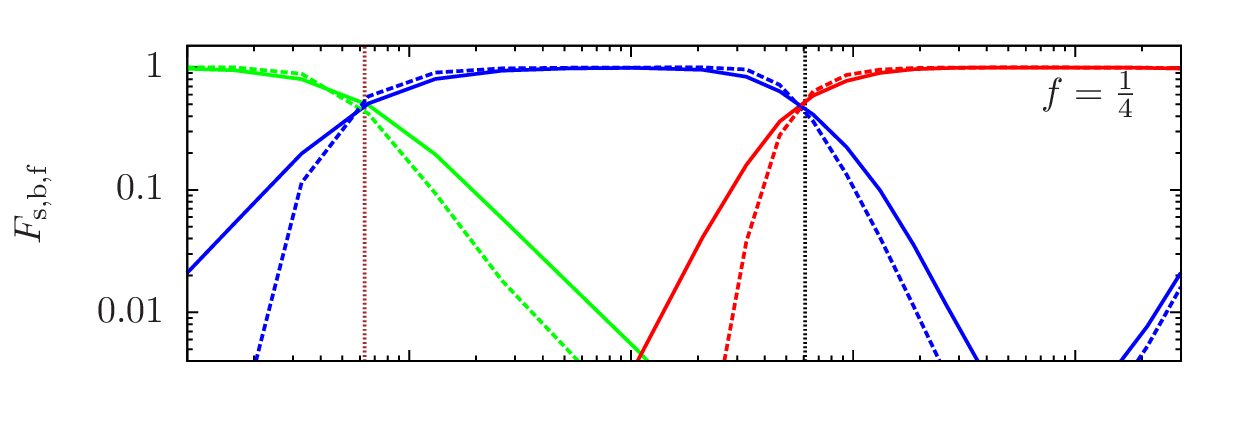}}
\vspace{-1.0cm}
\centerline{\includegraphics[width=3.9in]{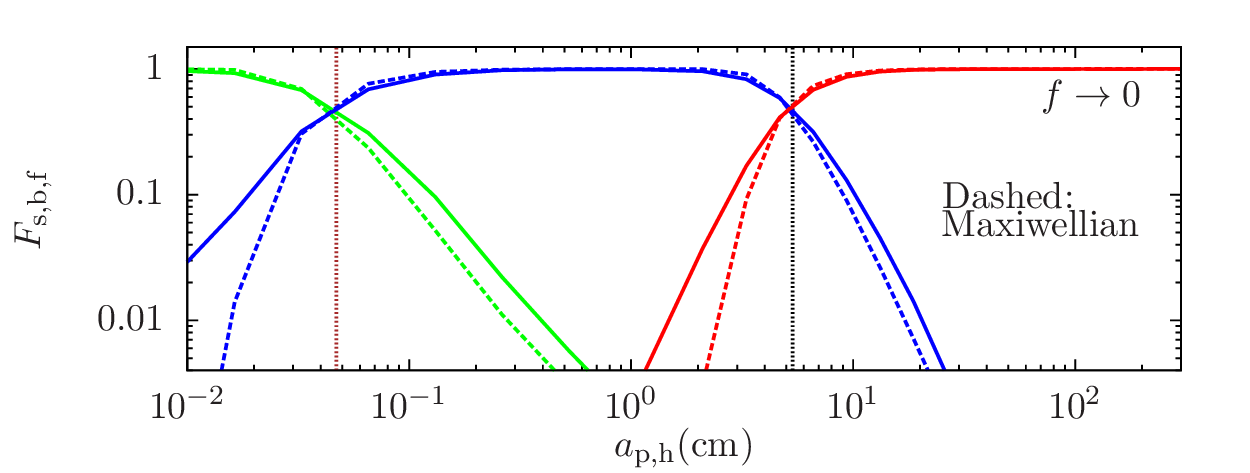}}
\caption{Sticking (green), bouncing (blue) and fragmentation (red) fractions as a function 
of the size, $a_{\rm {p, h}}$, of the larger  particle. The four panels show fixed Stokes number ratios at $f=1$, $\frac{1}{2}$, $\frac{1}{4}$, and $f\to 0$.   
Vertical brown and black dotted lines correspond to critical $a_{\rm {p, h}}$ at which the rms 
relative velocity reaches bouncing and fragmentation thresholds, respectively. Dashed lines 
assume a Maxwellian distribution. Results for $f\to0$ are based on the particle-flow 
relative velocity  (\S 3.1).  
}
\label{fractions} 
\end{figure}

Fig.\ \ref{fractions} shows the fractions of sticking, bouncing and 
fragmentation as a function of the size of the larger particle, $a_{\rm p, h}$, at 
different Stokes ratios, $f$. Each panel is plot in a similar way as Fig.\ 1 in Windmark et al.\ (2012b).  
As the particles grow, the collision velocity increases, and bouncing and finally fragmentation 
take place. The vertical brown and black dotted lines correspond to particle sizes, at which the 
rms collision velocities, $\langle w^2 \rangle^{1/2}$ , become equal to the bouncing ($w_{\rm b}$) and 
fragmentation ($w_{\rm f}$) thresholds.  If we  ignore the collision velocity distribution and assume that, for each size pair,
the collision velocity is single-valued with $P(|\bs w|; f, St_{h}) =\delta (|\bs w| -\langle w^2 \rangle^{1/2})$, these vertical lines mark 
instantaneous transitions to bouncing and fragmentation,
%If the distribution of the collision velocity is ignored (i.e., the collision speed is taken to be single-valued for each given size pair), the vertical brown and black 
%lines in Fig.\ \ref{fractions} 
which occur typically at millimeter and decimeter sizes, respectively.
%Typically, bouncing and fragmentation start to occur at, respectively. 
In this case, the growth would stop once the size reaches the brown 
line, a problem known as the bouncing barrier for planetesimal formation. 
If the particle growth somehow manages to pass the bouncing barrier, it would eventually 
be frustrated by the fragmentation barrier  (black dotted line). As $f$ decreases from 1 (top panel) toward 0 (bottom panel), 
the vertical lines shift toward smaller sizes because 
the rms relative velocity increases with decreasing $f$ (see Fig.\ 7 in Paper II).

Assuming a Maxwellian distribution for $|\bs w|$, Windmark et al.\ (2012b) and Garaud et al.\ (2013) 
examined the effect of the collision velocity PDF on particle growth.  As shown in Fig.\ 1 of Windmark et al.\ (2012b), 
accounting for the PDF leads to gradual transitions from sticking to bouncing and 
to fragmentation. The reason is clear:  With a probability distribution, there are always 
possibilities for sticking (or bouncing), corresponding to the low (or high) tail of the 
$|\bs{w}|$ PDF, even if the rms collision velocity 
is already above (or still below) the bouncing threshold, $w_{\rm b}$. The same applies to the bouncing-to-fragmentation transition. 
For comparison with the estimates based on our simulation data, the dashed lines in Fig.\ \ref{fractions} 
show the fractions computed from a Maxwellian PDF for $|\bs w|$ using the PP10 model prediction for the rms width. %These lines are qualitatively similar to the 
%central panel of Fig.\ 1 in Windmark et al.\ (2012b). 
An interesting result of Windmark et al.\ (2012b) was that, accounting for the $|\bs w|$ distribution, the bouncing 
barrier may be overcome. The sticking probability beyond the barrier size allows 
further growth, and the peak of the particle size distribution, $p(a_{\rm p})$, can move % toward larger $a_{\rm p}$ 
past the vertical brown line. Intuitively, a large particle can from  if, by ``luck", it kept encountering sticking events and enjoyed continuous growth. 

The transition from bouncing to fragmentation is also gradual, and of particular 
interest is the finite probability of fragmentation before $a_{\rm p}$ 
reaches the black dotted line, corresponding to the high tail of the PDF of $|{\bs w}|$. 
These fragmentation events %before the black dotted line 
would replenish particles of small sizes, increasing the size distribution, 
$p(a_{\rm p})$, at small $a_{\rm p}$ (Windmark et al.\ 2012b).  
Under the assumption of a Maxwellian PDF, the decrease of the sticking probability is rather fast (the green dashed lines), 
and the growth past the bouncing barrier turns out to be slow. Windmark et al.\ (2012b) showed that it takes $
\sim 10^4$ yr to reach centimeter size. The growth by sticking toward the fragmentation barrier (if not impossible) 
would take significantly longer than $10^4$ yr. 
%affect the particle growth process toward the fragmentation barrier size. 
%This was not discussed in details by  Windmark et al.\ (2012). 
%Apparently, the  
Note that, after the bouncing-to-fragmentation transition, the bouncing fraction (blue lines) 
in the top three panels start to increase at $a_{\rm p, h} \simeq 0.5-1$ m. The reason is 
that the rms relative velocity for a given Stokes ratio $f $ starts to decrease at $\tau_{\rm p, h}/T_{\rm eddy} \gsim 1/f$ (see Fig.\ 7 of paper II). 

The solid lines in Fig.\ \ref{fractions} are the fractions based on our simulation data for the collision 
velocity PDF. These lines show more gradual transitions than the dashed ones for a Maxwellian distribution. 
This is because the collision PDF measured in our flow is fatter than a Gaussian distribution. 
As mentioned earlier, in our terminology a fat PDF shape typically corresponds to a 
sharper inner part and a fatter tail part.  In Figures 14 and 15, we see that 
the measured PDFs have higher probabilities at both extremely small and large $|{\bs w}|$ than a 
Maxwellian distribution. Consider, for example, the switch from sticking to bouncing. The higher 
probability at small $|{\bs w}|$ is responsible for the slower decrease of $F_{\rm s}$ with the size 
of the larger particle, while the earlier rise of the bouncing curves, $F_{\rm b}$, at smaller $a_{\rm p, h}$ 
is due to higher probability of large $|{\bs w}|$ at the right tail. 

Interestingly, in the top panel for equal-size  particles, the decrease of  $F_{\rm s}$ past the bouncing barrier is quite slow, 
and it even extends to the fragmentation barrier (black dotted line). This can be understood from Fig.\ 14 in Paper I 
for equal-size particles, which shows that the PDF  at extremely small $|{\bs w}|$ 
is orders of magnitude larger than a Gaussian distribution. An immediate implication of the persistence of significant sticking fraction is that it further helps alleviate the problem of the bouncing barrier. With significantly higher sticking probabilities, 
the growth beyond the bouncing barrier would proceed faster than predicted with 
a Maxwellian distribution. The peak  size  of the distribution 
$p(a_{\rm p})$ would increase faster. In the $f=1$ and $\frac{1}{2}$ cases 
(top two panels), accounting for non-Gaussianity may accelerate the growth by a factor of $\gsim 10$.

Due to the fat non-Gaussian PDF tail,  the solid red lines for fragmentation rise earlier than the dashed ones, 
suggesting that fragmentation can occur at smaller sizes and at a higher rate.  
One consequence is  that it provides faster replenishment of smaller particles 
than the Maxwellian case, increasing the size distribution $p(a_{\rm p})$ at small $a_{\rm p}$. 
Together with the significantly higher sticking fraction, we speculate that this would lead to a broader 
size distribution $p(a_{\rm p})$ around the peak size. Another consequence of the earlier start of 
fragmentation is that it may begin to impede the growth at smaller sizes. 
As the peak of  $p(a_{\rm p})$ moves closer to the fragmentation barrier,  coagulational growth 
will finally end at a size where the fragmentation fraction, $F_{\rm f}$, 
exceeds the sticking fraction, $F_{\rm s}$. We see in Fig.\ \ref{fractions} that the solid green and red lines tend to cross at  
smaller particle sizes than the dashed lines, meaning that, with non-Gaussian PDFs, 
fragmentation may start to compete with sticking earlier than the Maxwellian case. 
However, we argue this is a secondary effect in comparison to 
much higher sticking/growth rate between the bouncing and fragmentation barriers. 
We expect that the non-Gaussianity effect makes the growth toward the fragmentation barrier size 
considerably easier. Once the fragmentation fraction dominate over sticking, the 
possibility of further growth of some particles toward or above the fragmentation barrier 
requires alternative mechanisms, such as the sweep-up process discussed below.   
 
As $f$ decreases from 1 toward 0, the solid lines becomes closer to the dashed ones. This is consistent with our results 
in \S 4.2 that, at a given $St_{h}$, the PDF shape becomes thinner with decreasing $f$. In particular, 
the degree of non-Gaussianity decreases very rapidly as $f$ goes from 1 to $\frac{1}{2}$, as can be seen by comparing Fig.\ 14 
of Paper I and Fig.\ 14 in \S 4.2.4.
% due to the increased acceleration contribution.  
%However, some degree of non-Gaussianity still exists at $f\to 0$, 
%and the transitions at bouncing and fragmentation barriers is slightly more gradual than the Maxwellian case.  
The effect of non-Gaussianity is thus weaker for collisions between particles 
of very different sizes. The faster particle growth due to the effect of non-Gaussianity in 
between the bouncing and fragmentation barriers may occur mainly for particles of similar sizes.  
%Also, the differences between the solid and dashed lines appear to decrease with increasing 
%particle size $a_{\rm p}$, and this is due to the decrease of the non-Gaussianity of the relative velocity 
%PDF with the Stokes number.
 
The bottom panel for $f\to 0$ in Fig.\ \ref{fractions} corresponds to particles of very different sizes. 
The computation is based on the particle-flow relative velocity discussed in \S 3.1. As the collision statistics change only slightly 
as $f$ decreases from $\frac{1}{16}$ to 0, this panel can be approximately used for any $f$ in the range $ 0\le f\lsim \frac{1}{16}$. 
The $f\to 0$ case is imporant for a particle growth mechanism called the mass transfer by fragmentation 
(see Windmark et al.\ 2012a). Laboratory experiments showed that, when colliding with a much larger particle, a small 
particle may break up and leave some fragments on the larger particle. %there is a mass  transfer to 
In effect, the larger particle sweeps up masses from small particles, and it 
can thus keep growing even if its size already exceeds the fragmentation barrier size\footnote{Windmark et al. (2012b) 
assumed that this mass transfer occurs only for particle mass ratio larger than 50. 
This  corresponds to a size ratio of $\gsim 3.7$, or a Stokes ratio of $\gsim14$ for particles with $a_{\rm p} \gsim  2.6$ cm.}. 
The sweep-up mechanism was suggested as a possible path to planetesimal formation (Windmark et al.\ 2012a, 2012b; Garaud et al.\ 2013). 
The solid red line in the bottom panel can be viewed as representing the probability of such mass transfer or 
sweep-up events.  In comparison to the Maxwellian case,  
the mass transfer can start earlier,  %but then increases slower after the vertical black line. 
%It is not clear whether and how 
suggesting that the sweep-up mechanism can operate at smaller $a_{\rm p, h}$. 
%But it is not clear by how much the difference between the red solid and dashed lines in the bottom 
%panel may affect the sweep-up process.  

We argue that, when the non-Gaussian collision statistics is included, the sweep-up growth 
is likely more efficient than the Maxwellian case because a larger number of seed particles 
of large sizes would be available to sweep up small particles. As discussed above, 
the formation of these particles is expected to be easier due to larger sticking probability in between the bouncing and fragmentation barriers.
%These large particles can provide seed particles  for the further growth by the mechanism of sweeping up masses from small particles, 
%as proposed by Windmark et al.\ (2012a, 2012b).    
%Which can provides lucky seeds for further growth through a so-called mass transfer by fragmentation 
Also, the earlier start and higher rate of fragmentation from collisions of similar-size particles between 
the two barriers could help provide more small particles for the seed particles to sweep up. However, we 
emphasize these are only speculations, and a definite answer 
%of whether and how much the non-Gaussianity collision statistics can improve the sweep-up mechanism 
requires including the non-Gaussian PDFs  into a coagulation model.  
%We speculate that the higher sticking b

\subsection{Discussions}

We have shown that the non-Gaussianity of the collision velocity leads to more gradual 
transitions from sticking to bouncing and to fragmentation. The probability of sticking past the 
mean bouncing barrier is significantly larger than expected from a Maxwell distribution, and we argued 
this would further help alleviate the bouncing barrier, and help provide seed particles for the 
sweep-up process in a possible mechanism for planetesimal formation.  

We point out that our assumption for the PDF of $|{\bs w}|$ underestimates the 
degree of non-Gaussianity for particles with $St_{\rm eddy}^{\rm d} \lsim 0.31$ 
($a_{\rm p} \lsim 26$ cm) in the disk. As discussed in \S 2, the relative velocity of smaller 
particles samples the flow velocity at smaller scales, $\ell$. As the flow structures are more intermittent 
at smaller $\ell$, the PDF of $|{\bs w}|$ would become fatter as $St_{\rm eddy}^{\rm d}$ decreases. 
Thus, using the PDF shape of $St_{\rm eddy}^{\rm s} = 0.31$ particles in our flow for all 
inertial-range particles in the disk underestimates the non-Gaussianity at 
$St_{\rm eddy}^{\rm d} < 0.31$. The PDF for particles in the dissipation range ($St^{\rm d} \lsim 1$) 
of the disk would also be fatter than the corresponding particles ($St^{\rm s} \lsim 1$) in our flow, 
because the degree of turbulent intermittency at dissipation-range scales increases with $Re$. 
Therefore, the non-Gaussianity effects discussed above on the particle growth
should be viewed as a lower limit. With more accurate PDF shape at $St_{\rm eddy}^{\rm d} \lsim 0.31$, the transitions 
to bouncing and fragmentation would appear to be more gradual. %This trend may 
%further help with the particle growth in between the bouncing and fragmentation barriers. 
In a future work, we will examine the $Re-$dependence of the collision statistics 
using a $1024^3$ simulation, extrapolate the measured PDFs 
%in the inertial range and provide better estimates that account for increasing flow intermittency with 
to large $Re$, and
%conduct in a future work a systematic and more quantitative study on the PDF shape 
%as a function of the Stokes number pair, 
provide fitting functions  %as a function of the Stokes number pair, 
%as well as the $Re-$dependence, 
that can be implemented in coagulation models.
%As the distribution of the collision velocity is crucial for modeling the evolution of dust particles, in a followup work 

%%The speculation is subject to tests by future coagulation models 
%%that include the non-Gaussian collision statistics.     

%How the more gradual transition between bouncing and fragmentation, and the more gradual rise of the 
%probability for mass transfer affects the particle growth is unknown, and 
%However, an exact answer for the maximum size to which the particles can grow  requires an implementation of non-Gaussian collision velocity in a coagulation model.     

We finally  discuss some caveats in our results. Unlike our simulated flow, turbulence 
in a rotating disk is anisotropic at large scales.  For example, in magneto-rotational disks, 
nearly axisymmetric structures, known as zonal flows (e.g., Johansen et al.\ 2009), emerge. 
Turbulent structures at large scales are highly elongated in the azimuthal (or zonal) direction. 
Based on Kolmogorov's similarity hypothesis (e.g., Monin \& Yaglom 1975), we expect the 
statistical isotropy to be restored at small scales  where the eddy time is significantly smaller than the 
rotation period.  Therefore, our results for the collision velocity PDF are applicable for small particles 
that couple to eddies far below the large zonal structures. 

On the other hand, large particles with friction time, $\tau_{\rm p}$, near or above 
the rotation period, $\Omega_{\rm K}^{-1}$, would be affected by the 
anisotropic zonal flow, where azimuthal turbulent motions are more intense than 
in the radial direction. 
Stronger zonal structures indicate that the particle collision velocity 
primarily lies in the azimuthal direction, and the separation of nearby 
particle pairs backward in time would also proceed mainly in that direction. 
From the PP10 picture, the relative velocity of these large particles 
%with $\tau_{\rm p}$ close to $\Omega_{\rm K}^{-1}$ 
would mainly sample the flow structures in the zonal direction. 
Therefore, to understand the collision velocity PDF, 
we need to know the scaling behavior and the probability distribution of the zonal turbulent 
structures as a function of length and time scales, which, to our knowledge, 
are currently unavailable in the literature.   
%However, we are aware of no existing studies for the quantitive behaviors of the scaling and probability distribution 
%of the zonal flow structures at different length scales, which are needed to understand the particle collision velocity PDF. 
It is thus unclear how the anisotropy in a rotating disk exactly 
affects the collision velocity PDF for particles with $\tau_{\rm p}$ close to $\Omega_{\rm K}^{-1}$. 
A direct comparison of the PDFs measured in our flow 
with simulation results for rotating disks in future studies would help reveal the effects of  
large-scale zonal flows on the particle collision statistics\footnote{If the zonal flow is driven by an inverse cascade,
as suggested by Johansen et al.\ (2009), the non-Gaussianity of zonal structures may be lower, 
due to the lower degree of intermittency in an inverse cascade (e.g., Paret \& Tabeling 1998). 
In that case, the collision PDF of large particles coupled to the zonal eddies may be less 
non-Gaussian. Johansen et al.\ (2007) showed simulation results for the particle collision 
velocity PDF in magneto-rotational disks. However, form their linear-linear plots, it is hard to tell the fatness or non-Gaussianity of the PDF.}. 
%with $\tau_{\rm p}$ close to $\Omega_{\rm K}^{-1}$.   

In addition to its effects on the flow structures, rotation could also directly affect the 
particle collision velocity. For example, the rotation velocities of particles of different sizes are 
different, and this gives rise to a mean azimuthal collision velocity, which may be important for very 
large particles (e.g., Fig.\ 3 of Testi et al.\ 2014). Differential rotation (or shear) can also contribute 
to the collision velocity. Although weaker than the zonal eddies, turbulent motions in the radial 
direction can cause a radial separation of nearby particle pairs backward in time. The 
particle pair would thus have a memory of different rotation speeds of the gas at different radii in 
the past, leading to a contribution to the collision velocity in the azimuthal direction. 
Combined with the radial separation, the differential rotation can enhance the backward 
separation of particle pairs in the azimuthal direction. This also tends to increase the collision 
velocity, as the particles' memory would sample the zonal turbulent structures at large 
scales. The strength of these effects depends on the amplitude of the radial turbulent 
motions and thus on the degree of anisotropy. A definite estimate for the effects 
%on these issues requires quantitative knowledge of the exact statistics for the anisotropic 
%disk turbulence, which 
is beyond the scope of this work.

%\section{The PDF with Collision-rate Weighting}
%the PDF shape becomes thinner monotonically as $St_1$ 
%increases from $0.38$ to $795$. This is because of the contribution of  the generalized acceleration term. 
%As discussed in the text, with a larger contribution 
%from the acceleration effect, the relative velocity PDF tails tend to be thinner. In particular, the PDF shape from the acceleration 
%term becomes less fat with increasing difference in the friction time difference, 
%$\tau_{\rm p2} -\tau_{\rm p1}$.  Clearly, for a fixed $f$, the friction time difference 
%increases with $St_1$, which tends to decrease the fatness of the relative velocity PDF. 
%Apparently, at $f =\frac{1}{4}$, this effect wins over the trend of the shear term to fatten the PDF 
%tails with increasing $St_1$ for $St_1 \lsim 1$, and thus the overall shape of the PDF becomes 
%thinner continuously with $St_1$. In fact, the monotonic thinning trend is  found for any $f \ge \frac{1}{4}$.  

\section{Summary and Conclusions}

Motivated by the important role of the collision velocity  distribution in modeling dust particle 
growth, we investigated the probability distribution function of turbulence-induced relative velocity of 
inertial particles, extending our earlier work on equal-size particles (Pan \& Padoan 2013; Paper I) to 
the case of different particles of arbitrary sizes. We used the numerical simulation of Paper I. 
The simulation evolved inertial particles with friction time, $\tau_{\rm p}$, ranging from $0.1\tau_\eta$ ($St =0.1$) 
to $54T_{\rm L}$ ($St=795$), with $\tau_\eta$ and $T_{\rm L}$ the Kolmogorov time 
and the Lagrangian correlation timescale of the flow, respectively. We computed the PDF 
for all Stokes number pairs $(St_1, St_2)$ available in the simulation, and interpreted the 
results using the physical picture of the Pan \& Padoan (2010) model for the bidisperse case. 

In the PP10 picture, the particle relative velocity consists of two 
contributions, named the generalized acceleration term and the generalized shear term.  
The generalized acceleration corresponds to different responses of particles 
of different sizes to the flow velocity, and is connected to the temporal flow
velocity difference, $\Delta {\bs u}_{\rm T}$, along individual particle trajectories.  The acceleration term 
thus ``inherits" the non-Gaussianity of the temporal velocity structures in turbulent flows. 
The generalized shear term  represents the particles' memory of the spatial flow velocity 
difference across the separation of the two particles at given times in the 
past.  As shown in Paper I,  the shear term does not simply ``inherit" 
the non-Gaussianity of the spatial flow velocity structures; instead it 
reaches a significantly higher level of non-Gaussianity by a self-amplification mechanism.   
Based on the statistics of temporal and spatial flow velocity structures, 
we analyzed the contributions of the acceleration and shear terms to 
the PDF of the particle relative velocity.  Our simulation results for the relative velocity PDF 
are successfully explained by the behavior of the two contributions. 
%We have also conducted a systematic analysis of the probability distribution function (PDF) of the relative velocity 
%as a function of the Stokes number pair.  
The main conclusions from our statistical analysis of the numerical simulation are the following:

%{\bf [I made the list below a summary of purely numerical results, removing all physical interpretation. Otherwise it is too hard to read and not much of a summary.]}

\begin{enumerate}

\item As a special bidisperse case, we studied the relative velocity, 
${\bs w}_{\rm f}$, between inertial particles and the flow velocity at the particle position. 
%The particle-flow relative velocity 
%%For particles with a friction time of $\tau_{\rm p}$,  ${\bs w}_{\rm f}$ can be roughly 
%%estimated as the temporal flow velocity difference, $\Delta {\bs u}_{\rm T}(\Delta \tau)$, 
%%along the particle trajectory at a time lag $\Delta \tau \simeq\tau_{\rm p}$. 
%A simple model is developed  for the rms of ${\bs w}_{\rm f}$, under the assumption 
%that the temporal flow velocity correlation on the particle trajectory can be approximated by the 
%Lagrangian correlation function, $\Phi_{\rm L}$. 
%Adopting a bi-exponential form for $\Phi_{\rm L}$, our model prediction is in good agreement with 
%the simulation data. In particular, it predicts that the rms of  ${\bs w}_{\rm f}$ increases linearly with $St$ for $St\ll1$,  
%scales as $St^{1/2}$ for particles in the inertial range, and approaches the flow rms velocity 
%for $\tau_{\rm p} \gg T_{\rm L}$.
 %and approaches constant for $\tau_{\rm p} \gg T_{\rm L}$. 
The PDF shape of  ${\bs w}_{\rm f}$ becomes continuously thinner 
with increasing $St$, and approaches Gaussian for $\tau_{\rm p} \gsim 7 T_{\rm L}$. 
This corresponds to the thinning trend of the distribution of $\Delta {\bs u}_{\rm T}$ with increasing 
time lag, as inferred from the PDFs of the Lagrangian and Eulerian temporal flow velocity differences. 
%The particle-flow relative velocity is an interesting limit 
%that helps confine the particle-particle relative velocity behavior in the bidisperse case.  

\item The PDF of turbulence-induced relative velocity of inertial particles is generally 
non-Gaussian, exhibiting fat tails.  We found that, at a fixed $St_1$, the PDF shape is the 
fattest for the monodisperse case, i.e., at $St_2 = St_1$. The PDF shape becomes thinner as $St_2$ 
increases above or decreases below $St_1$. As $St_2 \to 0$, the PDF approaches that of the particle-flow 
relative velocity ${\bs w}_{\rm f}$. In the limit $St_2 \to \infty$, the relative velocity is approximated by 
the one-particle velocity of particle (1), and its PDF approaches Gaussian.  
At a given $St_1$, the PDF behavior as a function of $St_2$ is confined by three useful limits, 
$St_2 \to 0$, $St_2 =St_1$, and $St_2 \to \infty$. 

\item We also examined the trend of the PDF shape with varying $St_{ h}$ at fixed 
Stoke ratios, $f$. At $f=\frac{1}{2}$, the PDF tails slightly  fatten as $St_{ h}$ 
increases to 0.78, and then become continuously thinner for 
larger $St_{ h}$. 
%%This trend is similar to the case of equal-size particles 
%%($f=1$) discussed in Paper I,  and is successfully explained using our physical picture.  
%for the generalized shear and acceleration contributions to the PDF.  
%%The shear contribution is responsible for the fattening  of the PDF with increasing 
%%$St_{ h}$ at $St_{ h} \lsim 1$.  
%At a given $f$, the minimum value of $St_{ h}$ available in our 
%simulation is $\simeq 0.1/f$,  which is larger for  smaller $f$. 
For $f \le \frac{1}{4}$, the PDF fatness is found to decrease continuously 
with increasing $St_{ h}$ for  all $St_{ h}$ ($\gsim 0.1/f$ ) available in our simulation.  
%%For $ f \le \frac{1}{16}$, the relative velocity is completely dominated by the acceleration contribution, and, 
For a given $St_{ h}$, the PDF shape is almost invariant with $f$ for  $0 \le f \le\frac{1}{16}$.
%The PDF behavior at $St_{ h} < 0.1/f$ remains to be 
%explored in future studies.  
If the friction time, $\tau_{\rm p,h}$, of the larger 
particle is close to $T_{\rm L}$, the PDF tails of both the radial and tangental relative 
velocities are fit well by a $4/3$ stretched exponential function for any value of $f$.    
%\item Even though the PDF shape 
%is generally thinner than the case of equal-size particles, 
%significant non-Gaussianity exists for particles of different sizes.  
The particle relative velocity PDF approaches a Gaussian distribution only if the 
friction timescale, $\tau_{\rm p,h}$, of  the larger particle is sufficiently large. 
For $0 \le f \le \frac{1}{16}$,  the PDF reaches Gaussian only if 
$\tau_{\rm p,h} \gsim 7T_{\rm L}$. The condition is stronger at larger $f$. 
For $f=\frac{1}{8}$, $\frac{1}{4}$, and $\frac {1}{2} \le f \le 1$, the PDF becomes 
nearly Gaussian at $\tau_{\rm p,h} \gsim 14 T_{\rm L}$, $27 T_{\rm L}$
and $54 T_{\rm L}$, respectively.

\item The $r$-dependence of the PDF for small particles is 
significantly weaker than in the monodisperse case, making it easier to achieve 
numerical convergence.
%The PDF shape  has a significantly weaker dependence on the particle distance, 
%$r$, than the monodisperse case.
The shape of the relative velocity PDF already converges at $r \simeq \frac{1}{2} \eta$ for 
all bidisperse cases in our simulation, except for the smallest two particles with $St=0.1$ and $0.19$. 

\item We discussed the implications of the non-Gaussianity of  the particle collision velocity on 
the dust growth in protoplanetary disks.  With some simplifying assumptions,  we calculated the fractions of collisions resulting in sticking, 
bouncing and fragmentation as a function of the particle size, and showed that, when non-Gaussianity is accounted for, 
the transitions from sticking to bouncing and to fragmentation become more gradual.   
In particular, the non-Gaussianity leads to much larger sticking probabilities 
past the bouncing barrier, which we argue could help further alleviate the bouncing 
barrier for dust particle growth. 

\end{enumerate}

%Despite extensive conclusions drawn in the current work for the particle relative 
%velocity in the general bidisperse case, significant future developments are 
%required to  further advance our understanding of particle collision statistics induced by turbulent motions. 
%In a future work, we will explore the collision kernel as a function of the Stokes 
%number pair, accounting for  turbulence-induced collision velocity and the effect of turbulent clustering 
%in the bidisperse case. Due to the limited resolution, the simulated flow used in the current work only 
%has a short inertial range, and our model prediction for particles in the inertial range remains to be tested and 
%validated.  Dust particles of sub-millimeter to meter size in protoplanetary turbulence  belong to the inertial range, 
%and modeling the collisions of these particles is crucial to assessing the viability of the planetesimal formation 
%mechanism by particle coagulation.  

The PDF of the particle collision velocity plays an important role in the modeling of dust particle growth, 
and the non-Gaussianity of turbulence-induced collision velocity needs to be accounted for 
in order to accurately predict the size evolution of dust particles in protoplanetary disks.  
In a followup work, we will provide fitting functions and tables of our results for the relative velocity 
PDF as a function of the Stokes pair, for a straightforward implementation in dust coagulation models. %Due to the importance of the collision velocity PDF on the dust particle growth, we will 
Our study could be improved in several aspects with larger simulations. For example, %including smaller particles with $St\ll 1$
%in the simulation, one can  have a clearer idea on the PDF trend with $St_{ h}$ at small values of 
%$f$. Also 
increasing the number of small particles per species would allow to measure 
the PDF at smaller particle distances and help resolve the issue of the convergence for the PDF of
small particles ($St \lsim 0.1$) of similar sizes.  With simulations at higher resolutions, we 
could also advance our understanding of the collision statistics for particles in the inertial range of the flow, 
and examine the possibility of a Reynolds number dependence of the velocity PDF. 

\acknowledgements 
Resources supporting this work were provided by the NASA High-End Computing (HEC) Program through the NASA Advanced 
Supercomputing (NAS) Division at Ames Research Center, and by the Port d'Informaci— Cient'fica (PIC), Spain, maintained by a 
collaboration of the Institut de F'sica d'Altes Energies (IFAE) and the Centro de Investigaciones EnergŽticas, Medioambientales y
Tecnol—gicas (CIEMAT).
LP is supported by a Clay Fellowship at Harvard-Smithsonian Center for Astrophysics. 
PP acknowledges support by the FP7-PEOPLE- 2010-RG grant PIRG07-GA-2010-261359.

\appendix

\section{The PDFs of Lagrangian and Eulerian temporal velocity differences}

\begin{figure*}[h]
\includegraphics[height=2.9in]{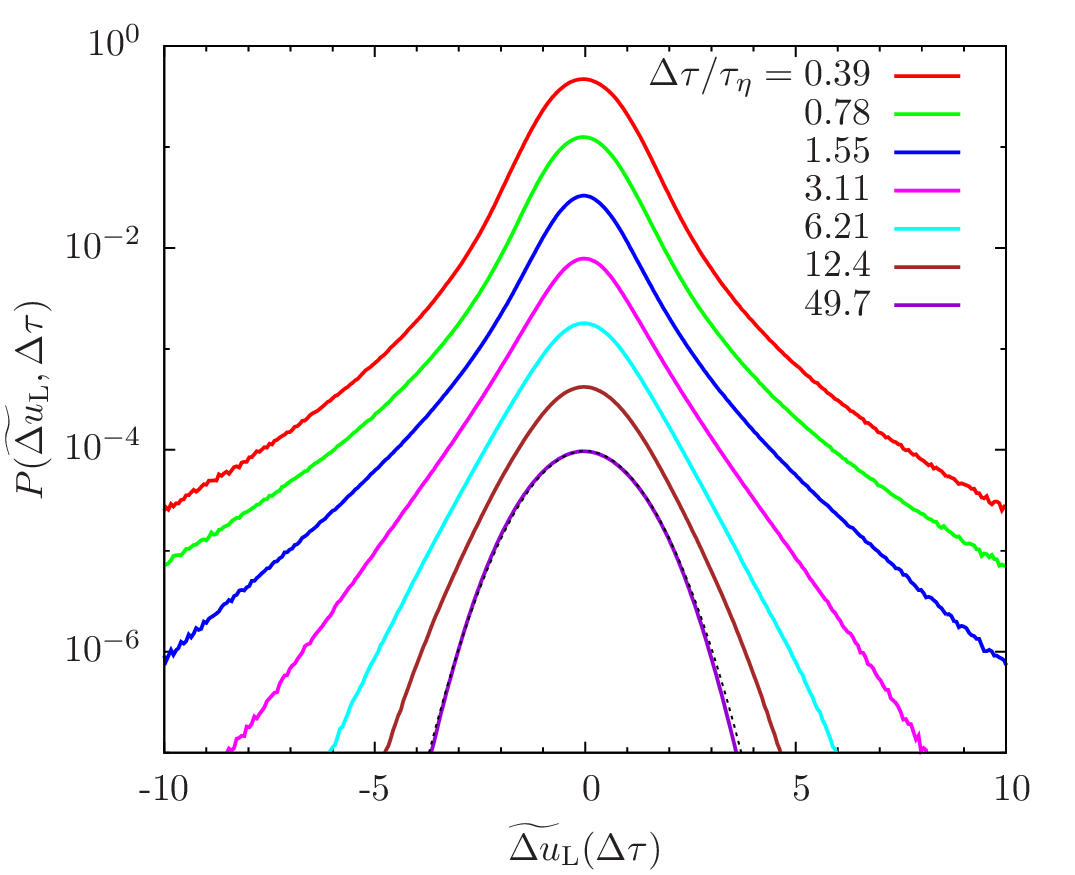}
\includegraphics[height=2.9in]{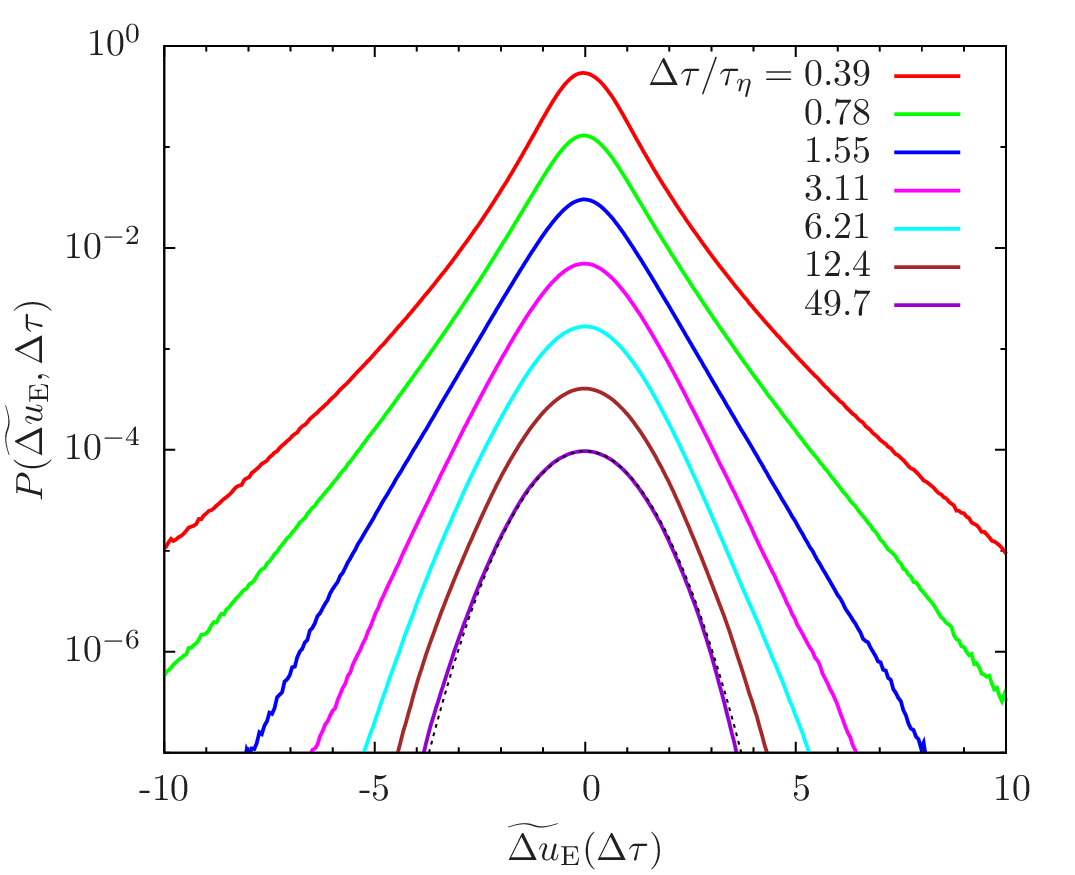}
\caption{The normalized PDFs of  the Lagrangian (left) and Eulerian (right) 
temporal velocity differences, $\Delta {u}_{\rm L,E} (\Delta \tau)$, as functions of the time 
lag. In each curve, the velocity difference is normalized to its rms value, i.e., 
$\widetilde{\Delta {u}}_{\rm L, E}  \equiv  \Delta_{\rm L, E} {u}/\langle (\Delta_{\rm L, E} {u})^2 \rangle^{1/2}$.
%Lines from top to bottom correspond to increasing $\Delta \tau$. 
In both panels, the top lines plot the actual values of the PDFs for $\Delta \tau=0.39 \tau_\eta$, while
each lower line for larger $\Delta \tau$ is shifted downward by a factor of 4 for clarity. The dotted black 
line corresponds to a Gaussian PDF.   
}
\label{dupdf} 
\end{figure*}

In Fig.\ \ref{dupdf}, we show the PDFs of the Lagrangian ($\Delta {\bs u}_{\rm L}$, left panel) and 
Eulerian ($\Delta {\bs u}_{\rm E}$, right panel) temporal velocity differences at different time lags, $\Delta \tau$.  
$\Delta {\bs u}_{\rm L} (\Delta \tau)$ is computed as $ {\bs u} ({\bs X}_{\rm L} (t+\Delta \tau), t + \Delta \tau) -   {\bs u} ({\bs X}_{\rm L} (t), t)$ 
by following the trajectories, $X_{\rm L}(t)$, of tracers, while  $\Delta {\bs u}_{\rm E} (\Delta \tau) = {\bs u} ({\bs x}, t + \Delta \tau) -   {\bs u} ({\bs x}, t)$ is measured 
at fixed points, ${\bs x}$.
Each line in the figure corresponds to the PDF of one component, $\Delta {u}_{\rm L}$ or $\Delta {u}_{\rm E} $, of the vector $\Delta { \bs u}_{\rm L} $ or  
$\Delta { \bs u}_{\rm E}$. All PDFs are normalized to have unit  variance. The bottom curves in 
both panels  correspond to $\Delta \tau = 49.7 \tau_\eta$ ($3.3 T_{\rm L}$), and, at this $\Delta \tau$, 
the PDFs of $\Delta {u}_{\rm L}$  and $\Delta {u}_{\rm E}$ are Gaussian (the dotted black lines). 
%for both the radial and transverse velocity difference. 
%This is consistent with the Gaussian 1-point statistics in developed turbulent flows. 

As $\Delta \tau$ decreases, the central part of the PDF 
keeps a Gaussian-like  shape,  but the PDF tails become fatter and fatter. The overall PDF 
shape is highly non-Gaussian at $\Delta \tau \sim \tau_\eta$. Fattening of the PDF of 
$\Delta {u}_{\rm L}$ corresponds to turbulent intermittency in the Lagrangian frame.  
A comparison of the two panels shows that, %the PDF fattening  for 
%$\Delta { \bs u}_{\rm L}$ proceeds faster than that for $\Delta {\bs u}_{\rm E}$.
%Clearly, 
at $\Delta \tau \lsim 6.21 \tau_\eta $, the PDF tails for $\Delta {u}_{\rm L}$ 
are significantly fatter than $\Delta {u}_{\rm E}$. This is consistent with the results 
of previous studies (e.g., Chevillard et al.\ 2005) that the degree of  intermittency in 
Lagrangian structures is higher than the Eulerian temporal velocity difference. %It is interesting to see that, for $\Delta {u}_{\rm L}$, 
%the PDF shape saturates at $\Delta \tau \simeq 0.4 \tau_\eta$, while 
%the PDF of $\Delta {u}_{\rm E}$ is still fattening.  

As the friction time $\tau_{\rm p}$ increases, the flow velocity seen by the particle may 
make a transition from Lagrangian-like to Eulerian-like. This transition is expected to
occur at $\tau_{\rm p} \sim T_{\rm L} \simeq15 \tau_\eta$. %Therefore, it is appropriate to approximate the PDF of the 
%flow velocity difference, $\Delta_{\rm T} u$, along the particle trajectory by that of $\Delta {u}_{\rm L}$ or $\Delta {u}_{\rm E}$ for 
%small ($\tau_{\rm p} \ll T_{\rm L}$) or large ($\tau_{\rm p} \gg T_{\rm L}$) particles, respectively. 
Interestingly, the PDF shapes of $\Delta {u}_{\rm L}$ and $\Delta {u}_{\rm E}$ are very 
similar at $\Delta \tau \gsim 12.4 \tau_\eta$.  This implies that one may use 
the PDF of $\Delta {u}_{\rm L}$ for all particles to understand 
the particle-flow relative velocity or the generalized acceleration 
contribution in the general bidisperse case.

In isotropic turbulence, $\Delta {\bs u}_{\rm E}$ can be related to the spatial velocity difference, 
$\Delta { \bs u} ({\bs \ell})$ ($\equiv { \bs u}({\bs x} + {\bs \ell}, t) -  { \bs u}({\bs x}, t) $), 
by the so-called random Taylor hypothesis (e.g., Tennekes 1975). In this hypothesis, $\Delta_{\rm E} {u} (\Delta \tau)$ 
is estimated by $\Delta {u} ({\ell})$ at a scale of $\ell \simeq u'  \Delta \tau$, 
where it is assumed that  the sweeping speed at which the energy-containing eddies 
advect small-scale eddies across a given point is given by the rms flow velocity, $\sim u'$. 
%In other words, $u'$ plays the role of the mean flow velocity in the original Taylor hypothesis. 
%In our flow, we find that the 
The hypothesis is supported by a comparison of the PDF shape of $\Delta_{\rm E} {u} (\Delta \tau)$ 
with the spatial velocity increment $\Delta u _{\rm t} (\ell)$ in the transverse direction 
at $\ell \simeq \sqrt{3} u' \Delta \tau$ in our flow. 
%As the PDF of $\Delta {u}_{\rm L}$ is fatter than $\Delta_{\rm E} {u}$ at 
%the same time lag $\Delta \tau \lsim T_{\rm L}$, 
%this also suggests that 
We also find that the intermittency of temporal structures in the Lagrangian frame 
is stronger than the Eulerian spatial structures, consistent with previous studies (e.g., Chevillard et al.\ 2005). 
The normalized PDF of $\Delta u _{\rm t} (\ell)$ has been examined in details 
in Appendix B  (right panel of Fig.\ 21) of Paper I.  
%\section{The PDF of the tangential component}

\section{The acceleration and velocity gradient PDFs  of the simulated flow}

\begin{figure}[t]
\centerline{\includegraphics[width=0.5\columnwidth]{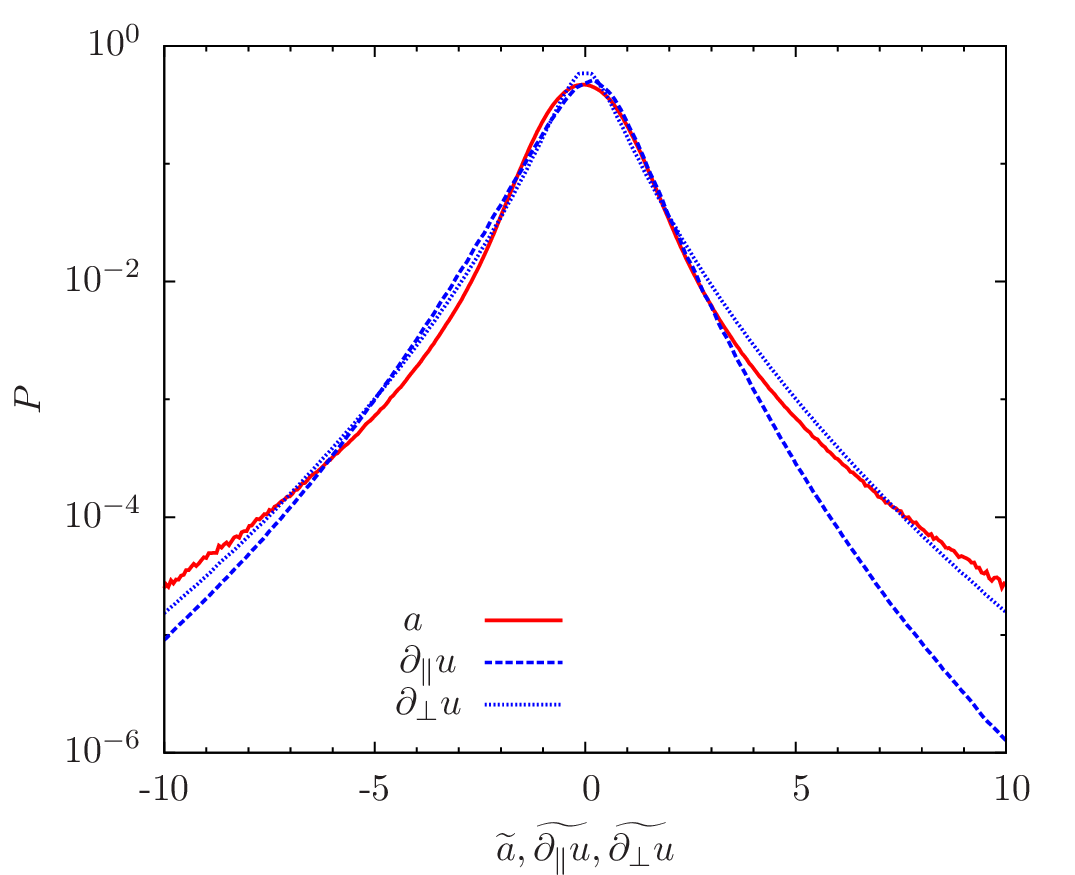}}
\caption{The normalized PDFs of the acceleration (red), 
and the longitudinal and transverse velocity gradients,  $\partial_{\parallel} u$ and $\partial_{\perp} u$.  
%at $\ell = 1.7 \eta$.
}
\label{ag} 
\end{figure}

In the small particle limit with $\tau_{\rm p} \to 0$, the generalized acceleration and the generalized shear 
contributions are related to the acceleration, ${\bs a}$, and the spatial gradient, 
$\partial_j u_i$, of the flow velocity, respectively (see \S 2). It is therefore of interest to 
compare the PDF shape of the acceleration and velocity gradients. 
Since ${\bs a} = \lim_{\Delta \tau \to 0} \Delta_{\rm L} {\bs u}/\Delta \tau$, the 
PDF of ${\bs a}$ can be obtained from the Lagrangian velocity difference $\Delta_{\rm L} {\bs u}(\Delta \tau)$ 
at small $\Delta \tau$. In our simulated flow, the PDF shape of $\Delta_{\rm L} {\bs u}(\Delta \tau)$ 
is found to be invariant at $\Delta \tau \le 0.19 \tau_\eta$, and we can 
thus measure the PDF of ${\bs a}$ at $\Delta \tau \simeq 0.19 \tau_\eta$. 
Similarly, the velocity gradient can be obtained from $\Delta { \bs u} ({\bs \ell})$  
at small ${\bs \ell}$. We consider the longitudinal ($\Delta u_{\rm r}$) and 
transverse ($\Delta u_{\rm t}$) components of  $\Delta { \bs u} ({\bs \ell})$ along 
and perpendicular to ${\bs \ell}$, and define the longitudinal and transverse velocity 
gradients as $\partial_{\parallel} u \equiv \lim_{\ell \to 0}\Delta u_{\rm r} (\ell)/\ell$ and 
$\partial_{\perp} u \equiv \lim_{\ell \to 0}\Delta u_{\rm t} (\ell)/\ell$, 
respectively. We compute the PDFs of the two gradients 
from  $\Delta u_{\rm r} (\ell) $ and $\Delta u_{\rm t} (\ell)$ across a 
distance of $\ell \simeq 1.7 \eta$, below which the PDF shape converges. In Fig.\ \ref{ag}, 
we show the measured PDFs of $a$ (red),  %measured at $\Delta \tau = 0.19 \tau_\eta$ (red), 
$\partial_{\parallel} u$  and $ \partial_{\perp} u$ %measured at $\ell = 1.7 \eta$ 
(blue), respectively. All the PDFs are normalized to have unit variance. The acceleration PDF 
shown here is for one component of ${\bs a}$. Clearly, the acceleration PDF is slightly fatter than the transverse 
velocity gradient. The longitudinal velocity gradient is skewed toward negative values,  
and,  as discussed in Paper I,  this asymmetry is related to the dissipative nature of turbulence. 
%The right PDF tail of the longitudinal velocity gradient is significantly 
%narrower than that of the acceleration.  

\section{The tangential PDFs of approaching and separating pairs at fixed Stoke ratios}

\begin{figure*}[h]
\includegraphics[height=2.9in]{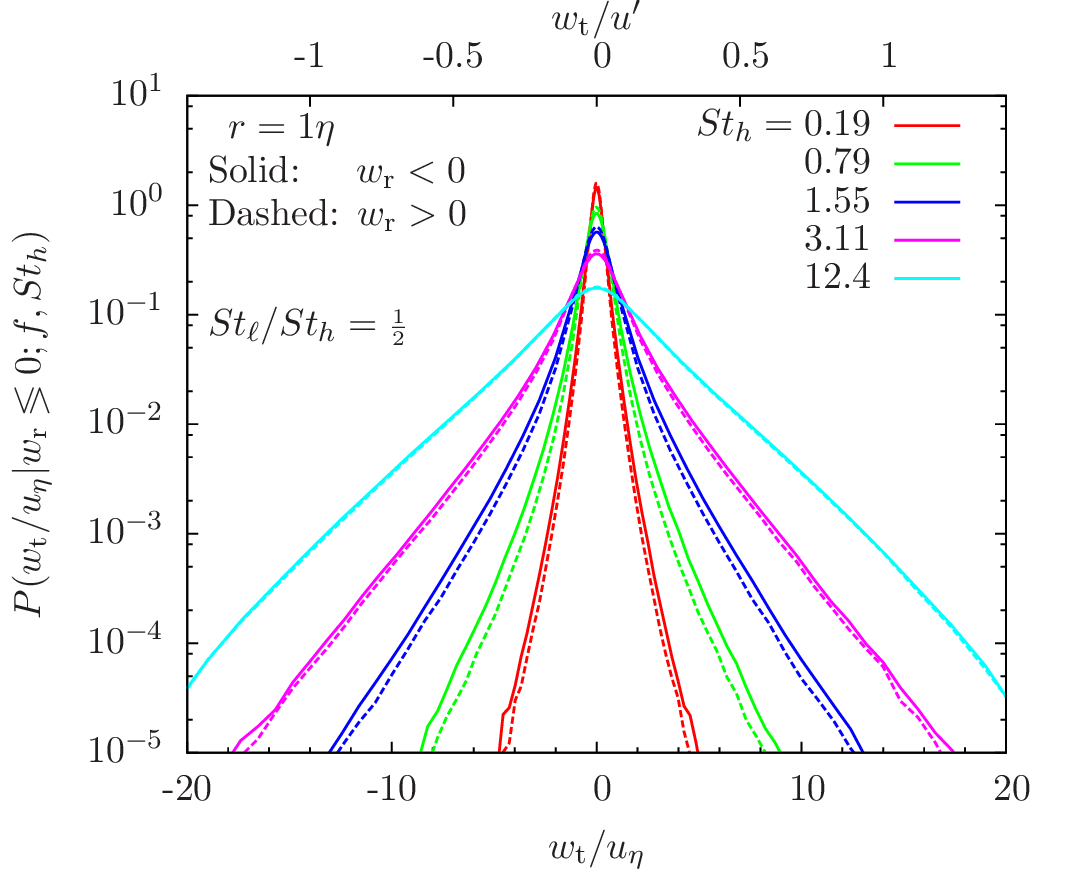}
\includegraphics[height=2.9in]{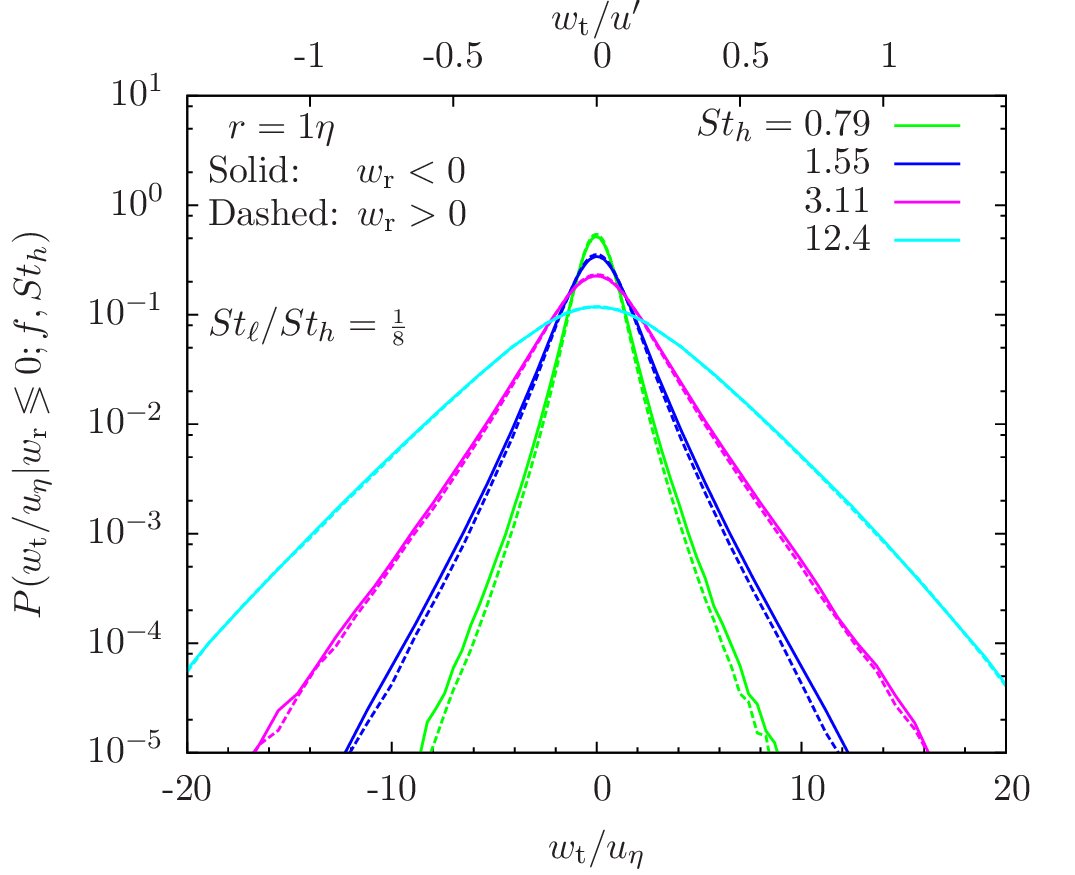}
\caption{The PDF of the tangential relative velocity, $w_{\rm t}$, for approaching (solid) and 
separating (dashed) particle pairs at a distance of $r=1\eta$. In the left and right panels, the 
Stokes number ratio is fixed at $f=\frac{1}{2}$ and $\frac{1}{8}$,  respectively. }
%{\bf  The normalizations of the particle friction time can be converted using $\Omega = St/14.4$ and $\Omega_{\rm eddy} =St/19.2$}.}
\label{tpmpdf} 
\end{figure*}

In Fig.\ \ref{tpmpdf}, we compare the PDFs, $P(w_{\rm t}|w_{\rm r} \lessgtr 0; f, St_{ h})$, 
of the tangential relative velocity for approaching (solid) and separating (dashed) particle 
pairs at a distance of $1\eta$. The left and right panels plot the results for $f=\frac{1}{2}$ 
and $\frac{1}{8}$, respectively. The two panels are plot in a similar way as the right panel 
of Fig.\ 13 in Paper I for equal-size particles ($f=1$), which 
showed that, for $St \lsim 6.21$, the PDF tails for approaching 
particles ($w_{\rm  r} <0$) are  broader than the separating ones ($w_{\rm  r} <0$). As discussed in Paper I,  
one reason for this asymmetry is that the particle distance for approaching pairs backward in 
time tends to be larger than that for separating ones, especially in the near past. 

As seen in Fig.\ \ref{tpmpdf}, the asymmetry is present also in the bidisperse case. %In \S 6.5, we pointed 
%out that the generalized acceleration contribution does not depend on the particle distance or the relative 
%motions of the two particles. Therefore, 
The difference between the tangential PDFs conditioned on $w_{\rm r} <0$ 
and $w_{\rm r} >0$ comes only from the shear contribution, as the acceleration 
contribution does not depend on the relative motions of the two particles (see \S 2). 
A comparison of the right panel of Fig.\ 13 in Paper I for the monodisperse case 
and the two panels in Fig.\  \ref{tpmpdf} here shows that, at the same $St_{ h}$, the 
difference in the tangential PDFs for $w_{\rm r} <0$ and $w_{\rm r} >0$ is smaller for smaller values 
of $f$. This is because the generalized acceleration term makes a larger contribution 
when the Stokes number difference increases. 
The presence of the acceleration contribution makes it easier 
for the tangential PDFs of approaching and separating pairs 
to equalize in the bidisperse case.  

%as the acceleration contribution completedominates. 
%This suggest that, a $St_2^{1/2}$ scaling  is expected for 
%a given $f$ if the acceleration term dominates the contribution to the relative velocity for $St_2$ in the inertial range.  
%CHECK
%We point out that, at exactly zero distance, the particle-flow relative velocity is 
%essentially 1-point statistics, and, from statistical isotropy, the PDF of $w_{\rm f}$ is 
%expected to symmetric for all particles. Since at $r=0$ the generalized acceleration term 
%is the only contribution to the particle-flow relative velocity, this suggests that the PDF from 
%the acceleration contribution is alway symmetric. 
%The fattening of the PDF tails of the temporal velocity difference with decreasing time lag has been 
%useful in explaining the trend of  the PDF shape for the particle-flow relative speed as a function of 
%the particle inertia (\S 5) and for the bidisperse  relative velocity (\S 6.2).

\end{document}